\documentclass[aps,twocolumn,prc,superscriptaddress,showpacs,nofootinbib,floatfix,amssymb,amsfonts,amsmath]{revtex4-1}
\usepackage{graphicx}
\usepackage{dcolumn}
\usepackage{bm}
\usepackage{amsmath}    
\usepackage{amsfonts}   
\usepackage{amssymb}
\usepackage{graphicx}   
\usepackage{color}
\usepackage{ltxcmds}
\pdfoutput=1

\begin{document}

\title{Hyperheavy spherical and toroidal nuclei: 
the role of shell structure.}

\author{S.\ E.\ Agbemava}
\affiliation{Department of Physics and Astronomy, Mississippi
  State University, MS 39762}
\affiliation{Ghana Atomic Energy Commission, National Nuclear Research
  Institute, P.O.Box LG80, Legon, Ghana.}

\author{A.\ V.\ Afanasjev}
\affiliation{Department of Physics and Astronomy, Mississippi
State University, MS 39762}

\date{\today}

\begin{abstract}
The properties of toroidal hyperheavy even-even nuclei and the role of toroidal shell structure 
are extensively studied within covariant density functional theory. The general 
trends in the evolution of toroidal shapes in the $Z\approx 130-180$ region of nuclear
chart are established for the first time. These nuclei are stable with respect of breathing
deformations. The most compact fat toroidal nuclei are located in the 
$Z\approx 136, N\approx 206$ region of nuclear chart, but thin toroidal nuclei become
dominant with increasing proton number and on moving towards proton and neutron
drip lines.  The role of toroidal shell structure, its regularity, supershell structure, shell
gaps as well as the role of different groups of the pairs of the orbitals in its formation 
are investigated in detail. The lowest in energy solutions at axial symmetry are 
characterized either by large shell gaps or low density of the single-particle states in the
vicinity of the Fermi level in at least one of the subsystems (proton or neutron). Related
quantum shell effects are expected to act against the instabilities in breathing and
sausage deformations for these subsystems. The investigation with large set of covariant 
energy density functionals reveals that substantial proton $Z=154$ and 186 and neutron $N=228$, 308 
and 406 spherical shell gaps exist in all functionals. The nuclei in the vicinity of the
combination of these particle numbers form the islands of stability of spherical hyperheavy
nuclei.  The study suggests that the $N=210$ toroidal shell gap plays a substantial
role in the stabilization of fat toroidal nuclei.
\end{abstract}


\maketitle

\section{Introduction}
\label{introduction}

\begin{figure*}[htb]
\centering
\includegraphics[angle=0,width=12cm]{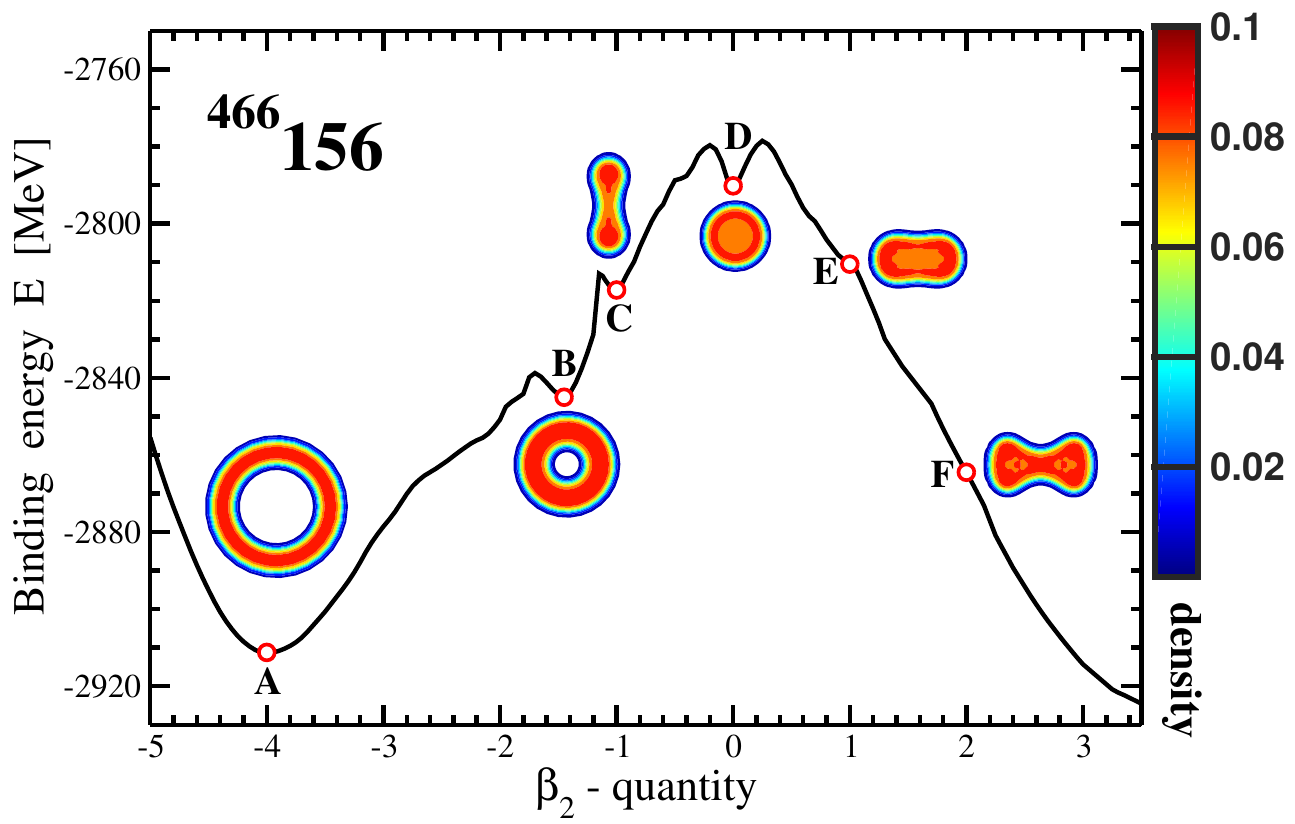}
\caption{
Schematic illustration of the physics of hyperheavy nuclei. Solid black line shows 
the deformation energy curve of the $^{466}156$ nucleus obtained in axial RHB 
calculations with CEDF DD-PC1 in Ref.\ \cite{AAG.18}.  Open red circles 
indicate selected points on 
this curve for which neutron density distributions $\rho_n$ are shown. The density
distributions in the minima A and B (C, E, and F) are shown in the plane which is 
perpendicular (along) the axis of symmetry. The density 
colormap starts at $\rho_n=0.005$ fm$^{-3}$ and shows the densities in fm$^{-3}$.
The density profiles reflect their relative sizes with respect of the spherical shape 
in the minimum D. See Fig.\ \ref{densit-tori} in the present paper and Fig.\ 2 in Ref.\ 
\cite{AATG.19} for these density profiles in their actual sizes. 
}
\label{156-466-pot-b}
\end{figure*}

   The studies of the nuclei at the limits are guided by human curiosity,
by the need to understand new physical mechanisms governing nuclear
systems in these extreme conditions and by the demand for nuclear
input in nuclear astrophysics. A number of questions related to the 
physics at the limits emerge. These are: What are the limits 
of the existence of nuclei? What are the highest proton number $Z$ at 
which the nuclear landscape and periodic table of chemical elements 
cease to exist? What are the positions of proton and neutron drip 
lines?  What types of nuclear  shapes dominate these extremes of nuclear 
landscape? They look deceivable simple but unique answers on most of
them are extremely difficult.

\begin{figure*}[htb]
\centering
\includegraphics[angle=0,width=18.0cm]{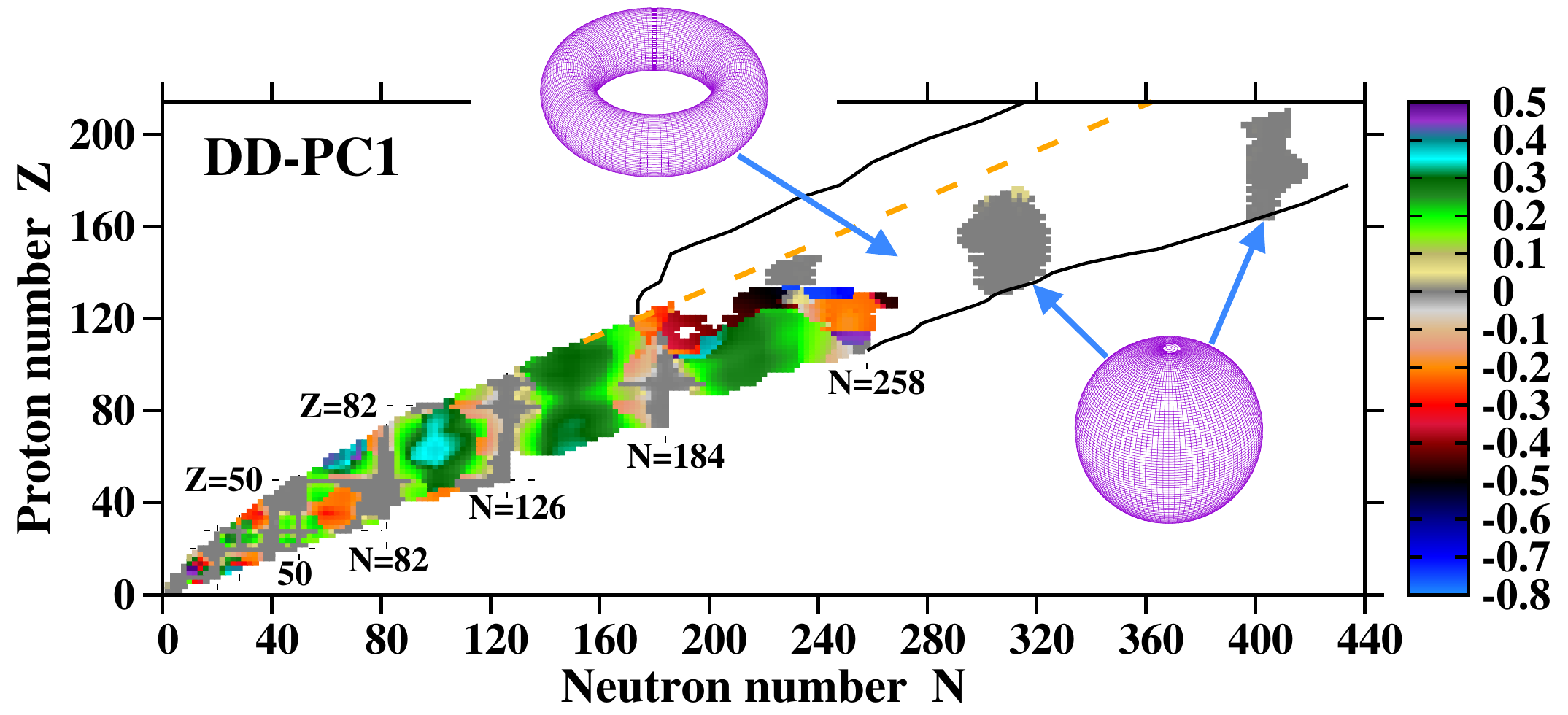}
\caption{
The distribution of ellipsoidal and toroidal shapes in the nuclear landscape obtained in 
the RHB calculations with CEDF  DD-PC1. The nuclei with ellipsoidal shapes are shown by the squares 
the color of which indicates the equilibrium quadrupole deformation $\beta_2$ 
(see colormap).  Note that ellipsoidal shapes with the heights of fission barriers 
smaller than 2.0 MeV are considered as unstable (see the discussion in Sect. III
of Ref.\ \cite{AAR.12} and in Sect.\ XI in Ref.\ \cite{AATG.19}).  Two-proton and two-neutron drip 
lines for toroidal nuclei are shown by solid black lines.  White region between them (as well as the 
islands inside this region shown in gray) corresponds to the nuclei  which have 
toroidal shapes in the lowest in energy minimum for axial symmetry (LEMAS).
The islands of relatively stable spherical hyperheavy nuclei in the  $Z>130$ nuclei, 
shown in light grey color, correspond  to the solutions which are excited in energy 
with respect of the LEMAS corresponding to toroidal shapes. Note that in the same nucleus 
two-neutron drip lines for spherical and toroidal 
shapes are somewhat different. This is the reason why some islands of stability of spherical 
hyperheavy nuclei extend beyond the two-neutron drip line for toroidal shapes. 
The extrapolation of the two-proton drip for ellipsoidal shapes, defined from 
its general trends seen in the $Z<120$ nuclei, is displayed by thick orange 
dashed lines. Similar extrapolation for two-neutron drip line of ellipsoidal
shapes is close to the two-neutron drip line of toroidal shapes (see Fig.\ 1 in Ref.\ \cite{AAT.20});
thus it is not shown. Partially based on Fig. 24 of Ref.\ \cite{AATG.19}.
}
\label{landscape}
\end{figure*}

  Recent systematic investigations of hyperheavy ($Z\geq 126$) nuclei
performed in Refs.\ \cite{AAG.18,AATG.19,AAT.20}  have allowed to shed 
some light on these questions. Emerging new physics is summarized in Figs.
\ref{156-466-pot-b} and \ref{landscape}. The increase of Coulomb interaction 
with increasing proton number $Z$ leads to the fact that compact nuclear shapes
such as spherical, prolate and oblate (further ellipsoidal-like shapes) become 
either unstable against fission or energetically unfavored in hyperheavy nuclei with 
high $Z$ values (see Fig.\ \ref{156-466-pot-b}). As a consequence, the lowest in 
energy  solutions in such nuclei are characterized by non-compact toroidal
shapes\footnote{Toroidal nucleus is represented by a thin cylinder which has the 
ends joining together \cite{Wong.73}.}.  As illustrated in Fig.\ \ref{landscape} the boundary between  ellipsoidal-like  
and toroidal shapes depend on the combination of proton and neutron numbers. 
However,  spherical  shapes  can be stable against fission in some hyperheavy  
nuclei (see  Refs.\ \cite{AAG.18,AATG.19} and Fig.\ \ref{156-466-pot-b}).
Although these states are  highly excited with respect to the lowest in energy states 
with toroidal shapes (as obtained in axial calculations), they will become the ground  
states if toroidal states are not stable with respect to multifragmentation.

     The state-of-the-art view on the nuclear landscape born out in Refs.\ 
\cite{AAG.18,AATG.19} is shown in Fig.\ \ref{landscape}.  Well known 
nuclear structure with pronounced spherical shell gaps at particle numbers
8, 20, 28, 50, 82 (and $N=126$) leading to the bands (shown by gray
color) of spherical nuclei in the nuclear chart along the vertical and 
horizontal lines with these particle numbers is seen for proton numbers 
below $Z\approx 120$. With increasing proton number  these classical features 
disappear and only toroidal shapes
are calculated as the lowest 
in energy in axial relativistic Hartree-Bogoliubov (RHB) approach. This 
region (shown in white color in Fig.\ \ref{landscape}) is penetrated only
by three islands (shown in gray color) of potentially stable spherical 
hyperheavy nuclei; note that spherical minima are highly excited with 
respect of the minima corresponding to toroidal shapes. Thus, the richness 
of nuclear structure seen in experimentally known part of nuclear landscape 
is replaced by more uniform structure of the nuclear landscape in the region 
of hyperheavy nuclei dominated by toroidal and spherical nuclei. Fig.\ 
\ref{landscape} also reveals a substantial increase (equal to the area 
between extrapolated two-proton drip line for ellipsoidal shapes and the 
two-proton drip line for toroidal shapes) of nuclear landscape caused by the 
shift of two-proton drip line towards more proton-rich nuclei on transition to 
toroidal shapes.
 This transition drastically modifies the underlying single-particle
structure and as a consequence lowers the energy of the Fermi level 
for protons (see Ref.\ \cite{AAT.20}). 
   
  It is necessary to recognize that the physics of toroidal shapes plays
an important role in classical and quantum physics, chemistry 
and biology. There are numerous examples but let us mention only some
of them. Stable toroidal structures (micelles) play an important role in 
the  amphiphilic polymers in large parts of the parameter space spanned
by the degree of amphiphilicity, the temperature, the density and the 
molecular stiffness with respect to bending \cite{JL.20}. The wave 
propagation on the surface of the torus represents a vivid
example of light behavior on curved surface of manifolds with
interesting topologies and has potential applications in photonic
structures \cite{WLLHZ.18}.  The stability of toroidal drop freely suspended
in another fluid and subjected to an electric field has been studied
in Ref.\ \cite{Z.17}; this feature can play a role in a number of phenomena 
and applications such as thunderstorm formation, microfluids, bioimaging 
and effective drug delivery.  Biology finds the toroidal shape at the cellular
level when the reproduction of cells up to the 16$^{th}$ cell division creates 
a hollow torus called the morula \cite{D-med.09}. On a more microsocopic
level, the DNA toroids are formed from individual DNA molecules of individual
lengths \cite{LL.09}.

   The question of potential stability of toroidal nuclei has first been 
raised  by J. A. Wheeler (see references in Ref.\ \cite{Wong.73}). Later 
the toroidal shapes in atomic nuclei have been investigated in a  number  
of the  papers (see, for example, Refs.\ 
\cite{Wong.73,Warda.07,StaW.09,SW.14,IMMI.14,KSW.17,NBCHKRV.02}
and references quoted therein). However, in absolute majority of the 
cases such shapes correspond to highly excited states  either at extreme 
values of angular momentum  in the
nuclei across the nuclear landscape \cite{SW.14,IMMI.14,SWK.17} or at spin 
zero in superheavy elements \cite{StaW.09,KSW.17}.
In the former case, calculated angular momenta at which toroidal shapes 
appear substantially exceed the values of angular momentum presently 
achievable at the state-of-art experimental facilities \cite{PhysRep-SBT}. 
So far, only the experimental excitation function for the $7\alpha$ de-excitation
of $^{28}$Si  nuclei, revealing the resonance structures, may indicate the 
population of toroidal high-spin isomers \cite{28Si-tori.exp}. In the latter case, 
such states are unstable in superheavy nuclei against returning to the shape of sphere-like geometry 
(Ref.\ \cite{KSW.17}). This is similar to shrinking instability of uncharged toroidal 
droplets which are unstable due to surface tension and transform into spherical 
droplets \cite{FPBSF.17}. The situation is different in atomic nuclei since this
shrinking instability is counteracted by Coulomb repulsion of the protons which
increases with proton number $Z$.  Thus, toroidal shapes become the lowest 
in energy solutions in hyperheavy nuclei with $Z>130$ 
\cite{Warda.07,AAG.18,AATG.19}.  
   
   The present paper extends our previous investigations of hyperheavy
nuclei reported in Refs.\ \cite{AAG.18,AATG.19} and focuses on a number of
issues which have not been studied so far.  The presence of local minima 
A, B, C and D in deformation energy curve of the $^{466}$156 nucleus (see
Fig.\ \ref{156-466-pot-b})  is clearly due to the shell effects. So far, the 
underlying single-particle structure has been investigated only for spherical 
shapes and only for four covariant energy density functionals (CEDFs)
(see Sec. V in Ref.\ \cite{AATG.19}).  To estimate theoretical uncertainties
in the predictions of shell closures in hyperheavy nuclei at spherical shape 
we  perform such studies with ten most widely used CEDFs. This also allows 
us to compare respective spherical shell gaps, leading to the islands 
of potentially stable spherical hyperheavy nuclei, with the ones seen in 
experimentally known nuclei as well as with those predicted for spherical 
superheavy nuclei. In addition, for the first time we perform the detailed
investigation of the single-particle structure of hyperheavy toroidal nuclei.

   The analysis of the single-particle structure presented in Figs. 5 and 8 of Ref.\   
\cite{AATG.19} indicates the presence of large spherical shell gaps at $Z=186$ and
$N=406$. However, the investigations of Ref.\ \cite{AATG.19} have been
restricted to the $Z\leq 180$ nuclei. Thus, to better map this region of potentially
stable spherical hyperheavy nuclei, to investigate the potential role of these 
shell gaps as well as to search for other regions of potentially stable 
spherical hyperheavy nuclei we extended the calculations mapping the nuclear
landscape from $Z=180$ to $Z=210$. 

   Finally, because of numerical limitations the  studies of toroidal shapes in 
hyperheavy nuclei have been with a single exception restricted to the 
$Z\leq 138$ nuclei in Refs.\ \cite{AAG.18,AATG.19}. Thus, 
we performed detailed investigation of toroidal shapes corresponding to 
the lowest in energy solution at axial symmetry in extremely large basis
 for isotopic chains with $Z=136$, 146, 156, 166 and 176. This allows us to 
better understand their evolution with particle numbers and 
to get some understanding about their potential stability with respect of different 
types of distortions.

   The manuscript is organized as follows.  The details of
theoretical calculations are discussed in Sec.\ \ref{theory}. 
Section \ref{shell-sphere} is devoted to the analysis of the role 
of shell structure and large shell gaps at spherical shape. The 
distribution of the shapes of toroidal hyperheavy nuclei  across 
the nuclear landscape and major features of their shell structure 
are discussed in Sec.\ \ref{shell-tori}. Finally, Sec. \ref{summary} 
summarizes the results of our work.

\section{The details of the theoretical calculations.}
\label{theory}

 The investigations of the properties of hyperheavy  even-even nuclei are 
performed within the axial reflection symmetric Hartree-Bogoliubov (RHB) 
framework (see Ref.\ \cite{AARR.14}).  Until specified otherwise, the calculations 
are  performed with the DD-PC1 covariant energy density functional (CEDF)
\cite{DD-PC1}.  This functional is considered to be one of  the best CEDFs
today based on systematic and global studies of different physical 
observables related to the ground state properties and fission barriers 
\cite{AANR.15,AARR.14,AARR.17,AA.16,PNLV.12,LZZ.12,AAR.16}.

  The constrained calculations in the RHB code perform the variation of 
the function
\begin{equation}
E_{RHB} + \frac{C_{20}}{2} (\langle\hat{Q}_{20}\rangle-q_{20})^2,
\end{equation}
where $E_{RHB}$ is the total energy and $\langle\hat{Q}_{20}\rangle$ denotes the expectation
value of the mass quadrupole operator, 
\begin{equation}
\hat{Q}_{20}=2z^2-x^2-y^2.
\label{Q_20-quad}
\end{equation}
Here $q_{20}$ is the constrained value of the multipole moment, and
$C_{20}$ the corresponding stiffness constant~\cite{RS.80}.
In order to provide the convergence to the exact value
of the desired multipole moment we use the method suggested in
Ref.~\cite{BFH.05}. Here the quantity $q_{20}$ is replaced by the
parameter $q_{20}^{eff}$, which is automatically modified during
the iteration in such a way that we obtain
$\langle\hat{Q}_{20}\rangle = q_{20}$ for the converged solution.
This method works well in our constrained calculations.

  The $\beta_2$ quantity is extracted from the quadrupole moments:
\begin{eqnarray}
Q_{20} &=& \int d^3r \rho({\bm r})\,(2z^2-x^2-y^2),
\end{eqnarray}
via
\begin{eqnarray}
\beta_2 &=&  \sqrt{\frac{5}{16\pi}} \frac{4\pi}{3 A R_0^2} { Q_{20}},
\end{eqnarray}
where $R_0=1.2 A^{1/3}$.
The $\beta_2$ values have a standard meaning of the deformations of ellipsoid-like density
distributions only for $|\beta_2|\lesssim 1.0$ values. At higher $\beta_2$ values they should
be treated as dimensionless and particle normalized measures of the $Q_{20}$ moments.
This is because of the presence of toroidal shapes at large negative $\beta_2$ values and
necking degree of freedom at large positive $\beta_2$ values.

\begin{figure*}[htb]
\centering
\includegraphics[angle=0,width=8.0cm]{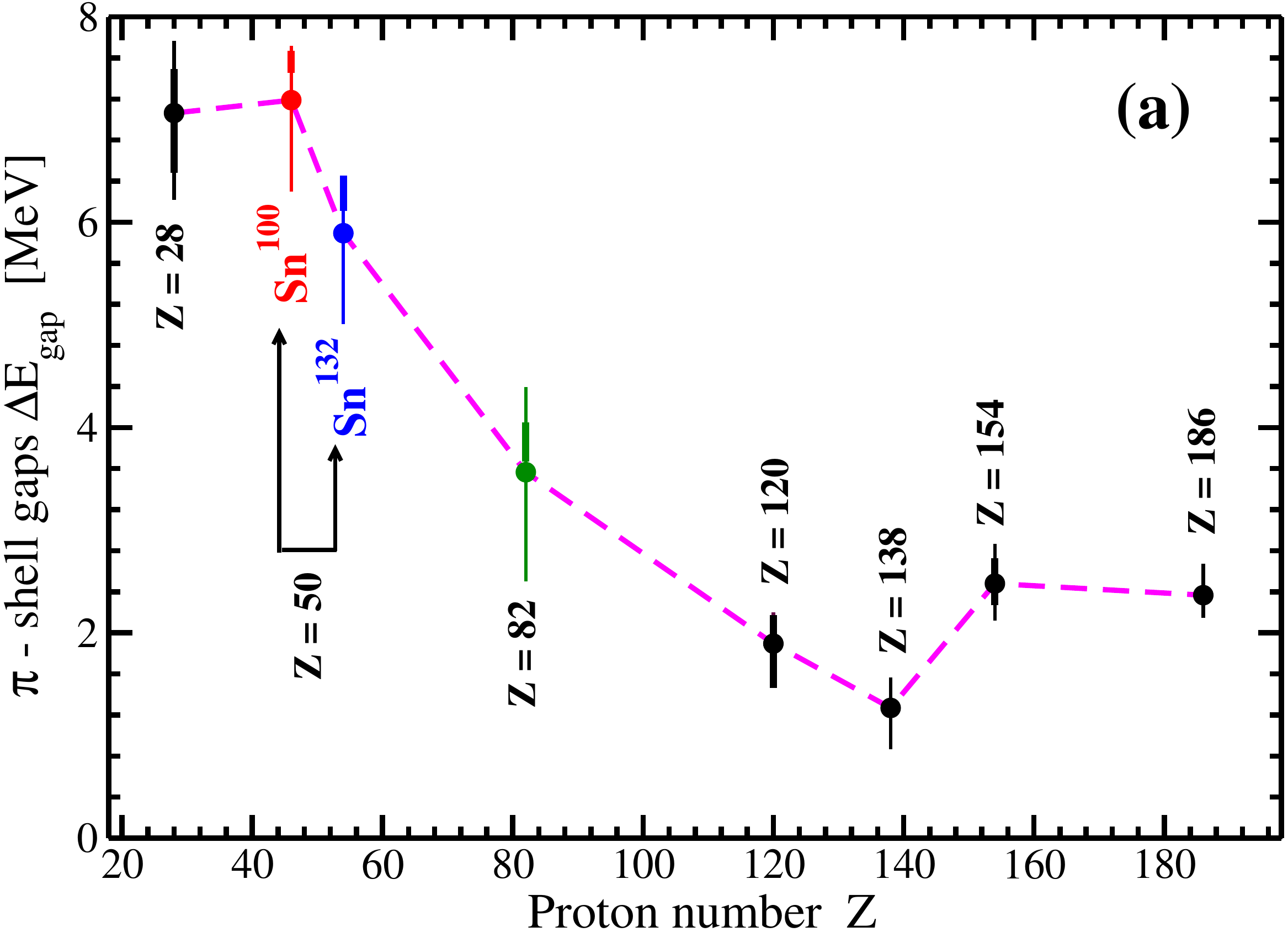}
\includegraphics[angle=0,width=8.0cm]{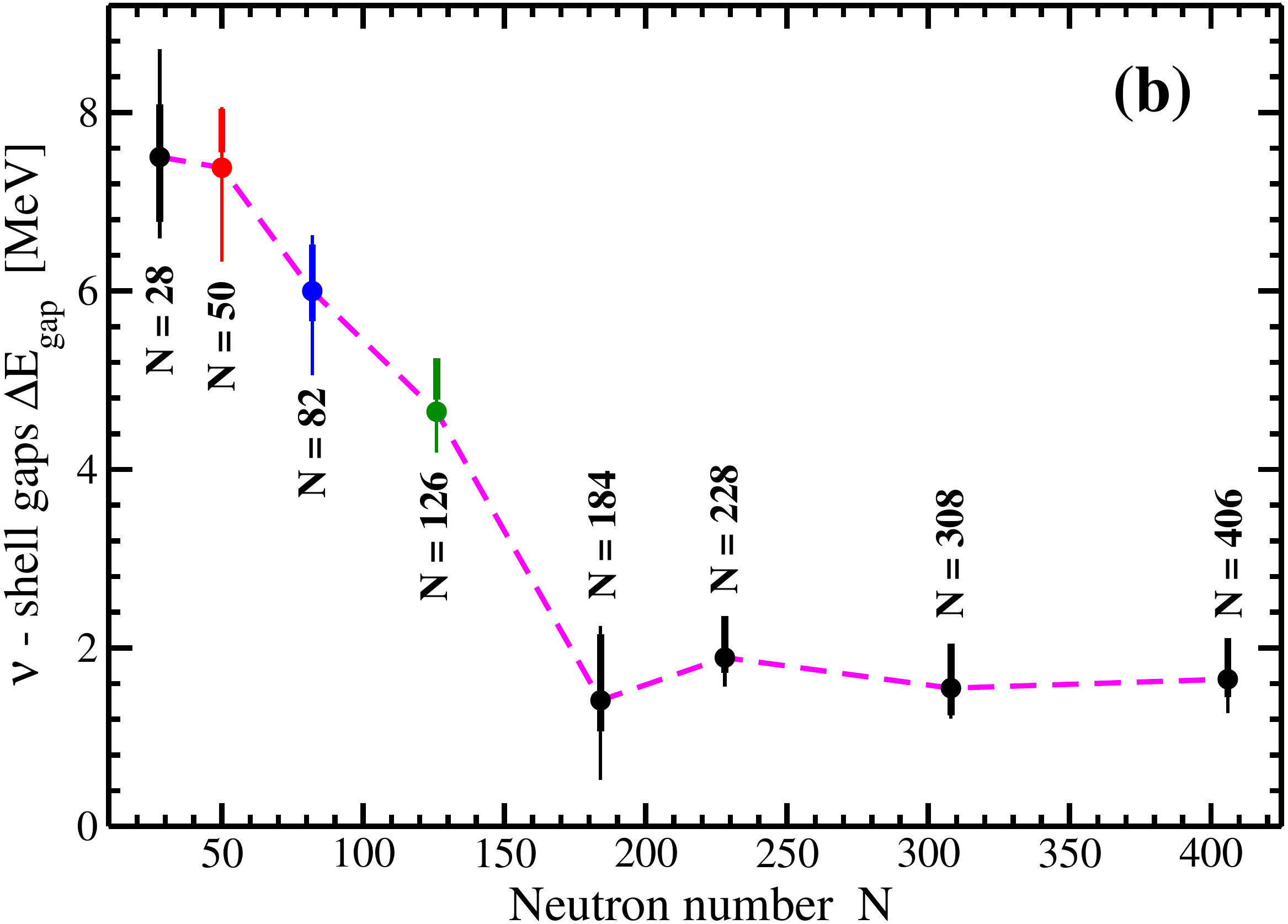}
\includegraphics[angle=0,width=8.0cm]{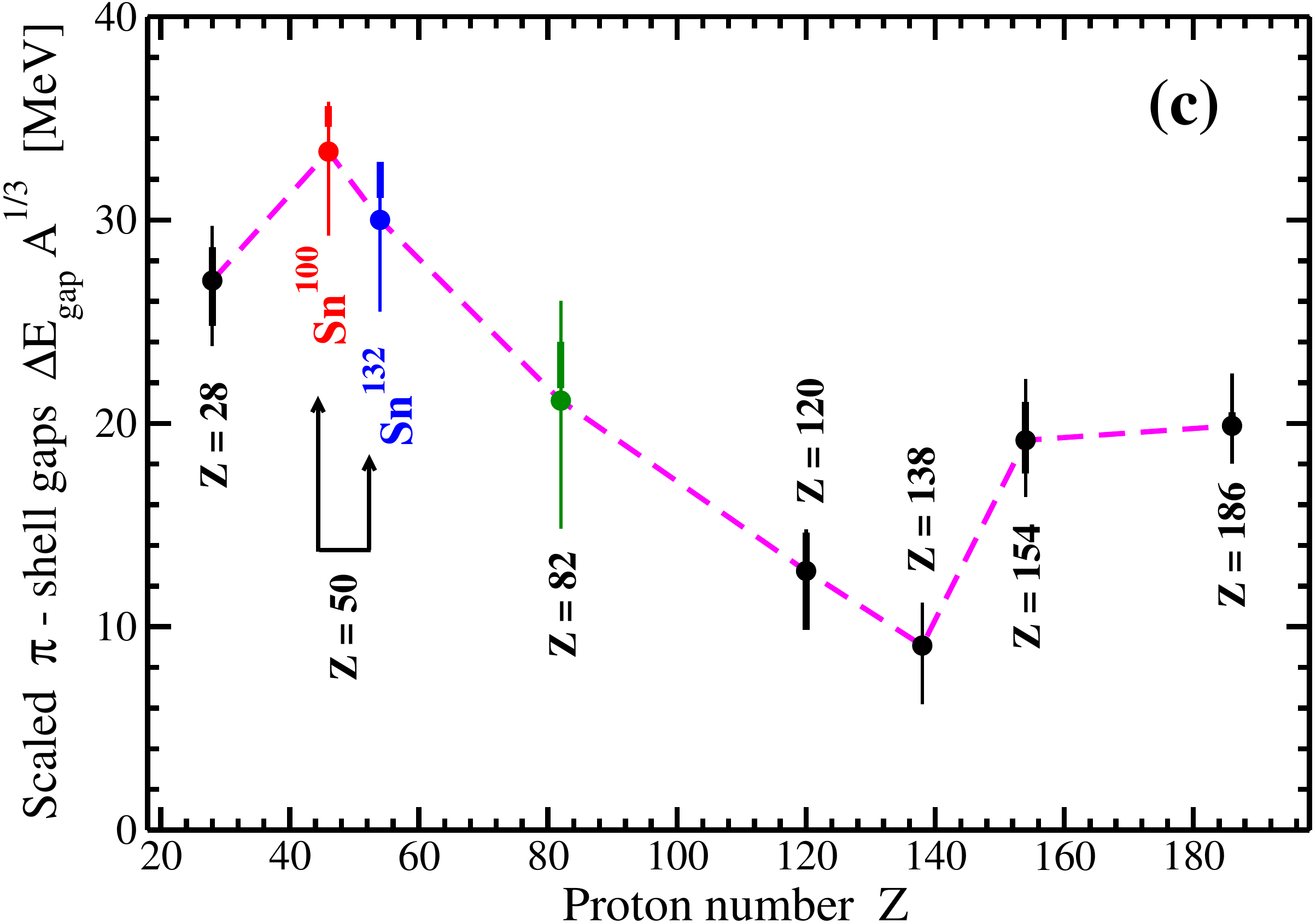}
\includegraphics[angle=0,width=8.0cm]{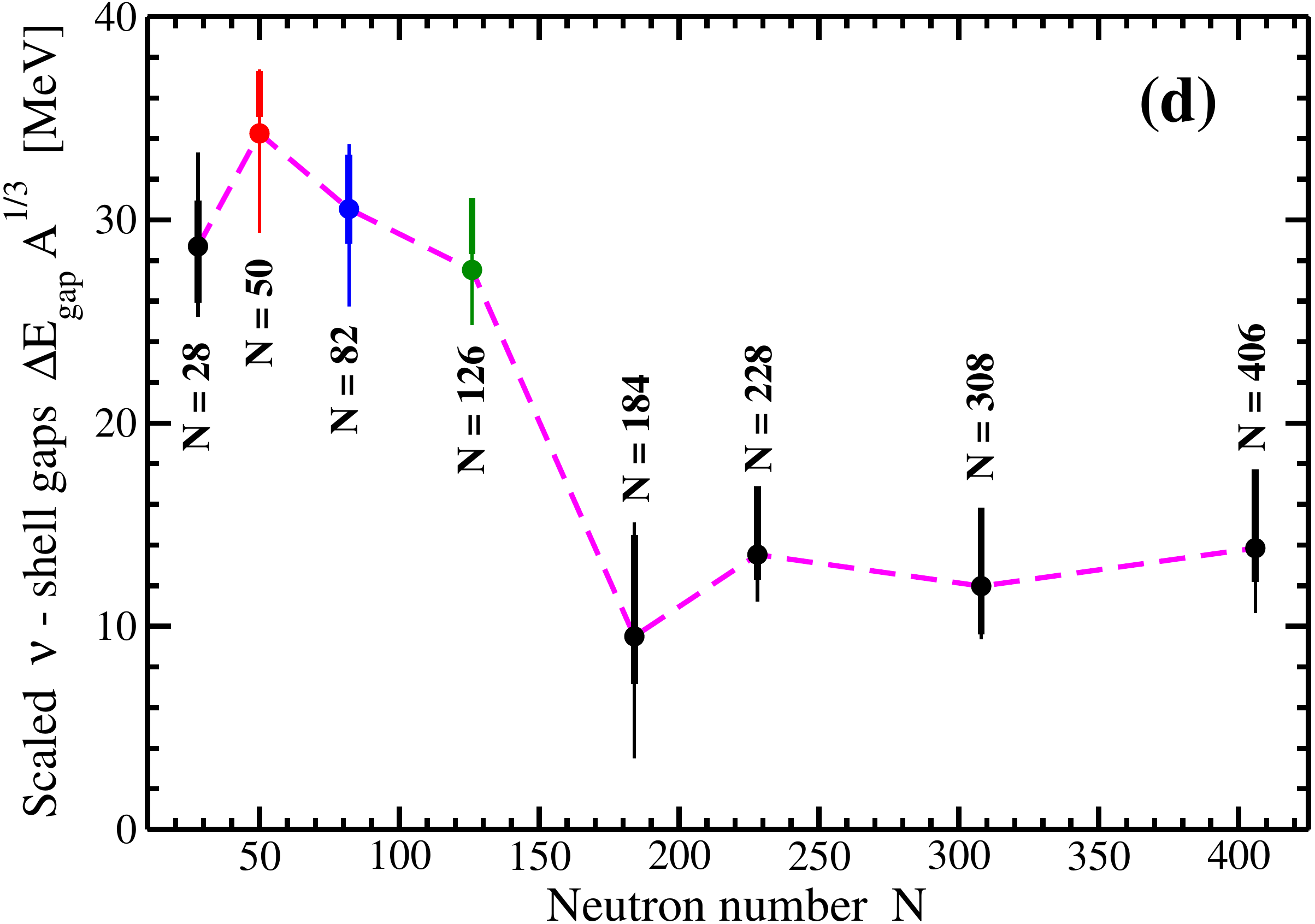}
\caption{(a) and (b) Calculated proton ($\pi$) and neutron  $(\nu)$ shell gaps
$\Delta {E_{gap}}$ in the doubly magic $^{58}$Ni, $^{100,132}$Sn and 
$^{208}$Pb nuclei and shell-closure superheavy $^{304}$120 and 
hyperheavy $^{366}$138, $^{462}$154 and $^{592}$186 nuclei.  Note that
last three nuclei are located in the centers of the islands of stability of
spherical hyperheavy nuclei. Particle numbers corresponding to shell 
gaps are indicated.  Ten most widely used  CEDFs, namely,
NL1 \cite{NL1}, NL3 \cite{NL3}, NL3* \cite{NL3*}, FSUGold \cite{FSUGold},
DD-ME2 \cite{DD-ME2}, DD-ME$\delta$ \cite{DD-MEdelta}, DD-PC1 \cite{DD-PC1}, 
PC-PK1 \cite{PC-PK1}, PC-F1 \cite{PC-F1} and TM1 \cite{TM1}
are employed in the calculations.
The average (among ten used CEDFs) size of the shell gap is shown by a solid circle
while the gaps obtained for individual functionals are summarized in Table
\ref{table-FB-gaps}.
Thin and thick vertical lines are used to show the spread of the sizes of
the calculated shell gaps; the tops and bottoms of these lines correspond
to the upper and lower shell gaps among the considered set of CEDFs.
Thin lines show this spread for all employed CEDF's, while thick lines are
used for the subset of four globally-tested CEDFs (NL3*, DD-ME2, DD-PC1, 
and PC-PK1). (c) and (d) The same as in panels (a) and (b) respectively, but 
with the sizes of the shell gaps and the spreads in their predictions scaled
with mass factor $A^{\frac{1}{3}}$.
}
\label{shell-gaps}
\end{figure*}

 For each nucleus under study, the deformation energy curves are calculated 
in the $-5.0 < \beta_2 < 3.0$ range; such large range is needed for a reliable 
definition of the type of shape (toroidal or ellipsoidal) representing
the lowest in energy minimum for axial symmetry (LEMAS). Two truncation 
schemes are used in the calculations based on the analysis presented
in Sec. III of Ref.\ \cite{AATG.19}  and additional analysis performed 
in this manuscript. All states belonging to major shells up to 
$N_F$=30 fermionic shells for the Dirac spinors are taken into account when 
detailed analysis of toroidal shapes in the $Z=136-176$ region and their 
underlying shell structure is performed. Note that these calculations are extremely 
time-consuming. As discussed in detail in Sec. III of Ref.\ \cite{AATG.19} on 
the example of the $^{466}$156 nucleus and verified by a similar analysis of a 
pair of the $Z=176$ nuclei, this basis provides sufficient numerical accuracy of 
the calculations of toroidal shapes. However,  the analysis
of numerical convergence in the $^{616}210$ nucleus reveals  that the description of higher 
$Z$ nuclei requires even large $N_F$ for a proper description of LEMAS corresponding 
to toroidal shapes. These facts were the reasons why we perform detailed study of 
toroidal shape only up to $Z=176$ and  for $Z>176$ nuclei we focus mainly on 
ellipsoidal-like shapes  which require smaller basis as compared with toroidal 
shapes (see Sec. III of Ref.\ \cite{AATG.19} for a detailed comparison of numerical
convergence for toroidal and ellipsoidal-like shapes). To save computational time the 
extension (as compared with the results presented in Ref.\ \cite{AATG.19}) of nuclear  
landscape to the $Z=182-210$ region is performed with $N_F=26$;
this truncation scheme allows accurate description of spherical and 
ellipsoidal shapes, reliable definition of toroidal shapes as corresponding 
to LEMAS but does not provide accurate enough description of their
energies and shapes in LEMAS.

   To avoid the uncertainties connected with the definition of the size of the pairing window
\cite{KALR.10}, we use the separable form of the finite-range Gogny pairing interaction
introduced in Ref.\ \cite{TMR.09}. Its matrix elements in r-space have the form
\begin{eqnarray}
\label{Eq:TMR}
V({\bm r}_1,{\bm r}_2,{\bm r}_1',{\bm r}_2') &=& \nonumber \\
= - G \delta({\bm R}-&\bm{R'}&)P(r) P(r') \frac{1}{2}(1-P^{\sigma})
\label{TMR}
\end{eqnarray}
with ${\bm R}=({\bm r}_1+{\bm r}_2)/2$ and ${\bm r}={\bm r}_1-{\bm r}_2$
being the center of mass and relative coordinates.
The form factor $P(r)$ is of Gaussian shape,
\begin{eqnarray}
P(r)=\frac{1}{(4 \pi a^2)^{3/2}}e^{-r^2/4a^2}.
\end{eqnarray}
The parameters of this interaction have been derived
by a mapping of the $^1$S$_0$ pairing gap of infinite nuclear
matter to that of the Gogny force D1S. The resulting
parameters are: $G=728$ fm$^3$ and $a=0.644$ fm\ \cite{TMR.09}.
This pairing provides a reasonable description of pairing properties in 
heaviest nuclei in which pairing properties can be extracted from 
experimental data \cite{AARR.14, AO.13, DABRS.15}.

\section{Spherical hyperheavy nuclei: the role of shell structure}
\label{shell-sphere} 
   
    Hyperheavy nuclei are stabilized by shell effects, i.e., by the large shell gap(s)
or at least a considerably reduced density of the single-particle states in the vicinity
of the Fermi level. To 
better understand the impact of shell gaps on the underlying structure of 
spherical nuclei in the context of global description of nuclear
structure,   Fig.\ \ref{shell-gaps} shows their evolution across nuclear chart.
It starts from well known gaps in doubly magic $^{56}$Ni, $^{100,132}$Sn 
and $^{208}$Pb nuclei and extends to the gaps in the hyperheavy nuclei. 
In addition, it provides the evaluation of theoretical uncertainties in their 
predictions by comparing the results obtained with ten most widely used CEDFs.

   Figs.\ \ref{shell-gaps}a and b  show  that the average sizes of proton $Z=154, 186$ and 
 neutron $N=228, 308$ and 406 gaps obtained in the calculations are larger than those 
 ($Z=120$ and $N=184$) in classical region of  superheavy nuclei\footnote{Note that the central 
 nucleus of the $Z\approx 138, N\approx 230$ island of stability of spherical hyperheavy nuclei 
 does not really show $Z=138$ shell gap in proton spectra (see discussion in Sect.\ V of Ref.\ 
 \cite{AATG.19}).}. This suggests that spherical hyperheavy nuclei may be more stable 
 as compared with spherical superheavy nuclei (see the discussion of fission barriers in 
 Refs.\ \cite{AAG.18, AATG.19}). It is also interesting that theoretical uncertainties in
 the sizes of shell gaps in hyperheavy nuclei are smaller than those in experimentally
 known nuclei and in classical region of superheavy nuclei.

   The absolute values of shell gaps do not tell full story about their
potential stabilizing effect since the single-particle level density increases 
with mass number $A$.  This is a reason why scaled shell gap 
$\Delta E_{gap} A^{1/3}$ provides a better measure 
(see discussion in Sect. III of Ref.\ \cite{AANR.15}).
 Scaled proton and neutron shell gaps are shown
in Figs.\ \ref{shell-gaps}(c) and (d). One can see that 
scaled proton $Z=154$ and 186 shell gaps are significantly 
larger  than scaled $Z=120$ shell gap  in superheavy 
nuclei  and that they are close to the scaled $Z=82$ shell gap in $^{208}$Pb
(see Fig.\ \ref{shell-gaps}(c)). On the contrary, scaled $N=228$, 308 and 406 
shell gaps are on average only slightly larger than scaled  $N=184$ gap in 
superheavy nuclei but they are smaller by a factor of approximately two than 
scaled $N=126$ shell gap in $^{208}$Pb (see Fig.\ \ref{shell-gaps}(d)).

Large uncertainties in the predictions of the $Z=120$ and $N=184$ 
shell gaps and softness of potential energy surfaces leads to substantial 
differences in the predictions of ground state properties of superheavy 
nuclei (see Ref.\ \cite{AANR.15}). For many nuclei it is even impossible to reliably predict whether 
the ground state will be spherical or oblate \cite{AANR.15}.  The situation
is different in hyperheavy nuclei where for ellipsoidal type shapes only 
potentially stable spherical minima appear in the calculations because 
of larger scaled spherical shell gaps seen in Figs.\ 
\ref{shell-gaps}(c) and (d).

\begin{figure*}[htb]
\centering
\includegraphics[angle=-90,width=16.0cm]{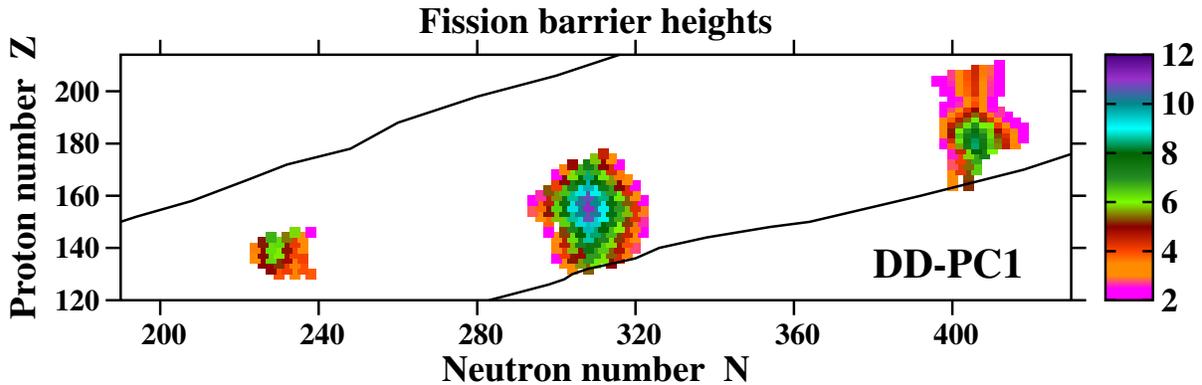}
\caption{The fission barrier heights $E_B$ [in MeV] as a function of proton and neutron 
numbers. Only the nuclei  with fission barriers higher than 2 MeV are shown.  Partially 
based on the results presented in  Fig. 6a of Ref.\ \cite{AAG.18}.
}
\label{Fiss-barr}
\end{figure*}

\begin{figure}[htb]
\centering
\includegraphics[angle=0,width=8.5cm]{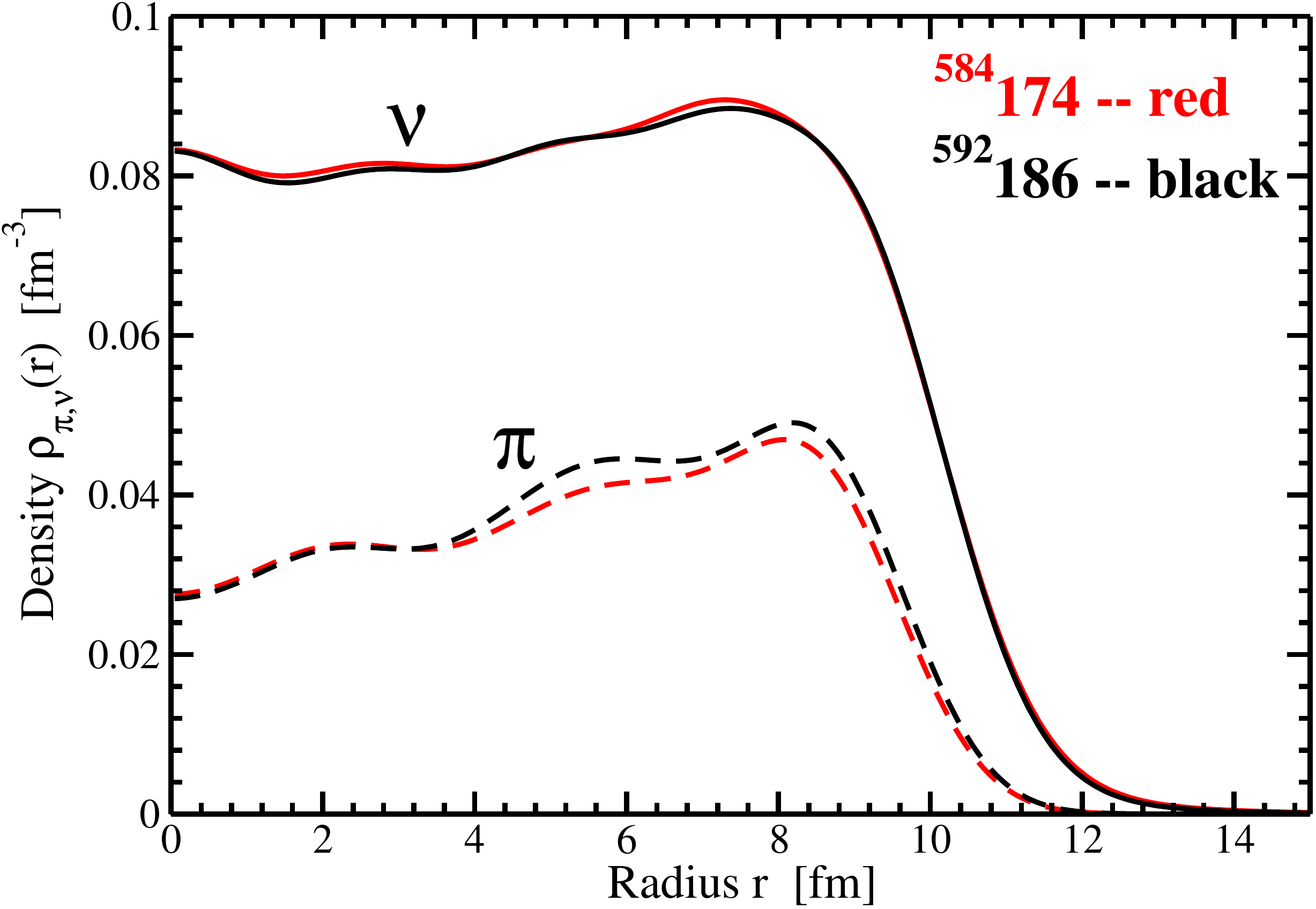}
\caption{Proton and neutron densities of the $^{584}174$ and $^{592}186$
nuclei. The figure is based on the results of spherical RHB calculations. 
}
\label{densit}
\end{figure}

  Fig. \ref{Fiss-barr} presents the extension of the map of the heights of fission 
barriers around spherical shape  from earlier published range of $Z=120-180$ 
(see Fig.\ 6a in Ref.\ \cite{AAG.18}) to the range of proton numbers from 
$Z=120$ up to $Z=210$. The value  of the fission barrier height $E_B$ is defined 
as the lowest value of the barriers located on the oblate and prolate sides with 
respect to spherical state in the deformation energy curves obtained in axial RHB 
calculations. One can see that the island of spherical hyperheavy
nuclei previously labeled as "$Z\approx 174, N\approx 410$ island" in Ref.\ 
\cite{AAG.18} has been considerably extended up to $Z\approx 206$. In a given 
isotope chain of this island, the maximum of fission barriers heights is located at $N=406$.
The highest fission barriers with the heights between $\approx 7.5$ and
$\approx 8.5$ MeV are found in the $Z=186$, 184, 182 and 180 isotopic
chains. They are higher than those obtained in the classical region of
superheavy nuclei (see Ref.\ \cite{AARR.17}).
Based on these results for fission barriers and for the sizes of the 
$Z=186, N=406$ spherical shell gaps, we relabel this island as  
"$Z\approx 186, N\approx 406$ island of spherical hyperheavy nuclei".
The extension of upper boundary of nuclear landscape from $Z=180$ to $Z=210$ 
does not reveal other islands of spherical hyperheavy nuclei.
    
  Similar to the results presented in Fig. 6 of Ref.\ \cite{AAG.18} the size of
the $Z\approx 186, N\approx 406$ island of spherical  hyperheavy nuclei and 
the stability of the elements in it are expected to depend strongly on employed 
functional. We have not attempted to map this region  with other than DD-PC1 
functionals but some insight on this issue can be obtained from the analysis of 
the heights of fission barrier $E_B$ of the central nucleus  ($^{592}$186) of this 
region  calculated with different functionals. These results are summarized in 
Table \ref{table-FB-gaps}.  
The  FSUGold and next three functionals (DD-ME2, DD-ME$\delta$ and DD-PC1) 
produce the highest  calculated fission barriers: at  10.66 MeV for FSUGold and
clustered around $E_B\approx 7.7$ MeV for other three functionals. These barriers are higher 
than those produced in the CDFT framework in  the classical region of superheavy nuclei 
(see Fig.\ 10 in Ref.\ \cite{AARR.17}). These functionals are also expected to produce 
the island of spherical hyperheavy nuclei which is comparable in size to that 
shown in  Fig.\ \ref{Fiss-barr}.  The next five functionals (PC-PK1, NL3, PC-F1, 
TM1 and NL3*) produce the cluster with $E_B\approx 4$ MeV (see Table
\ref{table-FB-gaps}); this value is not far away from what is obtained in the 
$Z\approx 116, Z\approx 180$ region of superheavy nuclei (see Fig.\ 10 in 
Ref.\ \cite{AARR.17}).  For these functionals  the $Z\approx 186, N\approx 406$ 
island of stability  of spherical hyperheavy nuclei is expected to be substantially 
smaller than the one shown in Fig.\ \ref{Fiss-barr}.  Finally, the lowest fission 
barrier is produced  by the NL1 functional; its value indicates the instability of 
spherical hyperheavy nuclei.  However, the predictions of this functional have to 
be considered as least reliable because of well known problems in its isovector 
properties (see Ref.\ \cite{NL3}). 
 
 The difference in the predictions of $E_B$ is in part related to the fact that
the first group of functionals predicts the $Z=186$ and $N=406$ shell gaps 
which are on average larger by $\approx 0.1$ MeV and $\approx 0.5$ MeV,
respectively, than those produced by the second group of CEDFs (see Table 
\ref{table-FB-gaps}). Note also that the nuclear matter properties and the density 
dependence are substantially better defined for density-dependent (DD*)
functionals as compared with non-linear (NL* and TM1)  and point-coupling
(PC-PK1 and PC-F1) ones \cite{AA.16}. As a consequence, in general, they are 
expected to perform better for large extrapolations from known regions.

\begin{figure*}[htb]
\centering
\includegraphics[angle=0,width=8.5cm]{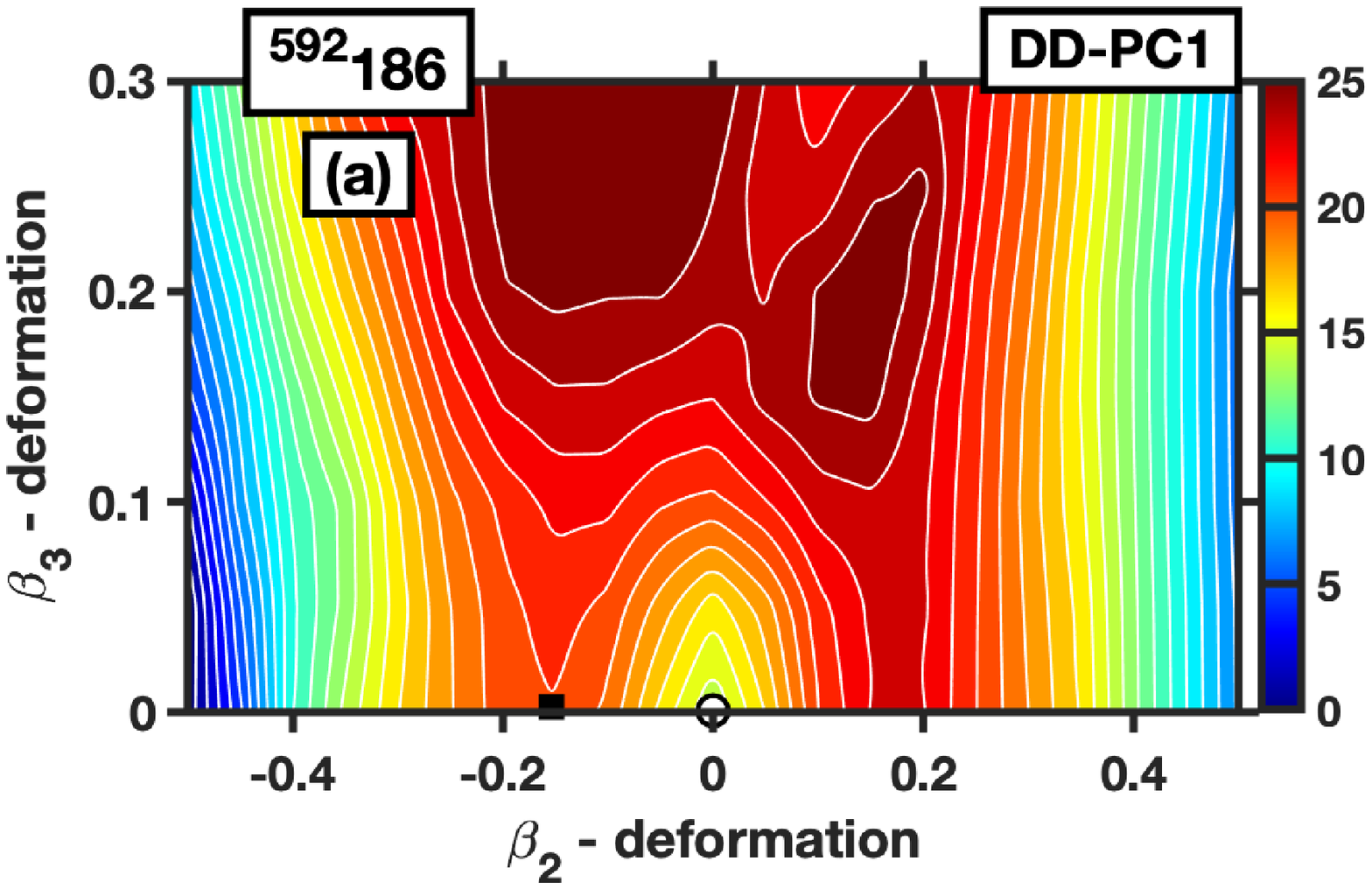}
\includegraphics[angle=0,width=8.5cm]{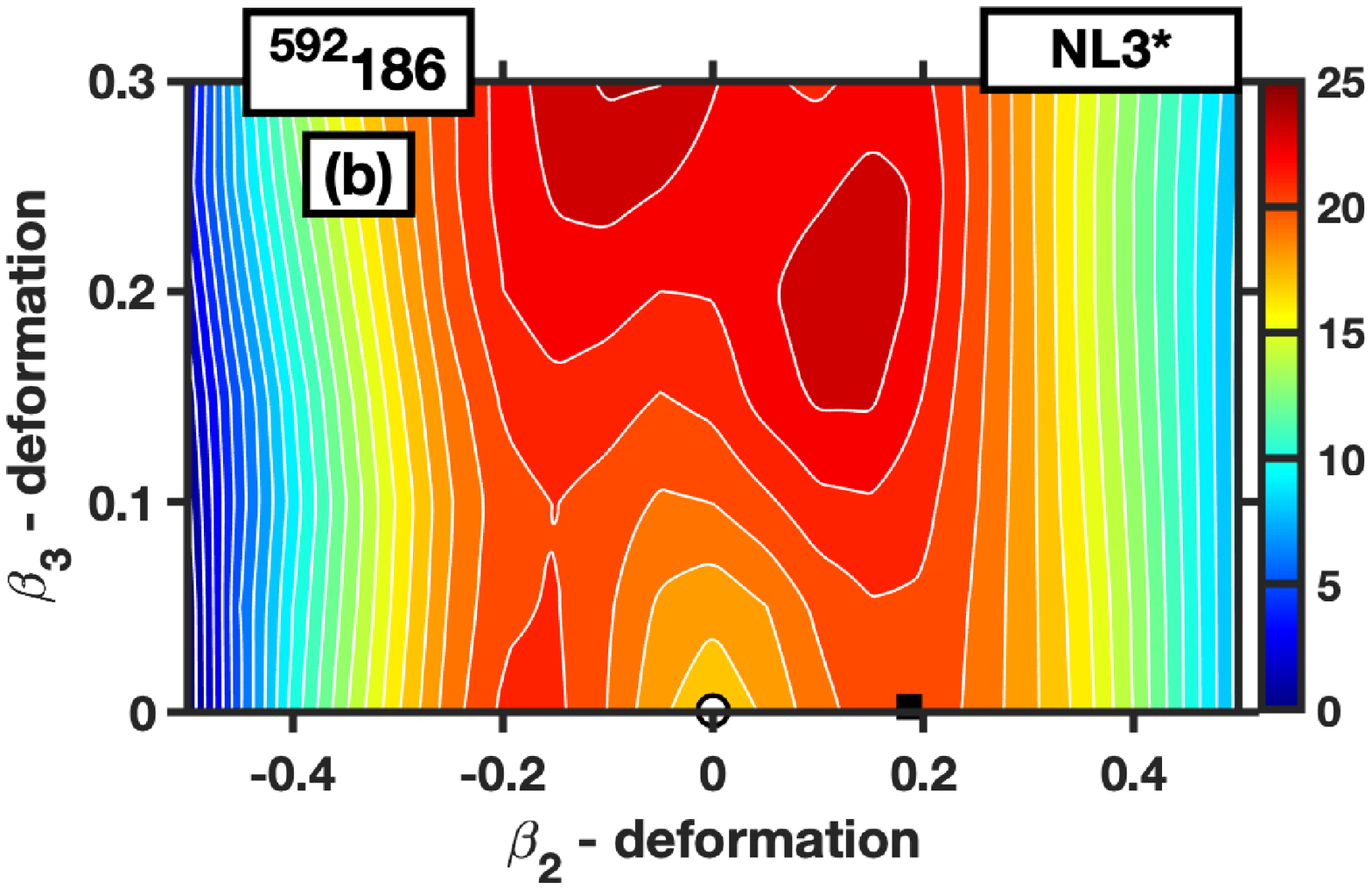}
\caption{
Potential energy surfaces of the $^{592}$186 nucleus obtained in the 
reflection asymmetric (octupole deformed) RHB calculations
with indicated CEDFs. The energy difference between two neighboring equipotential lines is equal to 
1.0 MeV. Spherical minimum is indicated by a circle and the saddle point of the barrier 
around  spherical minimum by solid black square. The colormaps show the excitation energies 
(in MeV) with respect to the energy of the deformation point with largest (in absolute value) 
binding energy. The calculations are performed with $N_F = 20$. Note that the topology of 
potential energy surfaces is almost the same in the calculations with $N_F = 20$ and $N_F = 26$. 
Thus, to save computational time these figures are plotted with $N_F=20$.
}
\label{oct-PES}
\end{figure*}

  Note that the axial RHB calculations for deformation energy curves 
in the vicinity of spherical minimum indicate nearly symmetric barriers with saddles 
at $\beta_2 \approx \pm 0.2$ (similar to Fig.\ \ref{Dif-energy-curves}(b) below).  
The experience in actinides and superheavy nuclei tells us that  octupole  deformation 
in fission barrier area typically does not develop for such low deformations 
\cite{AAR.12,PNLV.12,LZZ.14} [corresponding to inner fission barrier in actinides
and superheavy nuclei] and this result has been confirmed in octupole deformed 
RHB calculations with CEDF DD-PC1 for spherical minimum of several 
hyperheavy  nuclei in Ref.\ \cite{AATG.19}.  
The results presented in Fig.\ \ref{oct-PES} for the $^{592}$186 nucleus
are in line with these expectations; the saddle of fission barrier is located
at $\beta_3=0.0$ and octupole deformation does not affect the 
spherical minimum in the calculations with DD-PC1 and NL3* functionals.

\begin{figure*}[htb]
\centering
\includegraphics[angle=0,width=8.5cm]{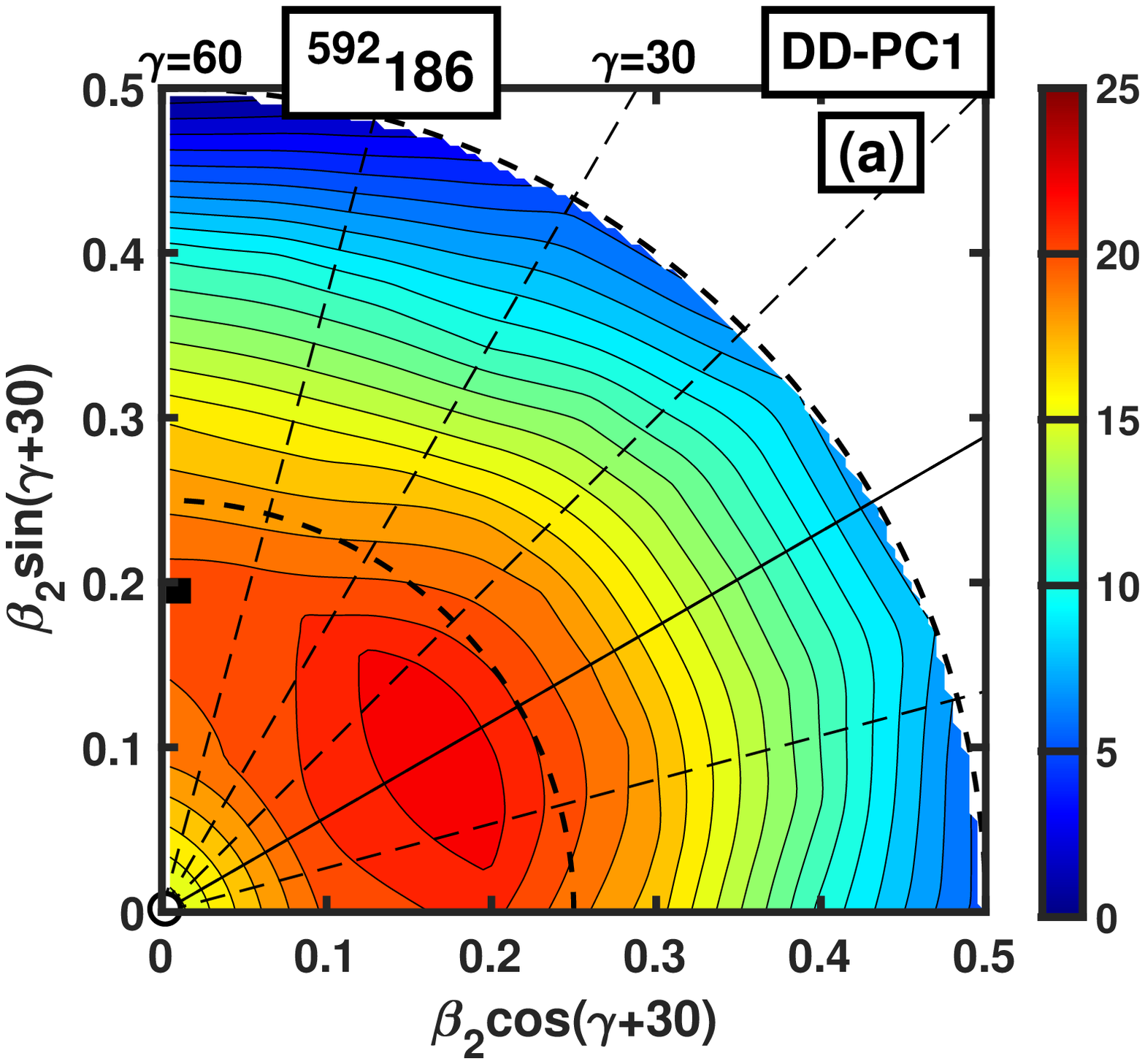}
\includegraphics[angle=0,width=8.5cm]{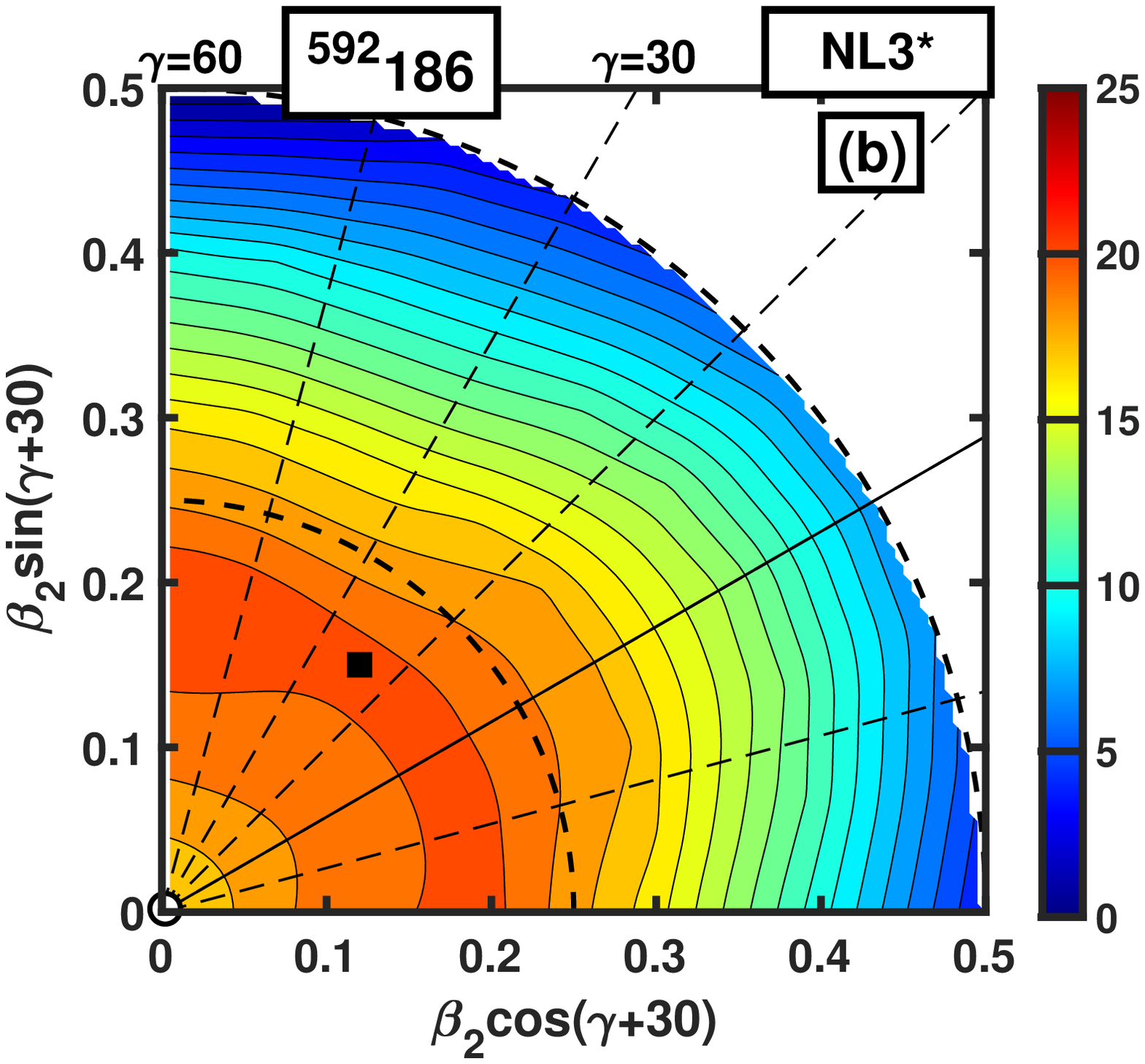}
\caption{
The same as in Fig.\ \protect\ref{oct-PES} but for potential energy surfaces obtained 
in triaxial RHB calculations.
}
\label{triax-PES}
\end{figure*}

   The analysis of Ref.\ \cite{AATG.19} indicates that the impact of triaxial 
deformation  on the fission barriers around spherical minima is relatively modest.     
This is the consequence of the topology of potential energy surfaces 
which is similar to those of volcanos (see Figs.\ \ref{triax-PES}). 
 The central area around spherical minimum is 
similar to caldera, the rim of which is represented by the fission barrier. The area beyond 
the rim (fission barrier) is fast down-sloping as a function of quadrupole deformation 
$\beta_2$. The saddles of axial fission barriers (on oblate and prolate sides of spherical 
minimum) are located at modest quadrupole deformation  of $\beta_2 \approx 0.2$.
As  a result, the distance between these two saddles in the $(\beta_2, \gamma)$ plane
plane is relatively small, so that large changes in binding energy due to triaxiality
for nearly constant $\beta_2$ values could not develop.  As a consequence, the lowest 
fission barrier around spherical minimum obtained in axial RHB calculations is a good 
approximation to the barrier obtained in the TRHB calculations. 
For example, this is a case in the calculations with CEDF DD-PC1 (see Fig.\ \ref{triax-PES}a). 
Even if the saddle of fission barrier is located at $\gamma \neq 0^{\circ}$ and $\gamma \neq 60^{\circ}$,
the energy lowering in fission barrier height as compared with the lowest
fission barrier at these $\gamma$ values is rather modest. For example, in the calculations 
with the NL3* functional the saddle of the fission barrier, located at $\gamma=22^{\circ}$, is lower 
than the fission barrier at $\gamma=0^{\circ}$ by only 50 keV (see Fig.\ \ref{triax-PES}b).
Note also that the TRHB results  clearly indicate that spherical minimum of the nucleus under study 
is relatively stable with  respect to triaxial distortions.

   Although the detailed studies have only been performed with two functionals, representing
one of the highest (DD-PC1) and one of the lowest (NL3*) fission barriers obtained in 
the calculations (see Table \ref{table-FB-gaps}), it is reasonable to expect that similar
situation will hold also for other functionals. This is because of the similarity of the
underlying shell structure. Thus, one conclude that the impact of triaxiality and octupole
deformation on $E_B$ of spherical hyperheavy nuclei is either very small or nonexistent (see 
also the discussion  in Refs.\  \cite{AAG.18,AATG.19}).

\begin{table}[htbp]
\begin{center}
\caption{The heights of the fission barriers $E_B$ and the sizes $\Delta E$
of spherical $N=406$ $(\Delta E_{N=406}$) and $Z=186$ 
($\Delta E_{Z=186}$) shell gaps  in the $^{592}$186 
nucleus obtained with indicated CEDFs. The functionals are ordered in such a way
that $E_B$ is decreasing. 
\label{table-FB-gaps}
}\begin{tabular}{|c|c|c|c|}\hline
  CEDF                   & $E_{B}$[MeV] & $\Delta E_{N=406}$ [MeV] &  $\Delta E_{Z=186}$ [MeV] \\ \hline
FSUGold               & 10.66              &    1.84                                  &       2.17          \\
DD-ME2                &   7.73              &    2.11                                  &       2.43          \\
DD-ME$\delta$     &   7.72              &    1.98                                  &       2.68          \\
DD-PC1                &   7.59              &    1.93                                  &       2.45          \\
PC-PK1                 &   4.35              &    1.61                                 &        2.37          \\
NL3                       &   4.28              &    1.43                                  &       2.15          \\
PC-F1                   &   3.87              &    1.41                                  &       2.45          \\
TM1                      &   3.86              &    1.38                                  &       2.29          \\
NL3*                      &   3.59              &    1.45                                  &       2.37          \\
NL1                       &   1.27               &    1.27                                  &      2.34          \\ \hline
\end{tabular}
\end{center}
\end{table}
    
   The density distributions at spherical shape for the nuclei representing
the centers of the islands of spherical hyperheavy nuclei have been 
compared and discussed in Sec.\ IV of Ref.\ \cite{AATG.19}. However, 
the $Z\approx 186, N\approx 406$ island (and, in particular, doubly magic 
$^{592}186$ nucleus corresponding to large shell gaps at $Z=186$ and 
$N=406$) has not been completely covered in that study because of 
the restriction to the $Z\leq 180$ nuclei.  To fill this gap in our knowledge, 
Fig.\ \ref{densit} compares proton and neutron density distributions of the 
$^{584}$174 nucleus (studied in Ref.\ \cite{AATG.19}) with those of doubly 
magic $^{592}$186 one.   Neutron densities of these two nuclei are very 
similar; they are slightly larger for the $^{584}$174 nucleus because of  
the occupation  of the $2j_{13/2}$ orbitals by four additional neutrons. 
The differences
are more visible for proton densities because 12 additional protons 
in the doubly magic $^{592}$186 nucleus (8 in the $2g_{7/2}$ and 4 in 
$1j_{13/2}$ orbitals) occupy the orbitals which fill the density either
in surface region (the $1j_{13/2}$ orbitals) or in-between central and 
surface regions (the $2g_{7/2}$ orbitals) (see Ref.\ \cite{AF.05}). 
The increase of the Coulomb repulsion in the $Z=186$ nucleus as compared with 
the $Z=174$ one also plays a role in an enhancement of proton 
density near the surface.  As a consequence, the semi-bubble structure 
becomes more pronounced in the proton densities of the
$^{592}$186 nucleus as compared with the $^{584}$174 one.

\begin{figure*}[htb]
\centering
\includegraphics[angle=0,width=18.0cm]{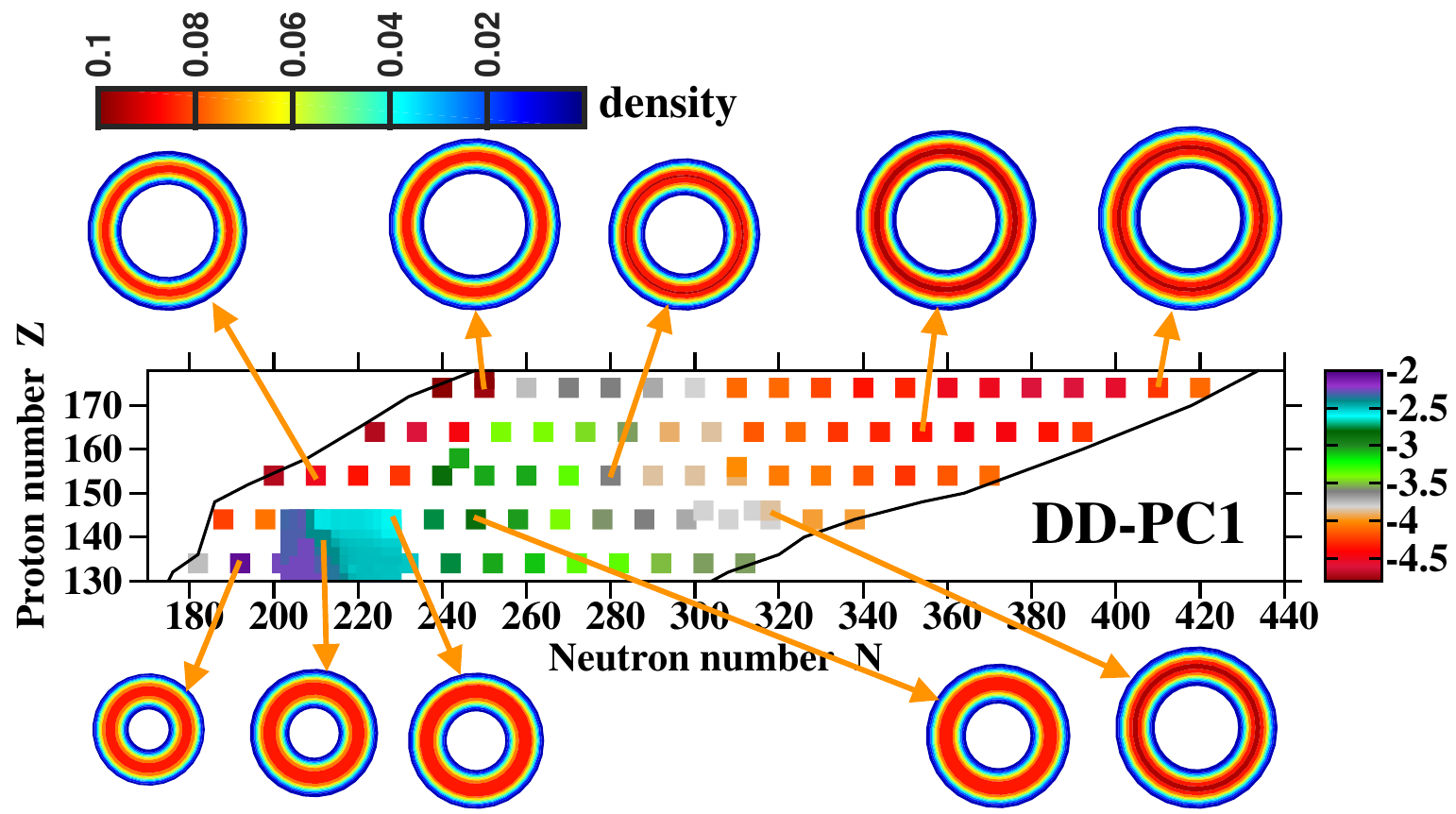}
\caption{
Proton $\beta_2$ values (see colormap on right) of the lowest in energy solutions  
of selected set of nuclei (see text for details). Solid black lines indicate two-proton and 
two-neutron drip  lines. Neutron density distributions of some nuclei are shown: orange 
arrows are used to indicate these nuclei.   The same colormap [shown in the upper left 
corner] as the one used in Figs.\ \protect\ref{156-466-pot-b} and \protect\ref{densit-tori}  is 
employed here for the densities. The density colormap starts at $\rho_n=0.005$ fm$^{-3}$ 
and shows the densities in fm$^{-3}$. The density of the $^{348}$138 nucleus is used 
here as a reference [see Fig.\ \protect\ref{densit-tori} for its actual geometrical size] with respect 
of which the geometrical sizes of the density distributions in other nuclei are normalized.
}
\label{torus-def}
\end{figure*}

\section{Toroidal nuclei}
\label{shell-tori}

\begin{figure}[htb]
\centering
\includegraphics[angle=0,width=4.25cm]{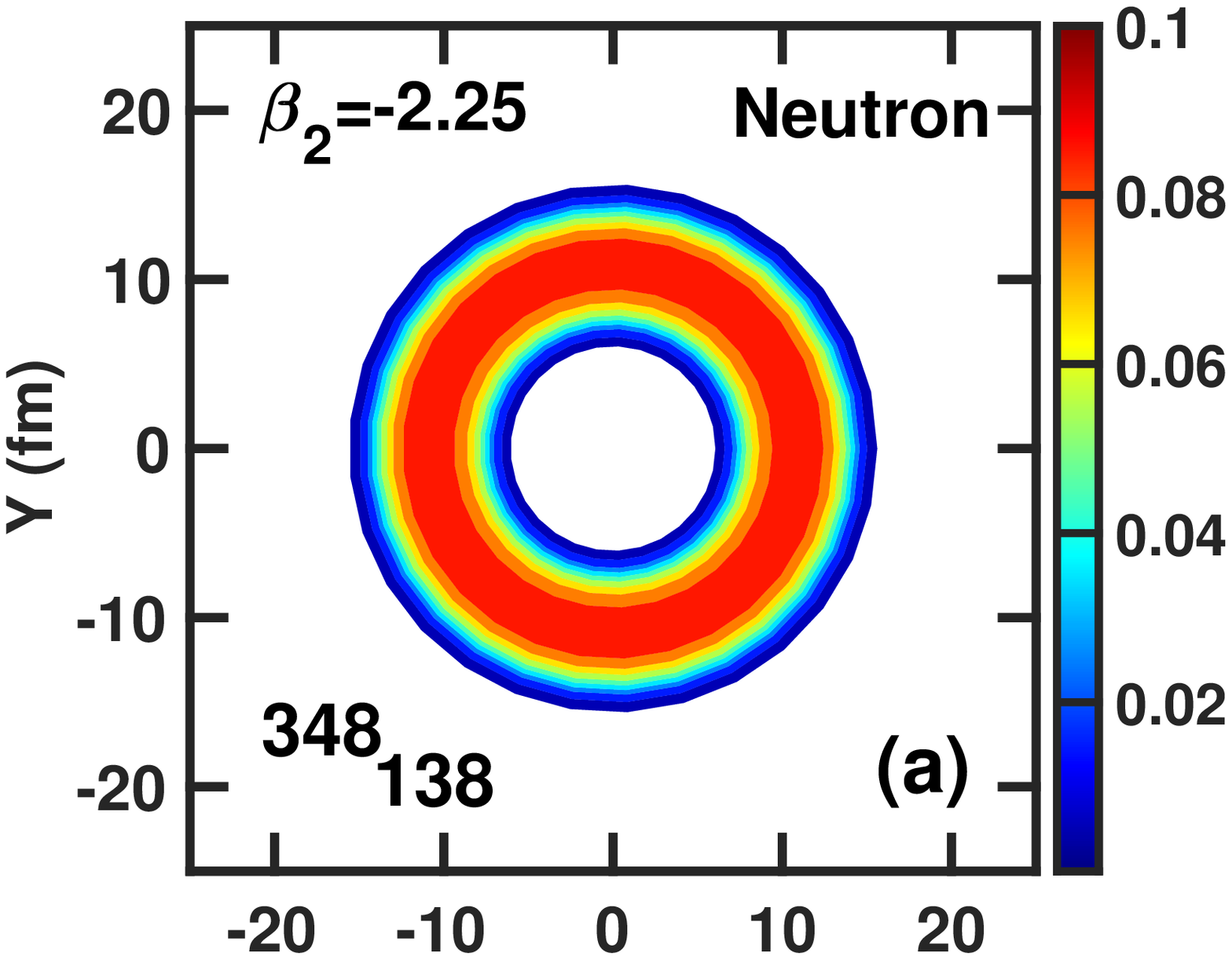}
\includegraphics[angle=0,width=4.25cm]{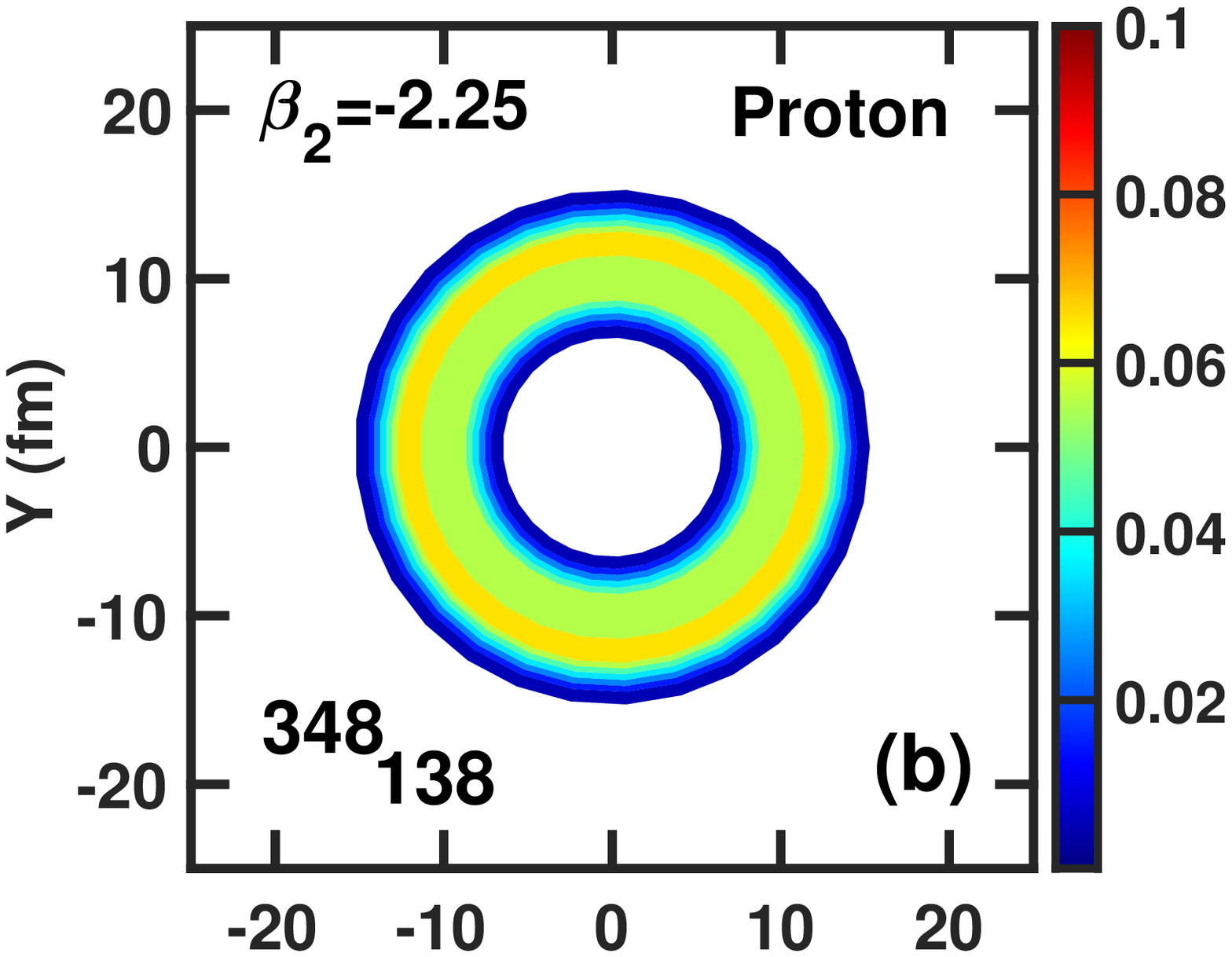}
\includegraphics[angle=0,width=4.25cm]{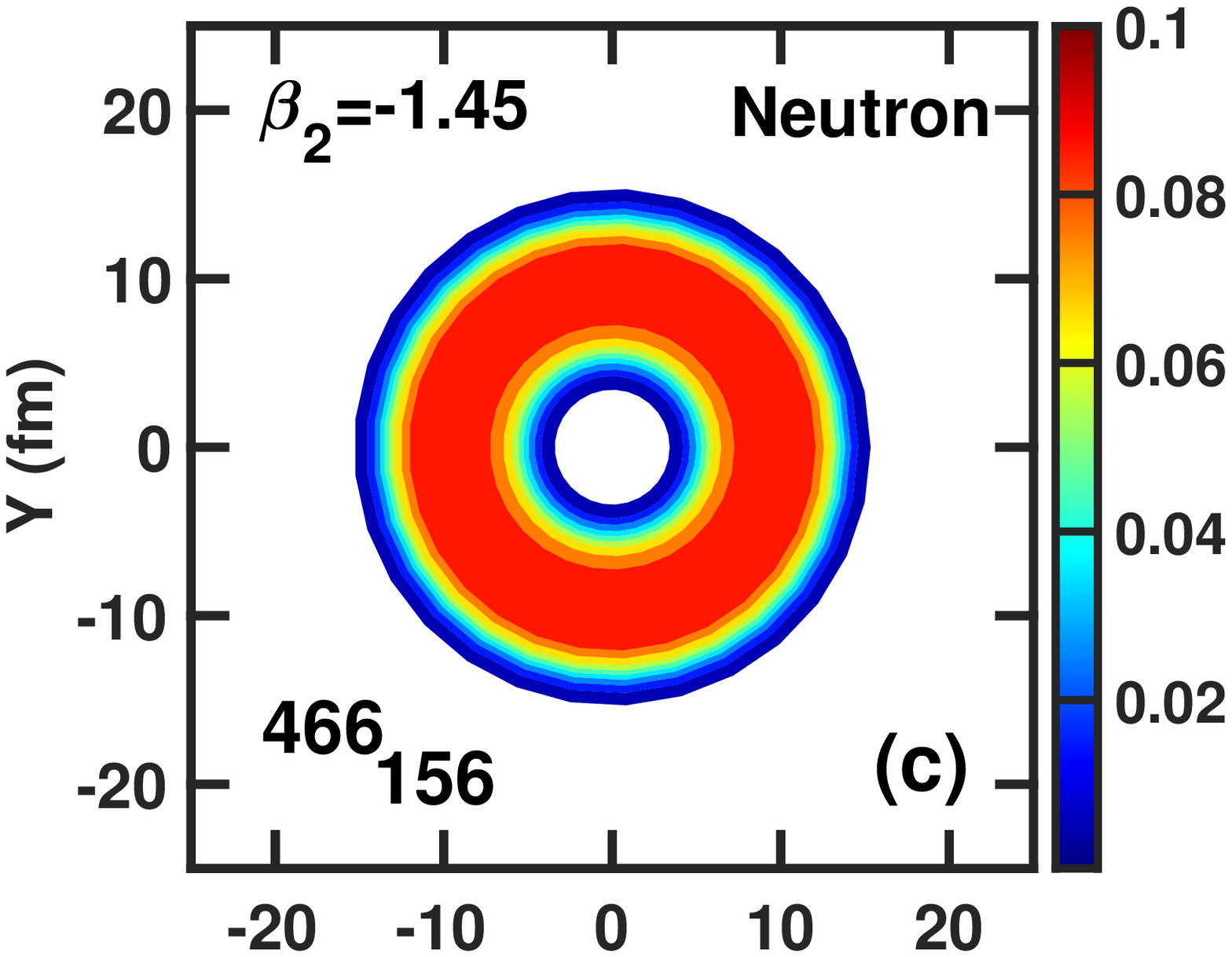}
\includegraphics[angle=0,width=4.25cm]{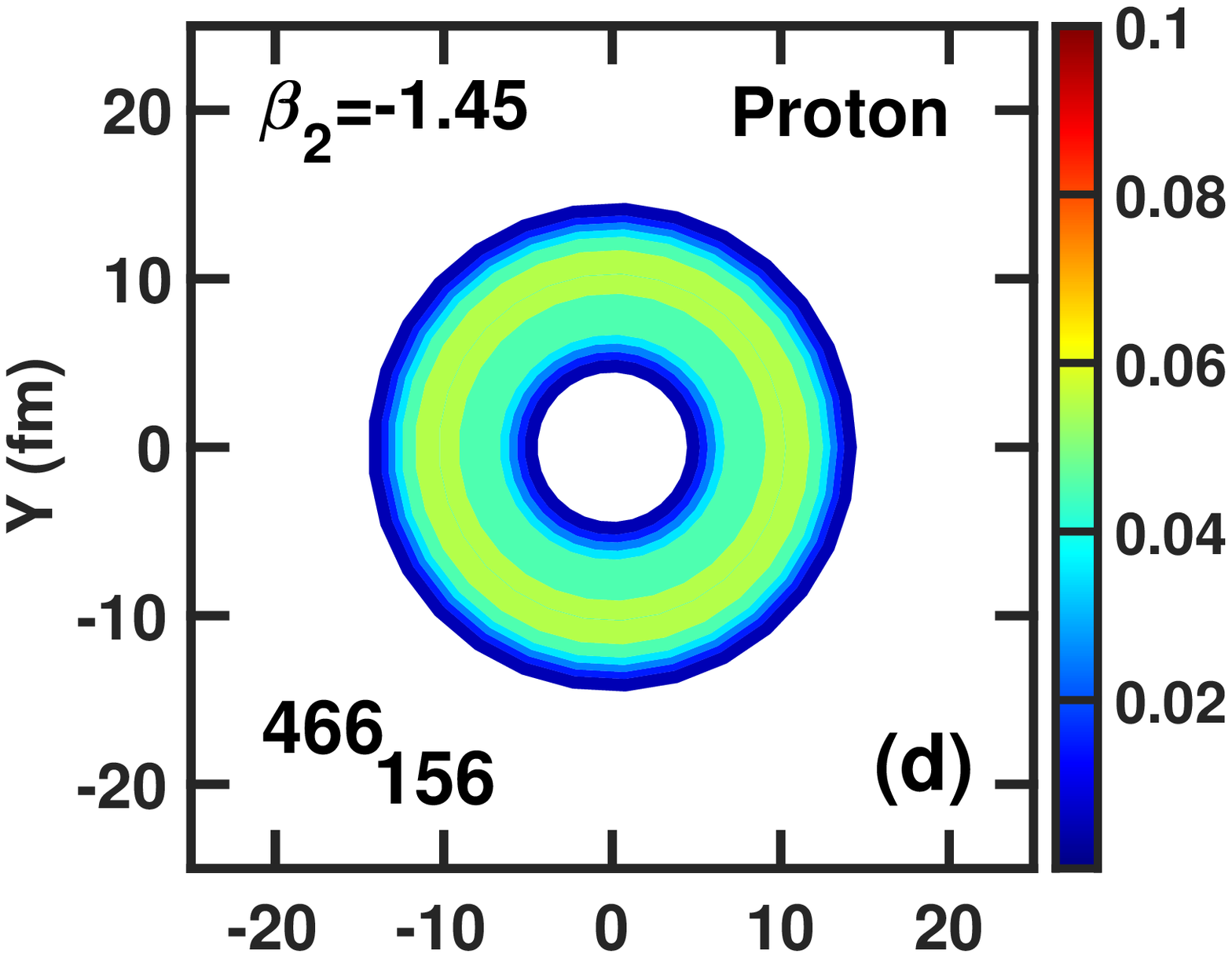}
\includegraphics[angle=0,width=4.25cm]{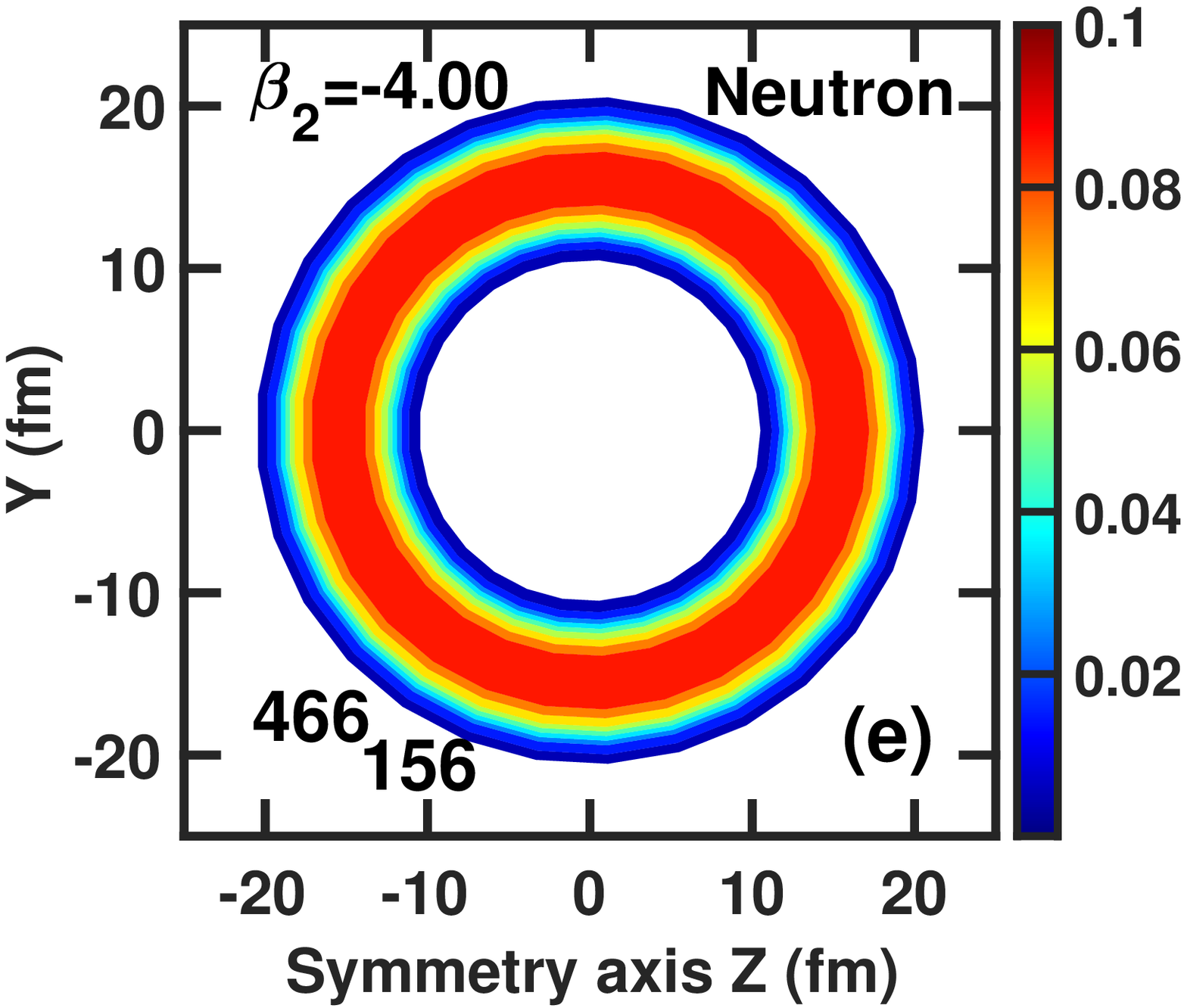}
\includegraphics[angle=0,width=4.25cm]{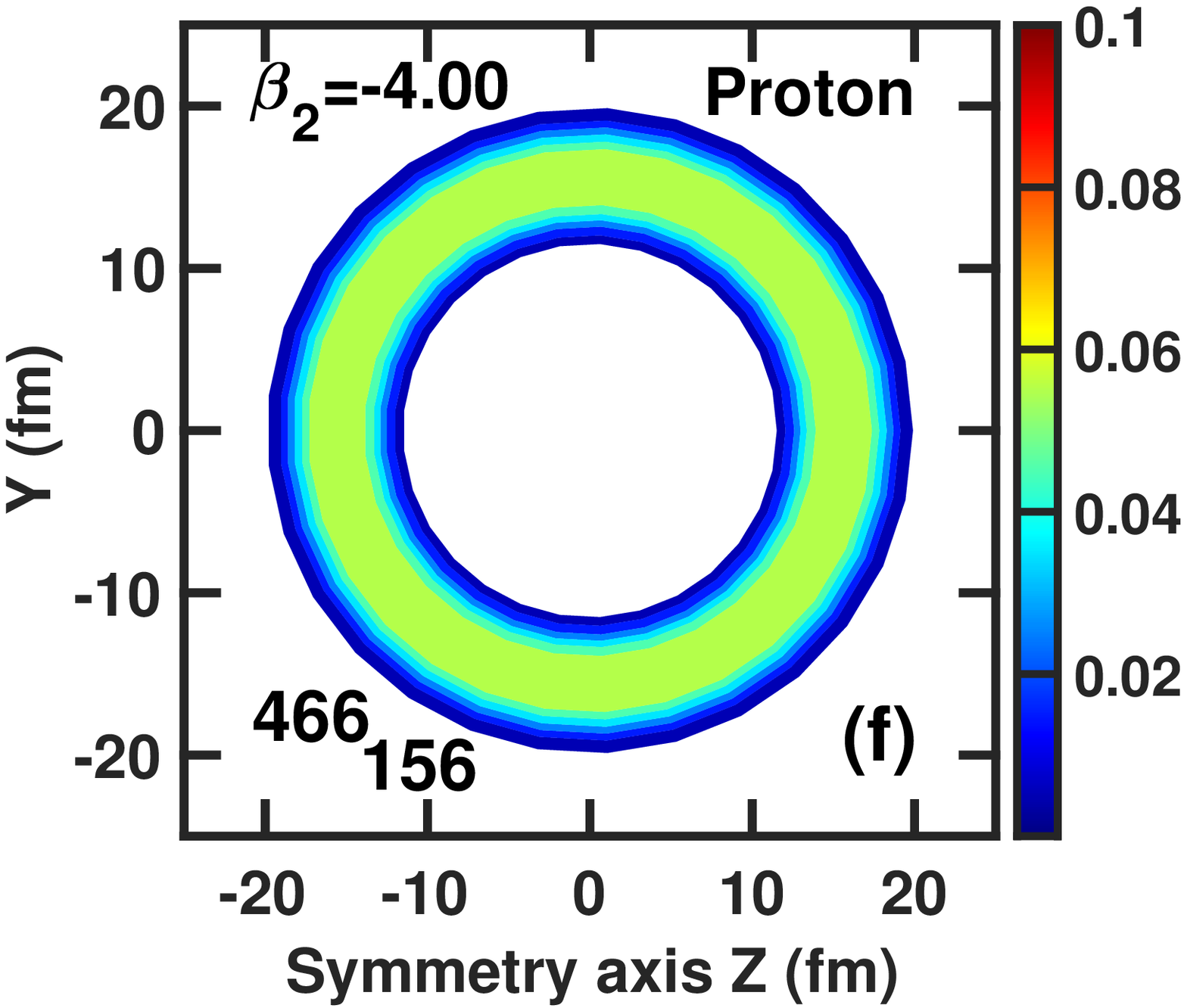}
\caption{Neutron $\rho_{\nu}$ (left column) and proton $\rho_{\pi}$ 
(right column) density distributions of toroidal configurations in the 
$\beta_2=-2.25$ minimum of the $^{348}$138 nucleus and in the
minima B and A of the $^{466}156$ nucleus (see Fig.\ 
\ref{156-466-pot-b}).}
\label{densit-tori}
\end{figure}

\subsection{Distribution of shapes of toroidal nuclei across the 
nuclear landscape} 
\label{toroid-distr}

  In our calculations the truncation of basis is performed in such a way
that all states belonging to the major shells up to $N_F$ fermionic shells
for the Dirac spinors are taken into account. Accurate calculations of 
LEMAS require extremely large fermionic basis  and its size, 
defined by $N_F$,  increases with the raise of proton and neutron numbers
(see discussion in Sect. III of Ref.\ \cite{AATG.19}). As a 
result, the $\beta_2$ values (and, thus, respective density distributions)  of the 
lowest in energy toroidal states have only been  partially mapped in the 
$Z=122-138$ region (see Fig. 3 in Ref.\ \cite{AAG.18}) in the
axial RHB calculations with $N_F=26$. For higher $Z$ nuclei, existing calculations 
only confirm that the lowest in energy solutions have always toroidal shapes (see
the discussion of Fig.\ 3 in Ref.\ \cite{AATG.19})  but do not provide accurate 
$\beta_2$ values. 

  To fill this gap in our knowledge, additional calculations are performed in the 
$N_F=30$ basis which provides quite accurate description of toroidal shapes in 
the $Z\geq 140$ hyperheavy nuclei (see Sect.\ III in Ref.\ \cite{AATG.19}).  Such 
calculations are extremely time-consuming  even in axial RHB framework and thus 
they are carried out only  for restricted set of nuclei displayed 
in Fig.\ \ref{torus-def}.  These are $Z=136$, 146, 156, 166 and 176 nuclei.   Apart of 
few  regions, the calculations are performed in step of $\Delta N=10$ to save 
computational time. Despite these limitations they allow to understand the general 
features of the distribution of toroidal shapes  as well as the evolution of underlying 
single-particle structure across the nuclear chart.

   The results of these calculations are presented in Fig. \ref{torus-def}. 
To facilitate the discussion we are using here the definitions of tori as {\it thin}
and {\it fat} employed in the physics of toroidal liquid droplets \cite{YB.11}. 
Large/small ratio of the radius $R$ of toroid (called as "major radius" in
some publications [see, for example, Ref.\ \cite{Wong.73}) to the radius 
$d$ of its tube (called as "minor radius" in Ref.\ \cite{Wong.73}) corresponds 
to thin/fat tori. The lowest $\beta_2$ values ($\beta_2 \approx -2.2$) are obtained in the 
$Z\approx 136, N\approx 206$ region (see Fig.\ \ref{torus-def}) and these 
nuclei can be defined as fat toroidal nuclei because of small aspect ratio $R/d$.
The absolute $\beta_2$ values increase on moving away from this region. 
Especially large values of $|\beta_2|$ are obtained in proton-rich nuclei with $Z>140$
in the vicinity of two-proton drip line. These toroidal shapes are characterized by 
very large radius of the torus and small radius of the torus tube and thus these
nuclei are described as thin toroidal nuclei. Slightly smaller values of $|\beta_2|$ 
are seen in neutron-rich $N\geq 310$ nuclei. The aspect ratios $R/d$ for these nuclei 
are slightly smaller as compared with the ones in proton-rich nuclei but these
nuclei are still the representatives of thin toroidal nuclei. Remaining nuclei shown by cyan,
dark and light green as well as grey colors in Fig.\ \ref{torus-def} are characterized by 
$\beta_2$ ranging from $-2.5$ to $-3.7$.  
A general trend of the increase of torus radius $R$ and the aspect ratio $R/d$ with increasing 
proton number  is seen in Fig.\ \ref{torus-def}. It is a consequence of Coulomb repulsion: 
toroidal shapes provide less compact distribution 
of charge as compared with spherical ones and thus the Coulomb energy is substantially 
reduced for toroidal shapes as compared with spherical ones (see discussion in Sect. XII 
in Ref.\ \cite{AATG.19})). The increase of proton number requires the increase of torus 
radius in order to minimize the Coulomb energy
by creating less compact distribution of charge. Observed features in the distribution of 
toroidal shapes, which  are the result of the competition of different energy minima similar to the 
minima A and B shown in Fig.\ \ref{156-466-pot-b} (see also Fig.\ 16 in Ref.\ 
\cite{AATG.19}), have a root in underlying shell structure of toroidal hyperheavy 
nuclei (see Sec.\ \ref{Tori-shell-structure}).

   To get a  better understanding of the relative properties of proton and neutron density 
distributions, we compare them in Fig.\ \ref{densit-tori} for the $^{348}$138 and 
$^{466}$156 nuclei. Similar to the situation at spherical shape (see, for example, Fig.\ 
\ref{densit}), the maximum of proton density distribution $\rho_{\pi}^{max}$ is significantly 
smaller ($\rho_{\pi}^{max} \approx \frac{2}{3} \rho_{\nu}^{max}$)  than the neutron one 
$\rho_{\nu}^{max}$ and those maxima do not necessary appear at the same distance
from the center of toroid.  The outer edges of the proton and neutron density distributions 
appear at approximately the same distances from the center of toroid. However, the 
diameter of the hole in the center of proton density distribution is visibly larger than the
one in the case of neutrons. This is most likely the consequence of the Coulomb repulsion
acting on protons. Thus, the diameter  of toroid tube is smaller in the case of protons
as compared with the one for neutrons.  Note also that the density distribution in toroid
tube is not necessary symmetric with respect of its geometrical axis of symmetry; this is
especially visible in the case of proton density distributions presented in Figs.\ 
\ref{densit-tori}(b) and (d). Detailed analysis reveals that this is a consequence of the
occupation of the single-particle orbitals characterized by different spatial distributions
of the single-particle densities.

  Because of the presence of well pronounced minima (similar to the minimum D
in Fig.\ \ref{156-466-pot-b}), the present axial RHB calculations in extremely large 
basis confirm for the first  time the stability of toroidal $Z\geq140$ nuclei shown in 
Fig.\ \ref{torus-def} with respect of so-called breathing deformations. The breathing 
deformation \cite{Wong.73} preserves the azimuthal symmetry of the torus and it is 
defined by the radius of torus and the radius of its tube. In our calculations, this type 
of deformation is related to the $\beta_2$ values (see discussion in Ref.\ 
\cite{AAG.18}). This result is clearly different as compared with the ones obtained 
for classical uncharged toroidal liquid droplets which are unstable with respect 
of shrinking instabilities \cite{YB.11,FPBSF.17,TPQC.13}. Because of surface 
tension such droplet starts from toroidal shape but then gradually shrinks by closing 
its interior hole and transforms into spherical droplet \cite{YB.11,FPBSF.17,TPQC.13}.  
In atomic  nuclei, this shrinking instability is counteracted by the Coulomb force: the 
transition to a more compact  spherical configuration leads to a substantial increase 
of the Coulomb energy and thus it is not energetically favored in hyperheavy nuclei 
\cite{AATG.19}. 

   Another class of potential instabilities of toroidal nuclei is related to so-called sausage  
deformations \cite{Wong.73}:  they make a torus thicker in one section(s) and thinner in 
another  section(s). This class of the instabilities is much more difficult to describe in the 
density functional theories since their consideration requires, in general, symmetry 
unrestricted computer codes.  This fact combined with the requirement for extremely large
basis in high-$Z$ systems makes this problem numerically intractable with existing
computer codes for absolute majority of toroidal nuclei. The only exception are
fat toroidal nuclei located in the $Z\approx 136, N\approx 210$ region for which 
(as illustrated by the examples of the $^{354}$134 and $^{348}$138 nuclei discussed
in Refs.\ \cite{AAG.18,AATG.19}) the calculations for even-multipole sausage  deformations 
within the triaxial RMF+BCS codes are possible \cite{AAG.18}. However, even such 
calculations are extremely time-consuming and can be performed only for a few nuclei. 

\begin{figure*}[htb]
\centering
\includegraphics[angle=0,width=8.7cm]{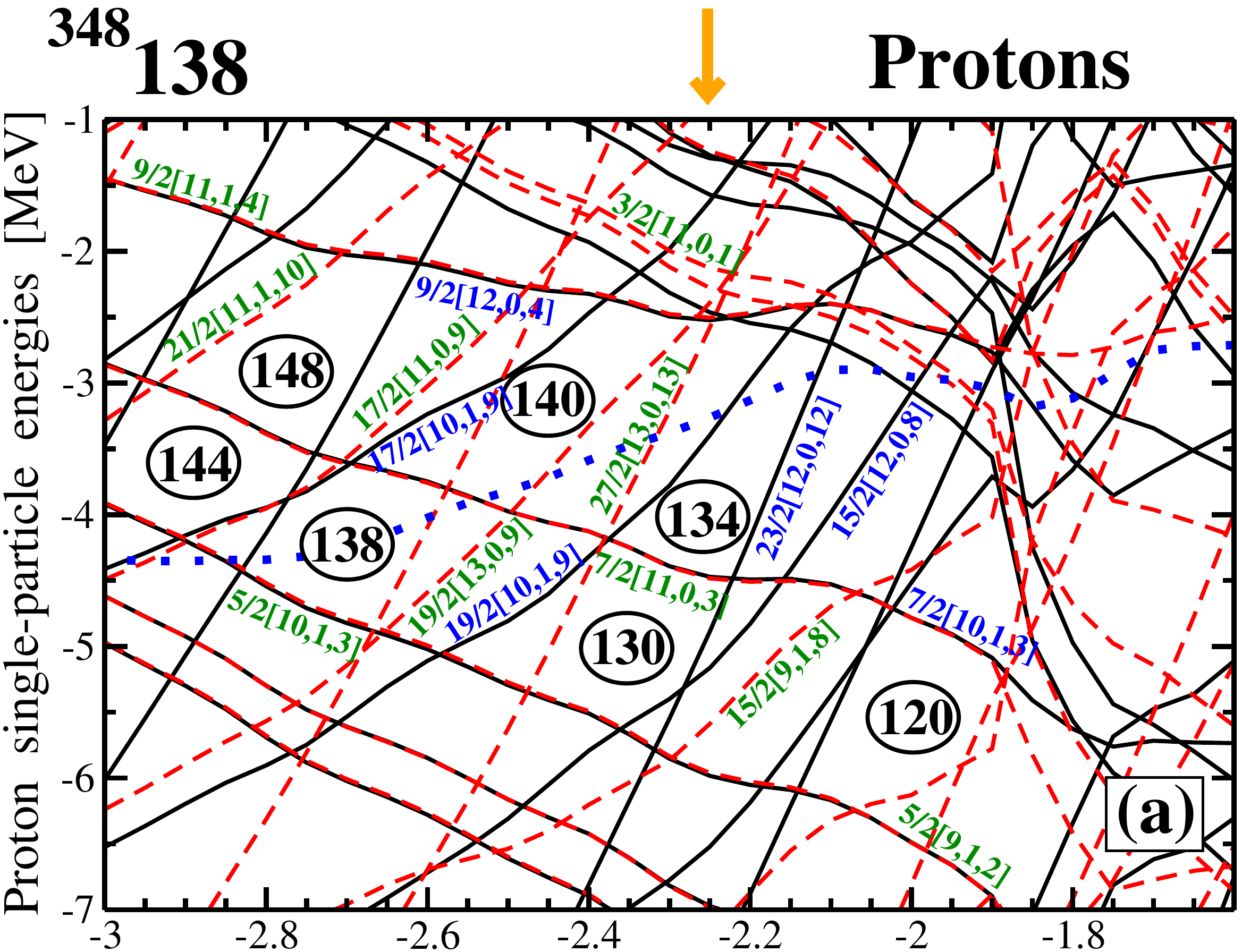}
\includegraphics[angle=0,width=8.7cm]{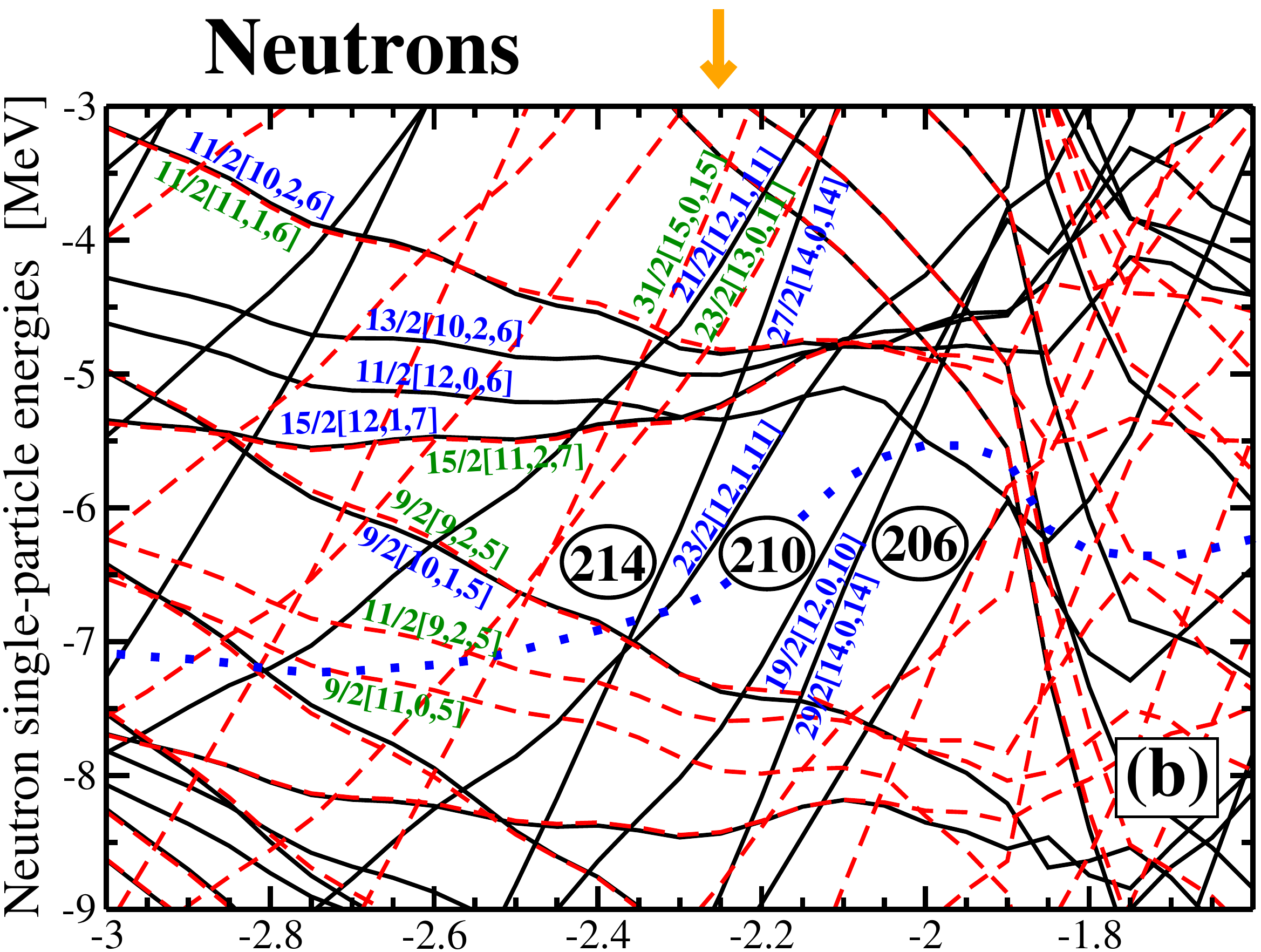}
\includegraphics[angle=0,width=8.8cm]{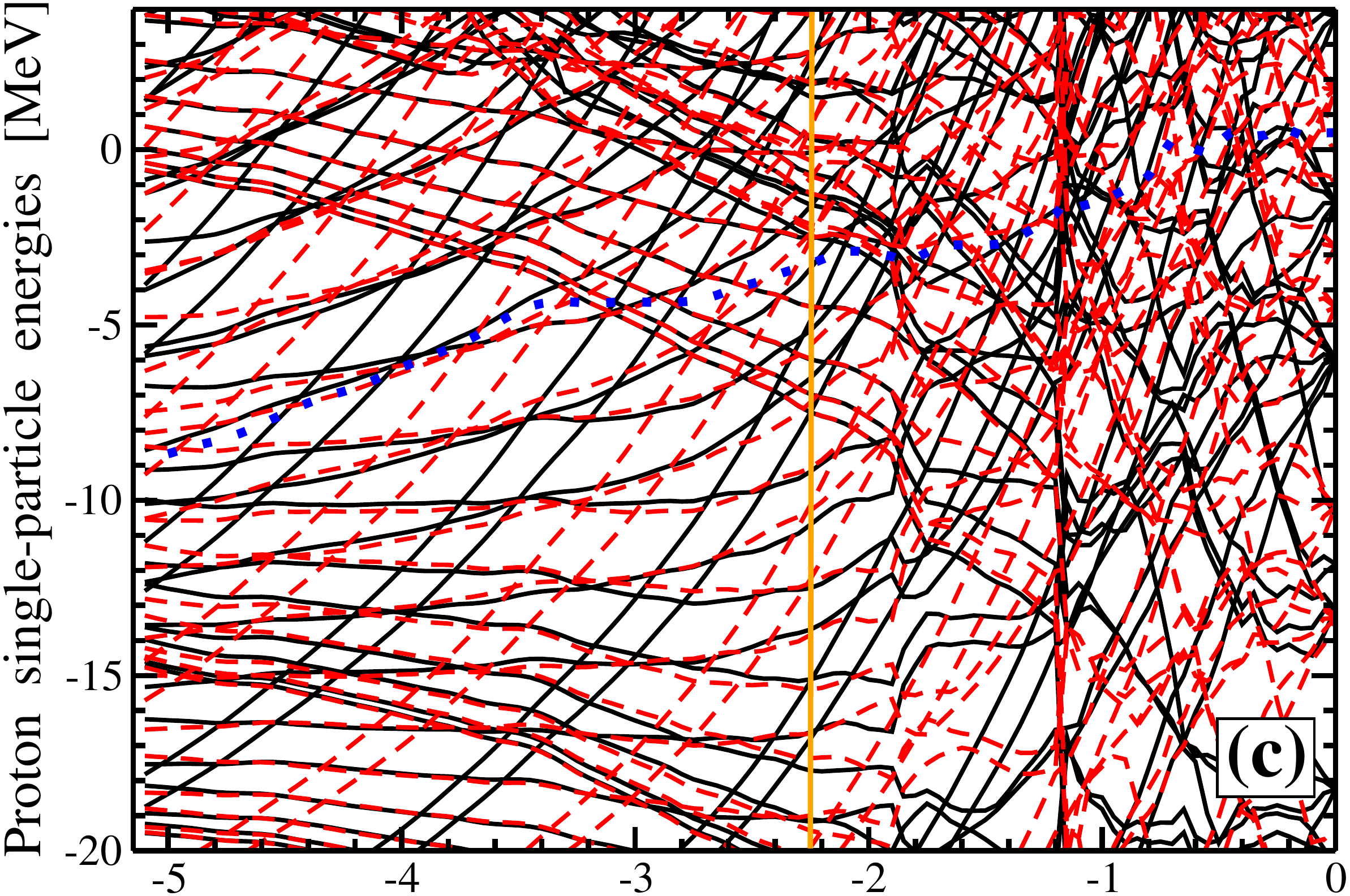}
\includegraphics[angle=0,width=8.8cm]{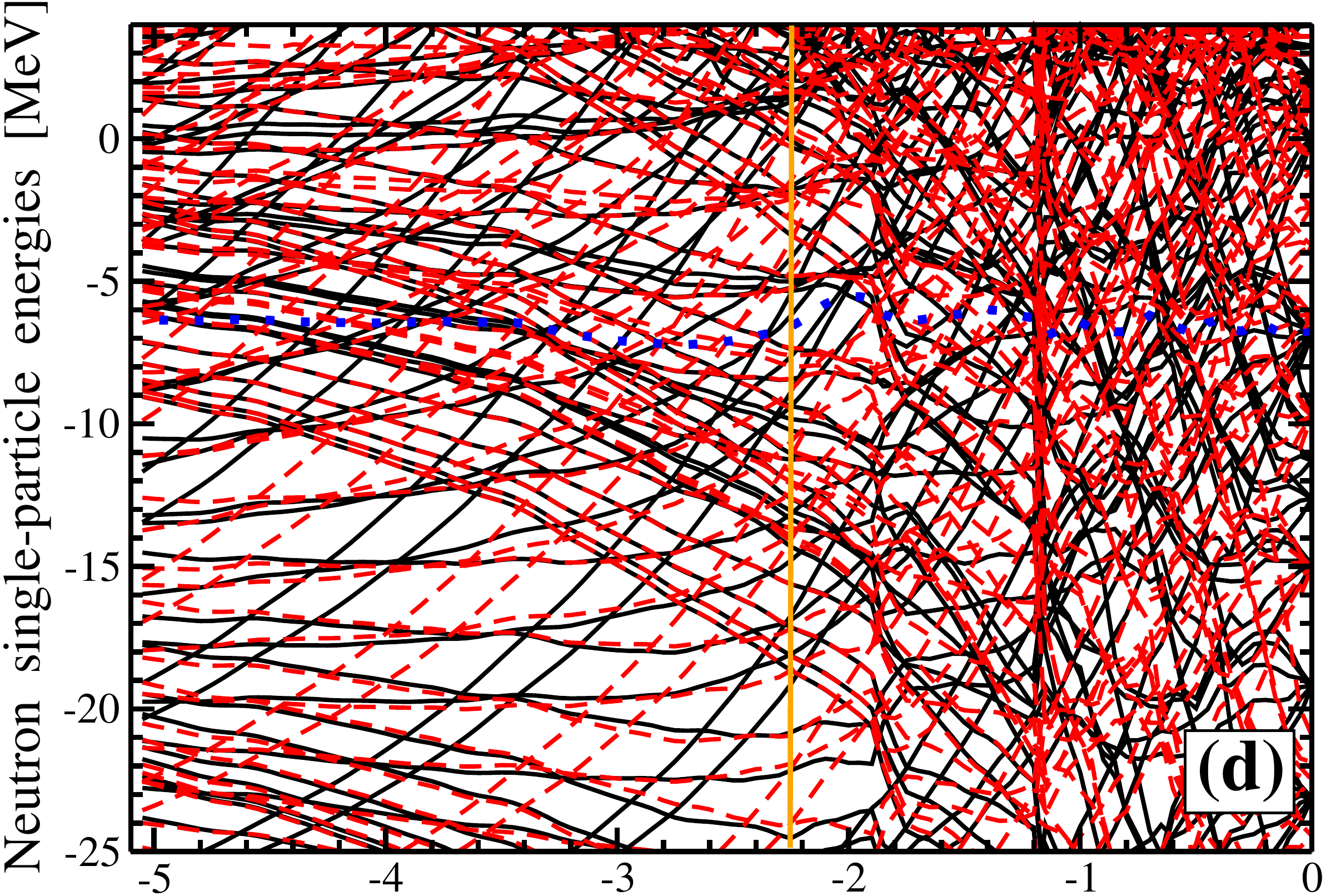}
\includegraphics[angle=0,width=8.8cm]{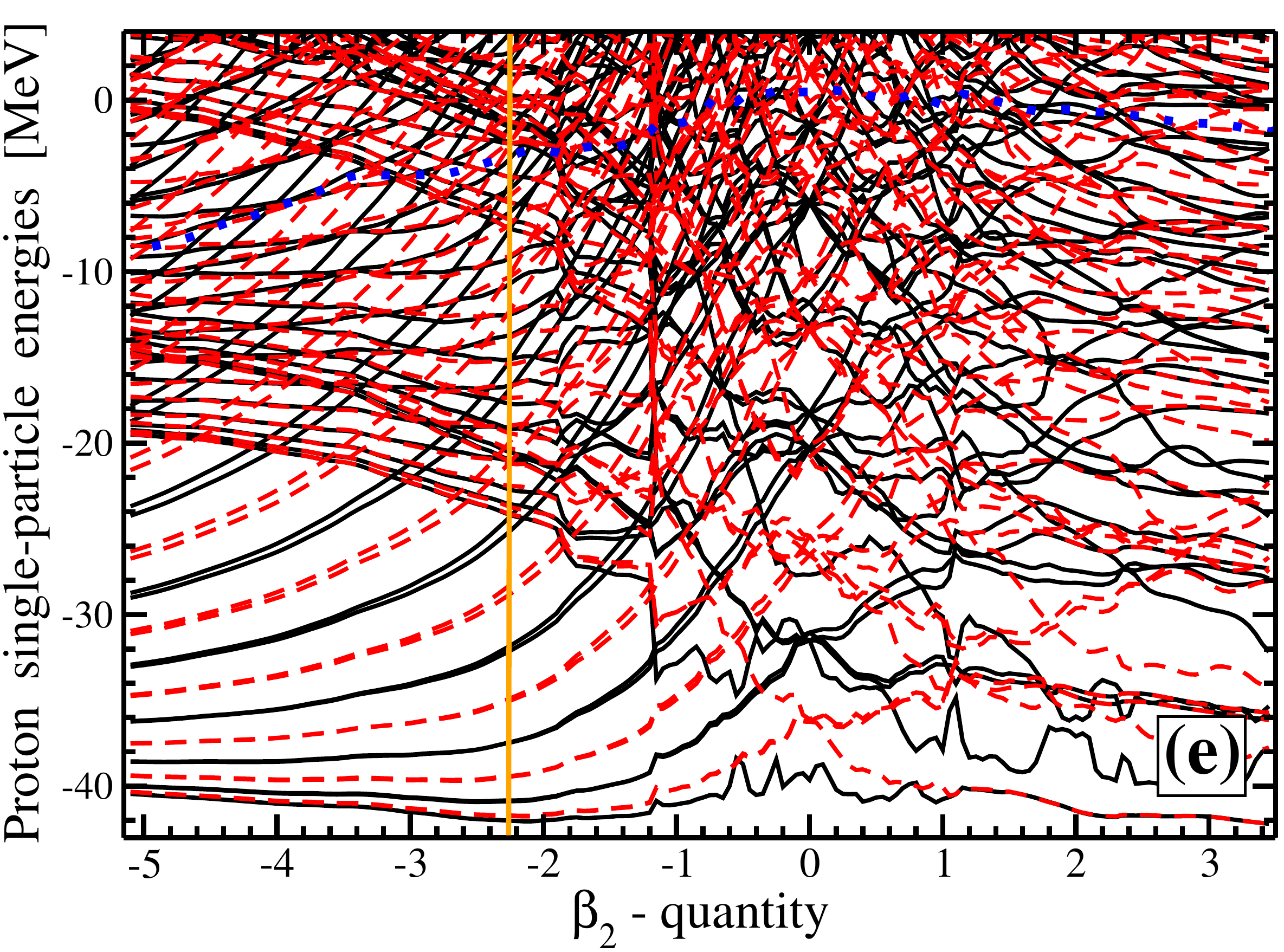}
\includegraphics[angle=0,width=8.8cm]{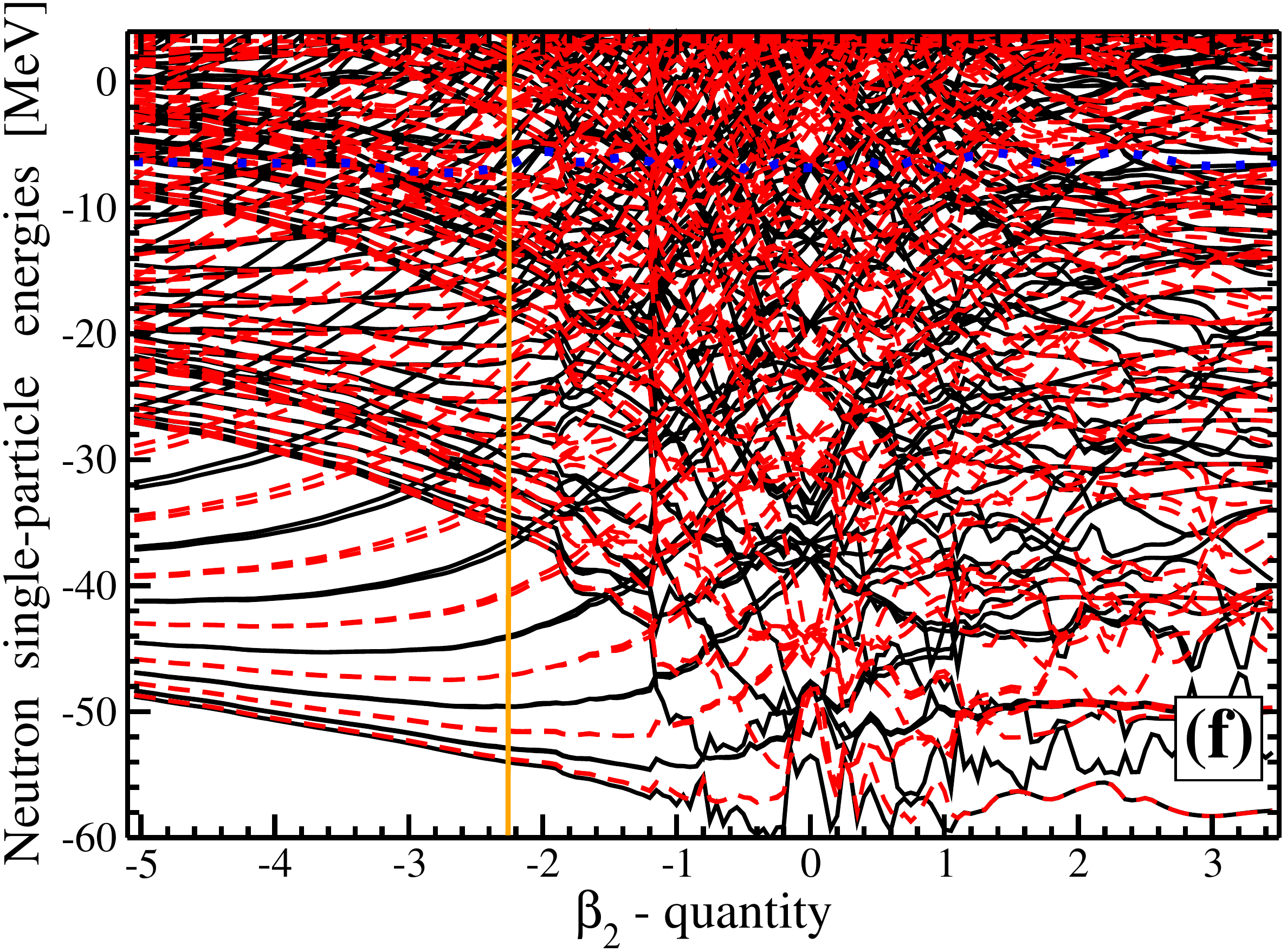}
\caption{Proton and neutron single-particle energies, i.e., the diagonal
elements of the single-particle Hamiltonian h in the canonical basis 
\cite{RS.80}, for the lowest in 
total energy solution in the $^{348}138$ nucleus calculated as a function 
of the $\beta_{2}$ quantity. Black solid  and red dashed lines are used
for positive- and negative-parity states, respectively. The dominant 
components $\Omega [N,n_z, \Lambda]$ of the wave functions (as
calculated at LEMAS) are shown by blue and green colors for the
positive- and negative parity orbitals, respectively. The energies $E_F$ 
of the respective Fermi levels are shown by blue dotted lines. The vertical 
orange lines and orange arrows are drawn at the $\beta_2$ value 
corresponding to LEMAS. Shell gaps are indicated by encircled numbers.
}
\label{138-348-nil}
\end{figure*}

\begin{figure*}[htb]
\centering
\includegraphics[angle=0,width=8.8cm]{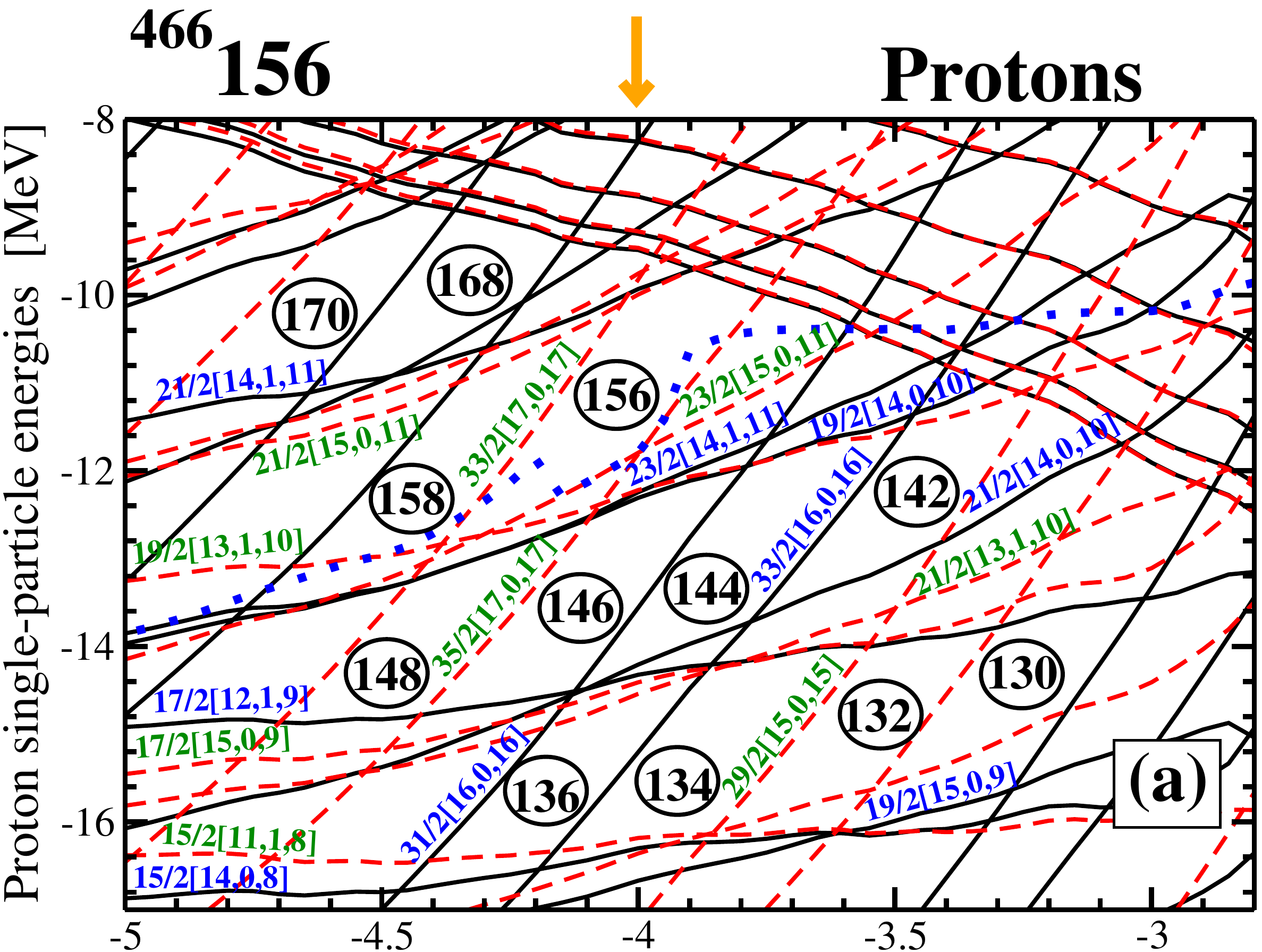}
\includegraphics[angle=0,width=8.7cm]{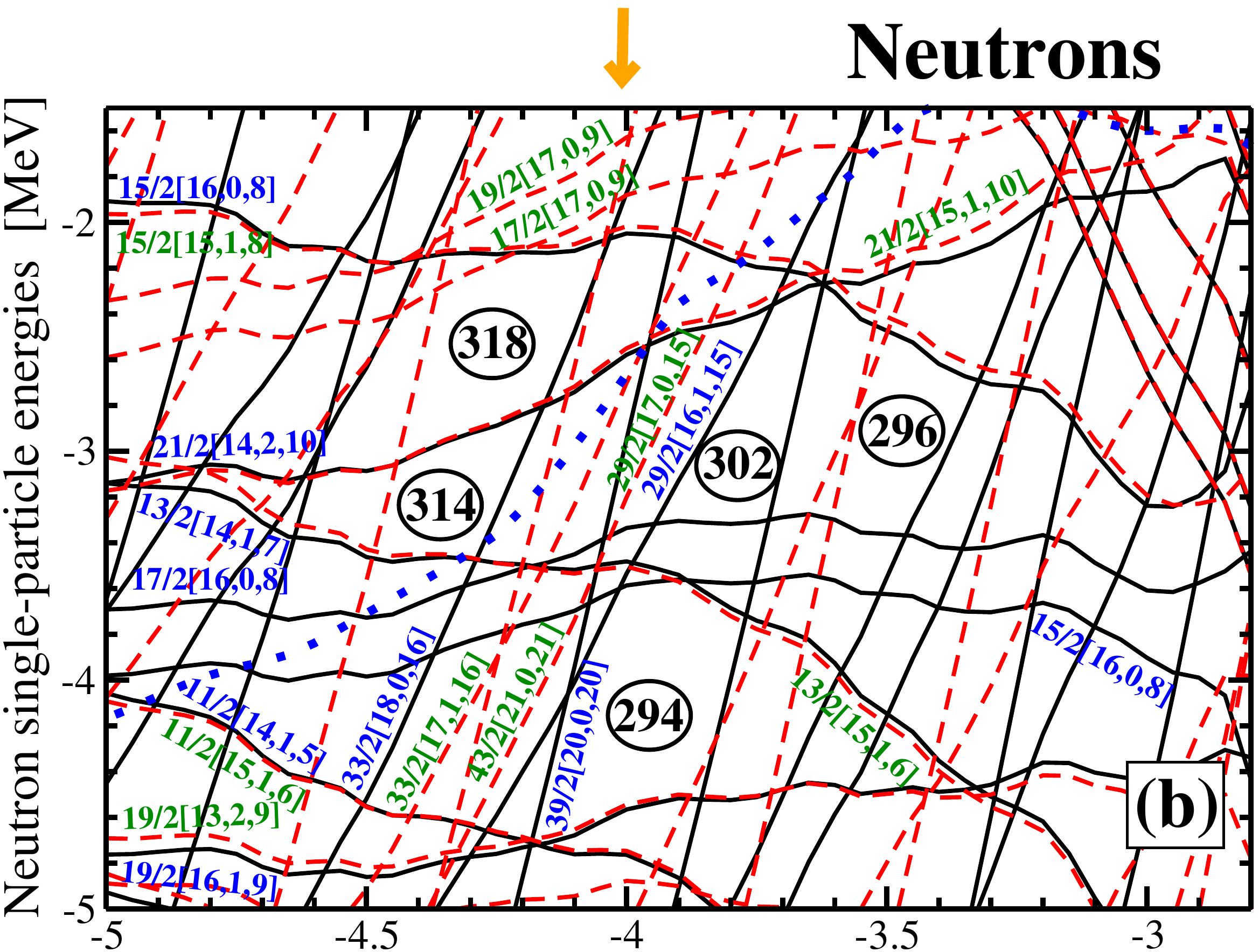}
\includegraphics[angle=0,width=8.8cm]{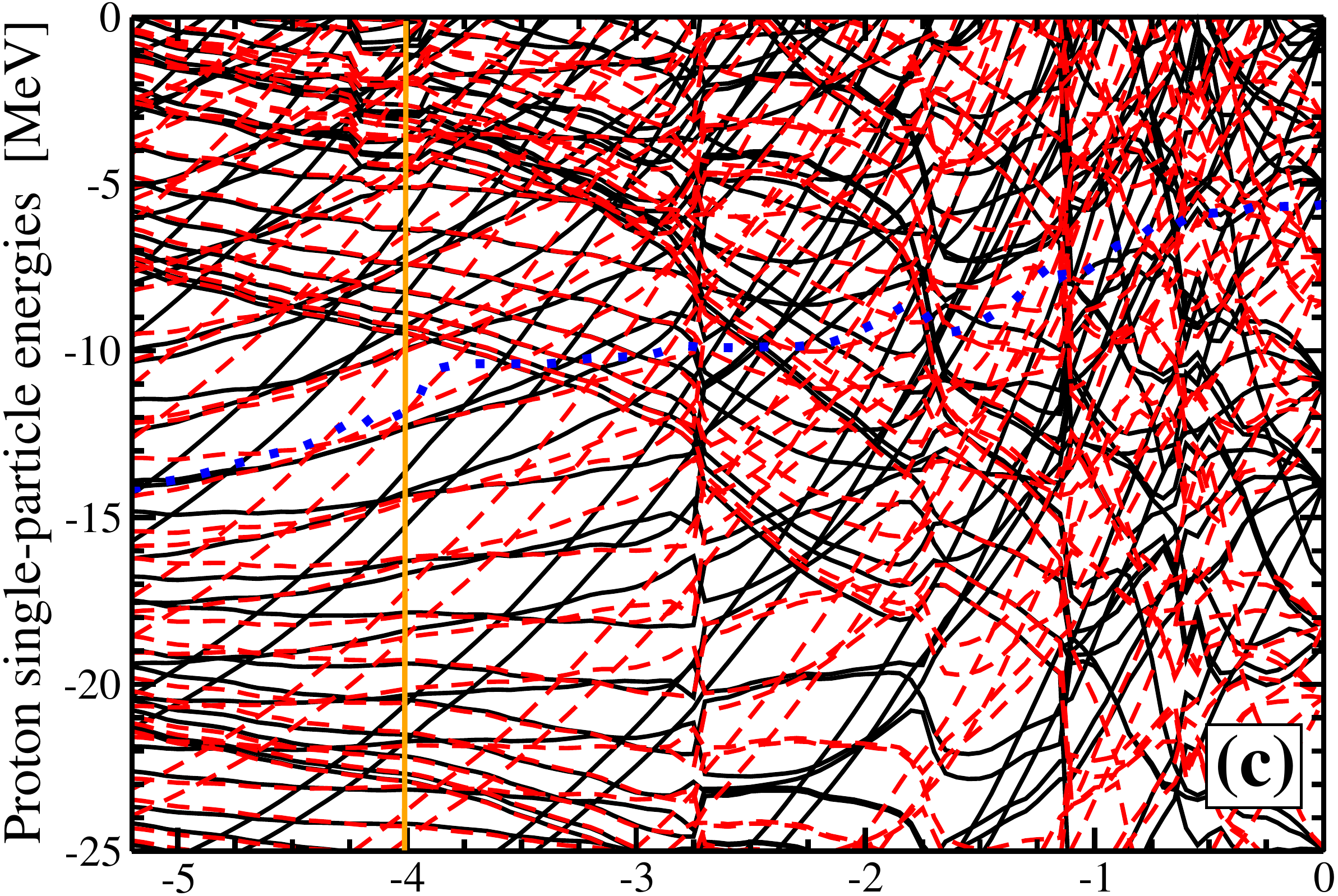}
\includegraphics[angle=0,width=8.8cm]{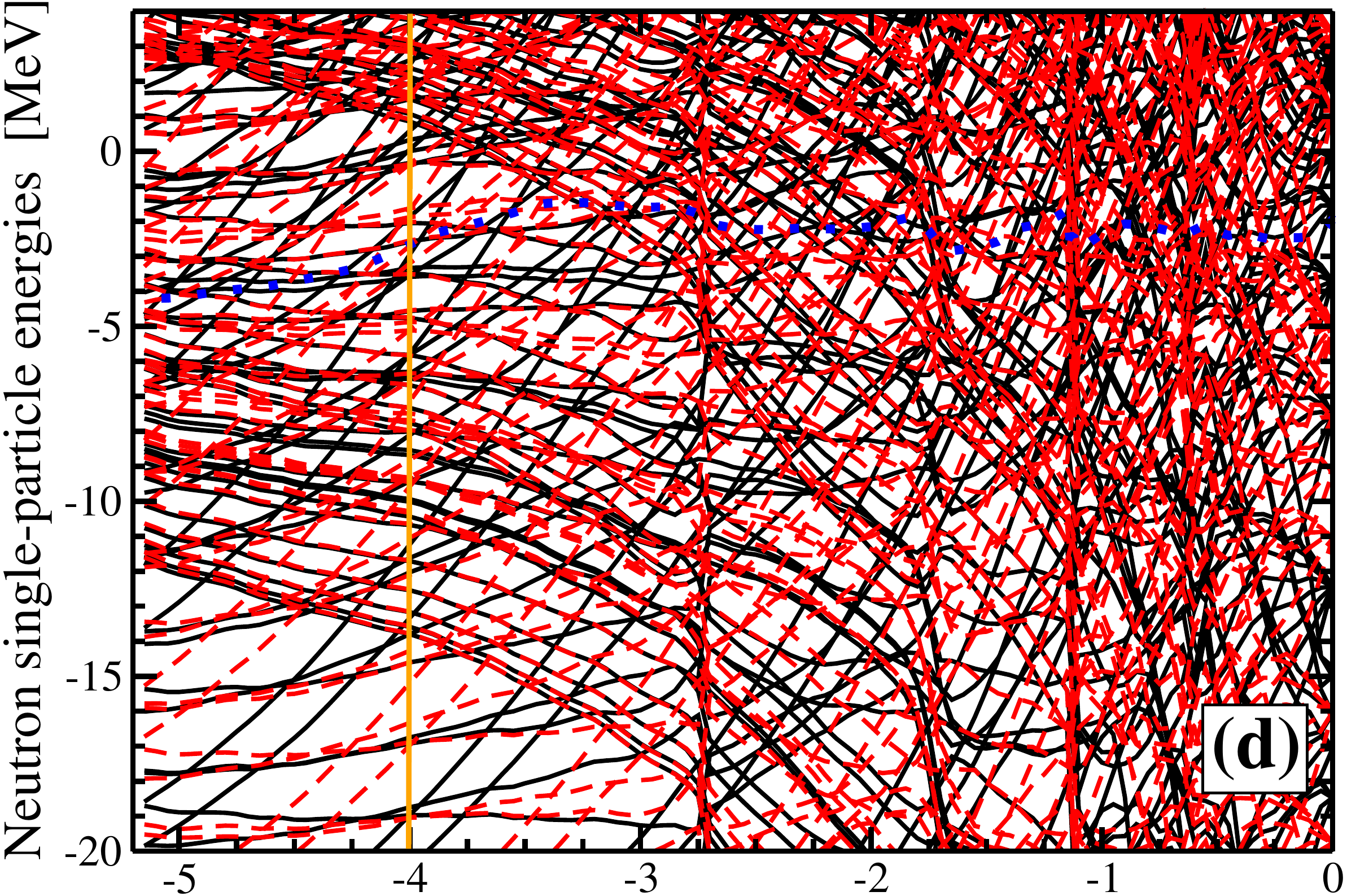}
\includegraphics[angle=0,width=8.8cm]{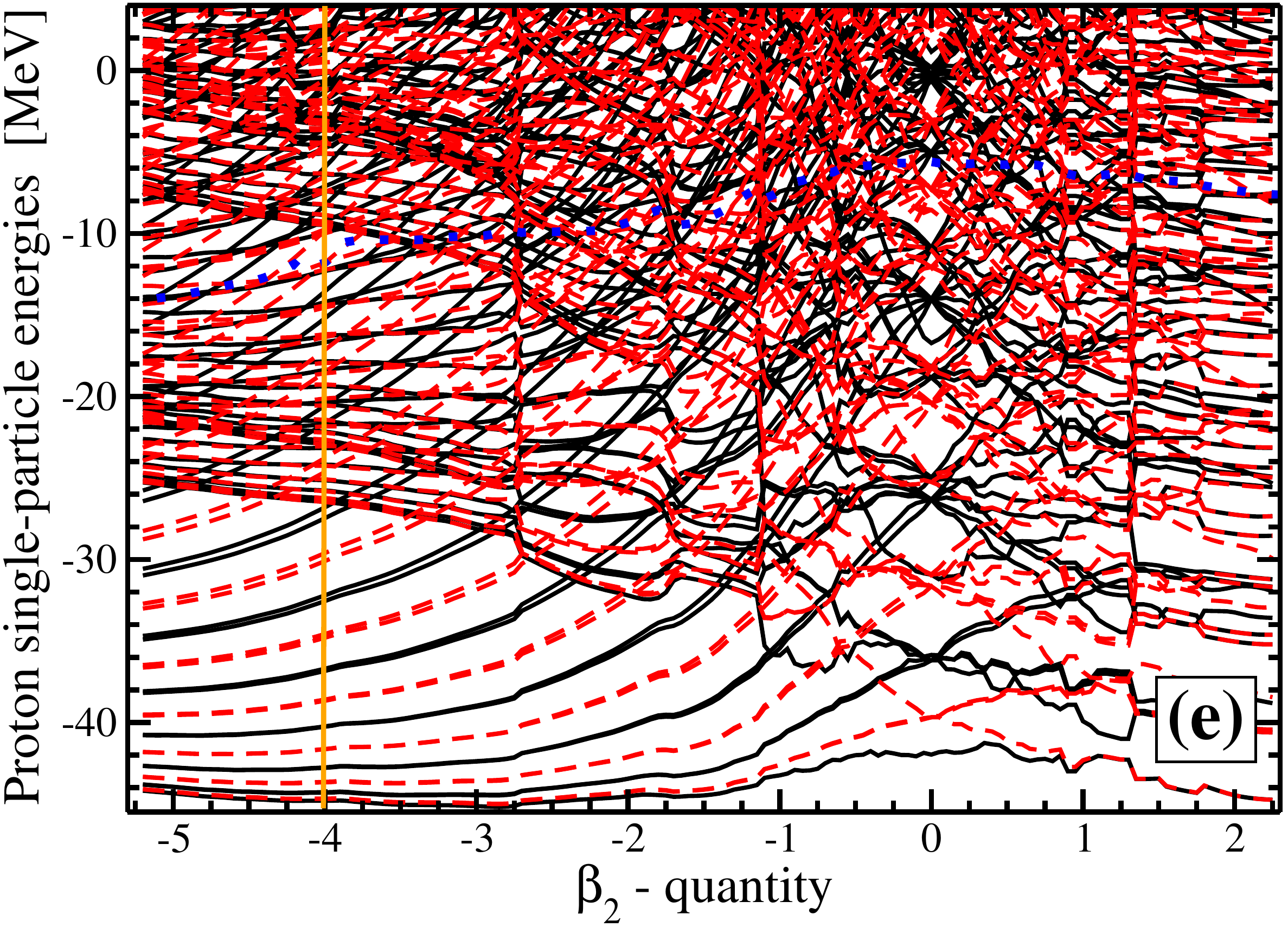}
\includegraphics[angle=0,width=8.8cm]{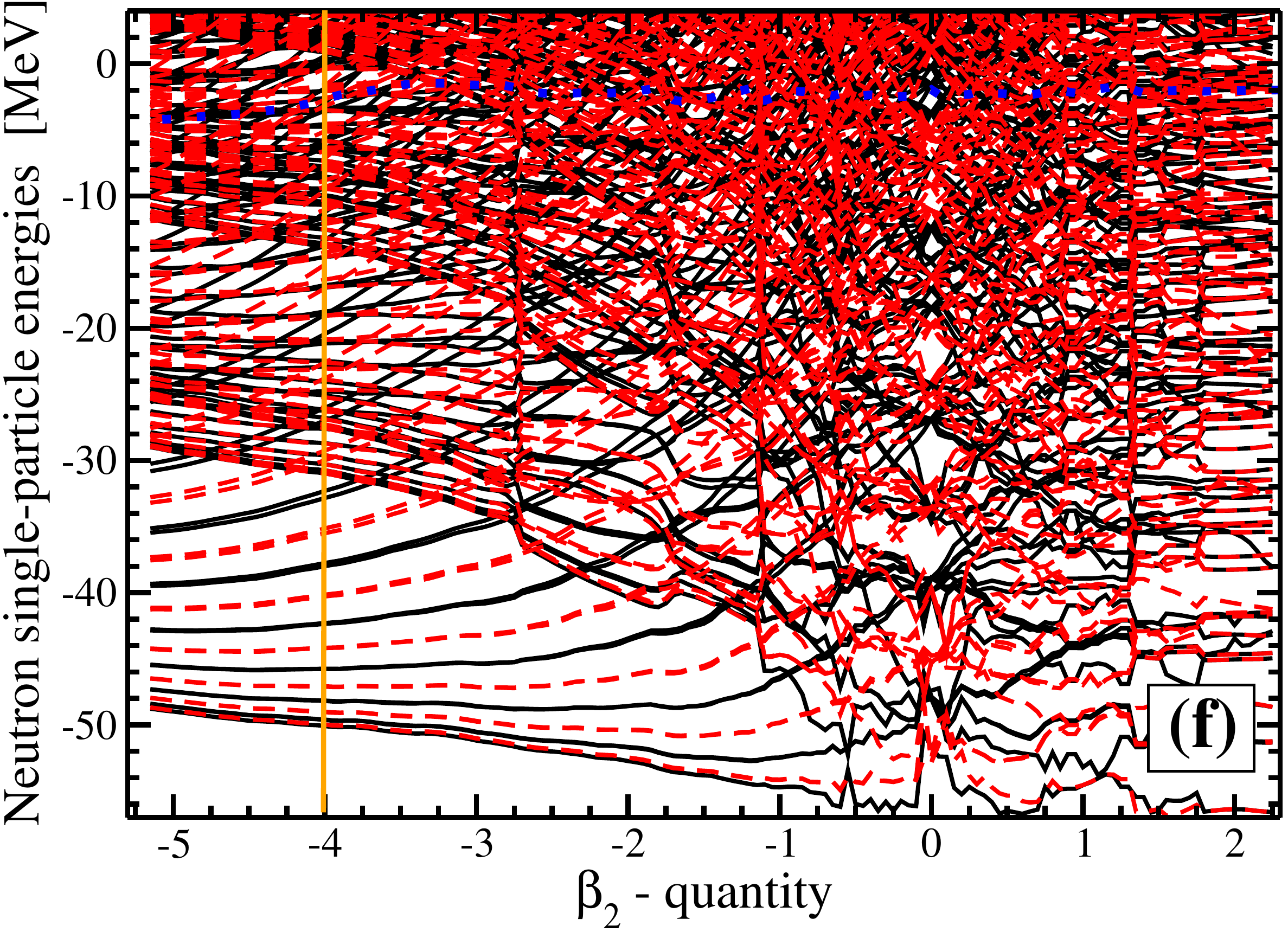}
\caption{The same as Fig.\ \protect\ref{138-348-nil} but for the $^{466}$156 nucleus.
}
\label{156-466-Nil}
\end{figure*}

  In such a situation it is useful to get some insight from the studies of classical 
liquid droplets. Thin toroidal droplets exhibit Plateau-Rayleigh instabilities: when 
the outer circumferences of toroid is equal to an integer ($n$) times of the wavelength 
$\lambda_c$ of unstable mode, the toroidal droplet will eventually fission into
$n$ spherical droplets \cite{YB.11} (see also Ref.\ \cite{MF.13} for the results obtained 
for liquid toroidal droplets suspended in another liquid). Note that in classical toroidal liquid droplets the 
Plateau-Rayleigh instability disappears for sufficiently fat tori ($R/d \leq 2$) while the 
shrinking mode is present for all aspect ratios \cite{YB.11}. These features have been confirmed 
in experimental studies of stability of both toroidal droplets in a viscous liquid \cite{PF-N.09} as 
well as melted polymer rings \cite{MLTSD.10}. The instability with respect 
of so-called sausage deformations \cite{Wong.73} in nuclear physics leading to 
multifragmentation\footnote{In this context it would be interesting to see whether the 
observed multifragmentation of high-spin configurations  of $^{28}$Si into 
7 $\alpha$-particles \cite{28Si-tori.exp}  represents the analog of Plateau-Rayleigh 
instabilities of toroidal droplets in nuclear physics.}  is an analog of the Plateau-Rayleigh 
instabilities. Thus, these results  suggest that such instabilities are less
important for fat toroidal nuclei [characterized by low (in absolute sense)  values
of $\beta_2 > -2.5$ and located in the  $Z\approx 134, N\approx 210$ region
(see Fig.\ \ref{torus-def})]  but become more critical (and probably fatal) for thin toroidal 
nuclei characterized by large (in absolute sense) values  of $\beta_2$. The latter
type of nuclei become dominant both with increasing proton number $Z$ and in 
proton- and neutron-rich nuclei  (see Fig.\ \ref{torus-def}). The former 
suggestion is in line with the results of triaxial RMF+BCS calculations for  the 
$^{354}$134 and $^{348}$138 nuclei, which have 4.4 and 8.54 MeV fission 
barriers for non-axial distortions, respectively (see Ref.\ \cite{AAG.18}).

   However, it is necessary to recognize that fully quantum mechanical calculations 
based on the density functional theory are needed for establishing the stability of 
toroidal nuclei with respect of sausage deformations. Toroidal liquid droplets
have a uniform density and the tube of torus has a cylindrical form \cite{YB.11}. On
the contrary, the DFT calculations paint much more complicated picture. First,
the density rapidly changes across the tube of the torus  with considerable 
mismatch between proton and neutron densities (see Fig.\ \ref{densit-tori} in 
the present paper, Fig. 2(c) and (d) in Ref.\ \cite{AAG.18} and Fig. 9 in Ref.\ 
\cite{SW.15}) which are defined by the occupation of underlying proton and 
neutron single-particle orbitals.  The description of such a situation on the level
of liquid-drop model would require the model based on two (proton and neutron) 
fluids with the specification of functional dependencies of their densities on the 
position in the tube of the torus. Second, not in all cases the tube of the torus is 
represented by a perfect cylinder (see Fig. 2 in Ref.\ \cite{AATG.19}). This
may lead to an enhanced stability against sausage deformations since experimental
studies of toroidal liquid droplets show that oblong cross section of the torus 
tube suppresses Plateau-Rayleigh instabilities as compared with circular one 
\cite{TPQC.13}. Because of above mentioned reasons the analysis of Ref.\ 
\cite{Wong.73} indicating the instability of toroidal nuclei with respect of sausage 
deformations in the liquid drop model should not be taken at face value. Note
also that this analysis considers only the nuclei with $Z<120$ in which toroidal 
shapes are formed at high excitation energies with respect of the ground states 
while the toroidal shapes in the majority of hyperheavy nuclei are expected to
be the ground states. Moreover, the quantum shell
effects can counterbalance the potential instabilities towards sausage 
deformations at some combinations of proton and neutron numbers and 
deformations \cite{Wong.73,AAG.18}.

\subsection{Shell structure of toroidal hyperheavy nuclei}
\label{Tori-shell-structure}

\begin{figure*}[htb]
\centering
\includegraphics[angle=0,width=8.5cm]{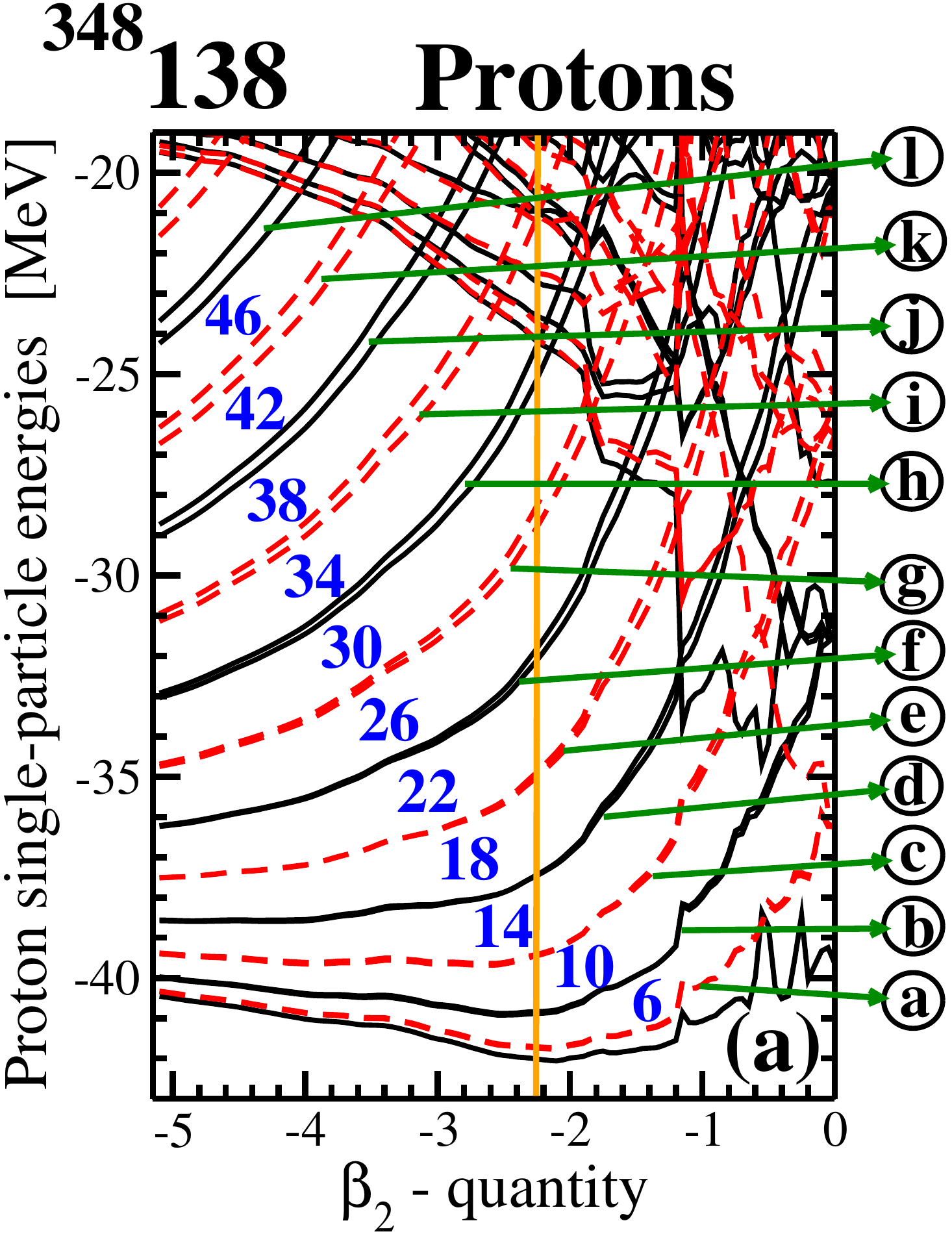}
\includegraphics[angle=0,width=8.5cm]{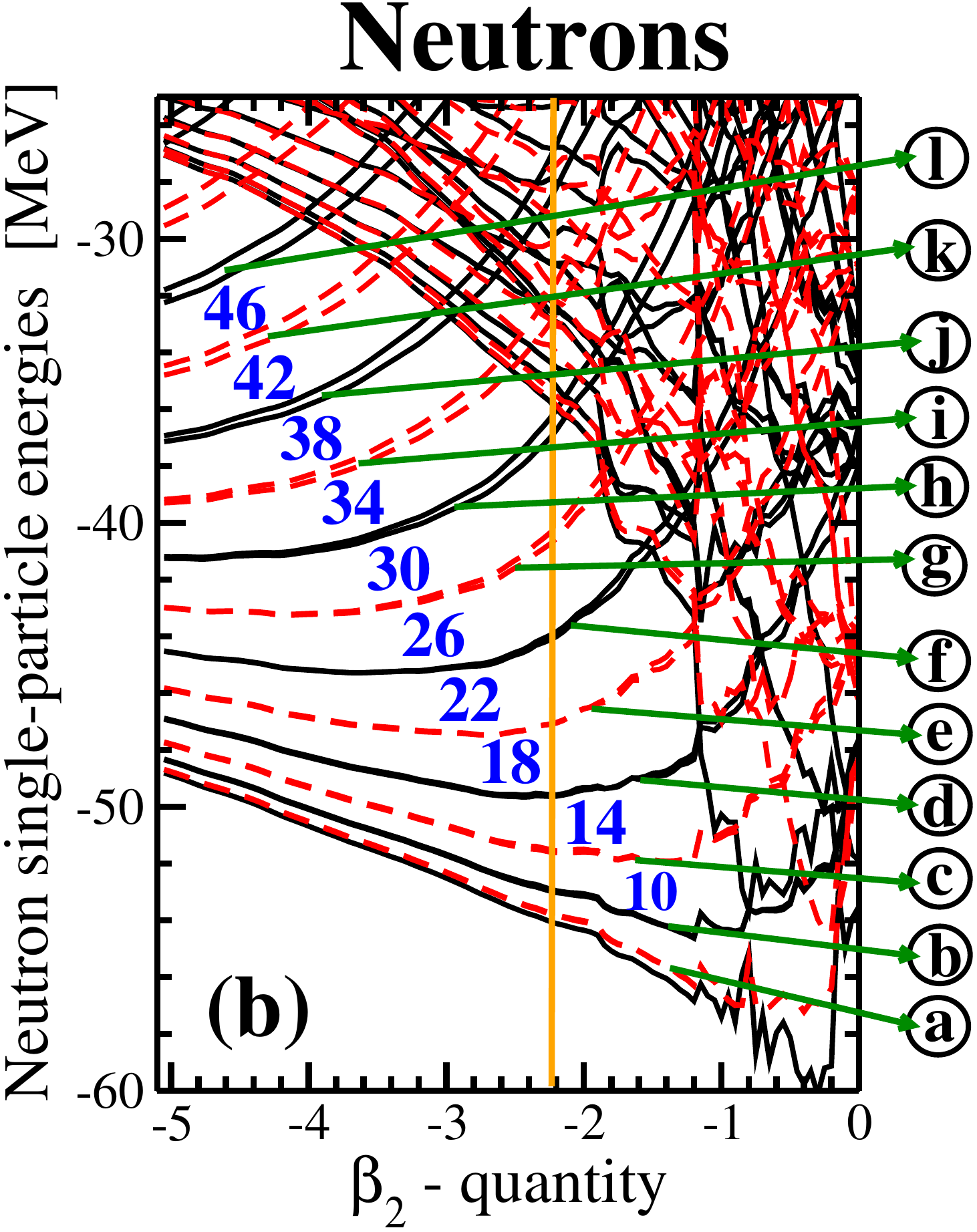}
\caption{The same as Fig.\ \ref{138-348-nil} but for the single-particle
states located in the bottom part of nucleonic potential and in the $\beta_2$ 
range from $-5.1$ up to 0.0. Toroidal shell gaps are shown by bold blue 
numbers. Encircled letters and green arrows are used to indicate the pairs 
of the single-particle states which are almost degenerate in energy. The structure 
of these states are shown in Table \ref{138-348-structure-bottom-pot}.}
\label{138-348-bot-poten}
\end{figure*}

  It is well known that the presence of large gaps in  proton and neutron  
single-particle energies leads to an extra stability of nuclear systems.  So far, 
the analysis of toroidal shell structure at spin $I=0$ has been performed 
in light nuclei \cite{Wong.73,SW.15},  in the intermediate mass region nuclei
\cite{WS.18} and  in superheavy $Z\approx 120$ nuclei \cite{KSW.17,SWK.17}. 
Such an analysis was based either on phenomenological toroidal single-particle 
potential (see Refs.\ \cite{Wong.73,SW.15,WS.18}) or on Skyrme DFT calculations
(see Refs.\ \cite{SW.15,KSW.17,SWK.17}).  Large gaps in the single-particle
energies have been found at toroidal shapes in all these regions. For example,
in light nuclei  
these energy  gaps give rise to “toroidal shells” at  “magic” nucleon 
numbers $N=2(2m+1)$ with $m$ being integer satisfying the condition 
$m\geq 1$ \cite{Wong.73}. The extra stability associated with toroidal shells leads 
to local energy minima at toroidal  shapes in many  nuclei either at spin
zero \cite{Wong.73,Wei_2010} or in some high spin isomer states
\cite{SW.15}. However, in all these nuclei such minima are
located at high excitation energies with respect of ellipsoidal-like ground state.

   However, the situation changes in hyperheavy nuclei in which the 
ground states are expected to have toroidal shapes. Thus, it is very 
important to investigate  shell structure of toroidal hyperheavy 
nuclei. In particular, it would be interesting to see whether there are
large shell gaps or reduced density of the single-particle states
at specific particle numbers which
could provide an extra stability with respect of potential instabilities 
originating from sausage deformations. One should also remember that even 
if hyperheavy nuclei are unstable with respect of sausage deformations 
in the liquid drop model, they can be stabilized by quantum shell corrections.
The best known example of such a situation are superheavy nuclei: they 
are unstable in the liquid drop model but are relatively stable in fully quantum 
mechanical picture which includes shell corrections \cite{SP.07,GMNORSSSS.19}.
  
    The analysis presented in Sec.\ \ref{toroid-distr} suggests that it is 
more likely to get  potentially stable toroidal nuclei when their shapes in
corresponding minima are characterized by small absolute $\beta_2$ values
(or small aspect ratio $R/d$).
The toroidal  $^{354}$134 and $^{348}$138 nuclei are representative 
cases of such shapes (see Fig. 1 in supplemental material to Ref.\ 
\cite{AAG.18} and  Fig. 19 in Ref.\ \cite{AATG.19}). Triaxial RMF+BCS
calculations of Refs.\ \cite{AAG.18,AATG.19} suggest that these  two
nuclei are expected  to be relative stable with respect of non-axial 
distortions (even-multipole sausage deformations) with calculated
fission barriers being equal to 4.4 and 8.54 MeV,  respectively.
Enhanced stability of the $^{348}$138 nucleus is a reason why we start the 
analysis of toroidal shell structure from this nucleus which is characterized
by moderately compact toroidal shapes [see Figs.\ \ref{densit-tori}(a) and (b)].  
We also consider toroidal shell structure in the  $^{466}$156 nucleus. The LEMAS 
of this nucleus is characterized  by non-compact toroidal shapes with large $R/d$ 
aspect ratio [see Figs.\ \ref{densit-tori}(e) and (f)], but there is also an excited 
minimum B (see Fig.\ \ref{156-466-pot-b}) which is characterized by very compact 
toroidal shapes with very small hole in the center [see Figs.\ \ref{densit-tori}(c) and
(d)].

\begin{table}[h]
\begin{center}
\caption{
The dominant components of the wave functions of  nearly-degenerate pairs 
of the single-particle states of same parity indicated by the letters in Fig.\ 
\ref{138-348-bot-poten}. They are defined at the $\beta_2$ value corresponding 
to LEMAS. The states forming the pair are shown in the columns labelled as 
"1$^{st}$ state"  and  "2$^{d}$ state". 
\label{138-348-structure-bottom-pot}
}
\begin{tabular}{|c|c|c|} \hline
        &   1$^{st}$ state   &   2$^{d}$ state     \\ \hline  
  (a)  & 1/2[ 9,0,1]    & 3/2[ 9,0,1] \\
  (b)  & 3/2[10,0,2]  & 5/2[10,0,2] \\
  (c)  & 5/2[ 9,0,3] & 7/2[ 9,0,3] \\
  (d)  & 7/2[10,0,4] & 9/2[10,0,4]  \\
  (e)  & 9/2[ 9,0,5] &11/2[ 9,0,5]      \\
  (f)  &11/2[10,0,6] &13/2[10,0,6]   \\
  (g)  &13/2[11,0,7] &15/2[11,0,7]    \\
  (h)  &15/2[10,0,8] &17/2[10,0,8]  \\
  (i)  &17/2[11,0,9] &19/2[11,0,9]      \\
  (j)  &19/2[12,0,10] &21/2[12,0,10]  \\
  (k)  &21/2[13,0,11] &23/2[13,0,11]   \\
  (l)  &23/2[12,0,12] &25/2[14,1,12]  \\ \hline
\end{tabular}
\end{center}
\end{table}

   The Nilsson diagrams for these nuclei are shown in Figs.\ 
\ref{138-348-nil} and \ref{156-466-Nil}. 
In order to illustrate the differences between shell structure of toroidal and
ellipsoidal-like nuclei,  bottom panels display proton and neutron single-particle
states in the very large energy and $\beta_2$ ranges. They are shown from 
the bottom of respective potentials up to 4 MeV energy above the
continuum threshold and from $\beta_2=-5.1$, corresponding to toroidal
nuclei with large $R/d$ aspect ratio, up to $\beta_2=+3.5$ in the $^{348}$138 
nucleus and up to $\beta_2=2.25$ in the $^{466}$156 nuclei. These large
positive $\beta_2$ values correspond to
pre-fissioning configurations  with well pronounced neck (see, for example,
density distribution at the position F of Fig.\ \ref{156-466-pot-b}). Middle and top panels
of Fig.\ \ref{138-348-nil} show the regions of interest in blown-up scale.
The analysis of these figures reveals  the general features which are discussed
below.

  Toroidal shell structure (especially the one for the shapes with large 
$R/d$ aspect ratio) has much more pronounced regular features 
as compared with the shell structure of ellipsoidal-like shapes in the
range of the $\beta_2$ values from $\approx -1.15$ up to $\approx 1.5$ 
which looks quite chaotic for deformed shapes [see Fig.\ \ref{138-348-nil}(e) and (f)
and  Fig.\ \ref{156-466-Nil}(e) and (f)]. At higher $\beta_2$ values typical 
features of shell structure of two-center shell model (see, for example,  
Ref.\ \cite{Gher.03}) are seen.

\begin{figure*}[htb]
\centering
\includegraphics[angle=0,width=8.7cm]{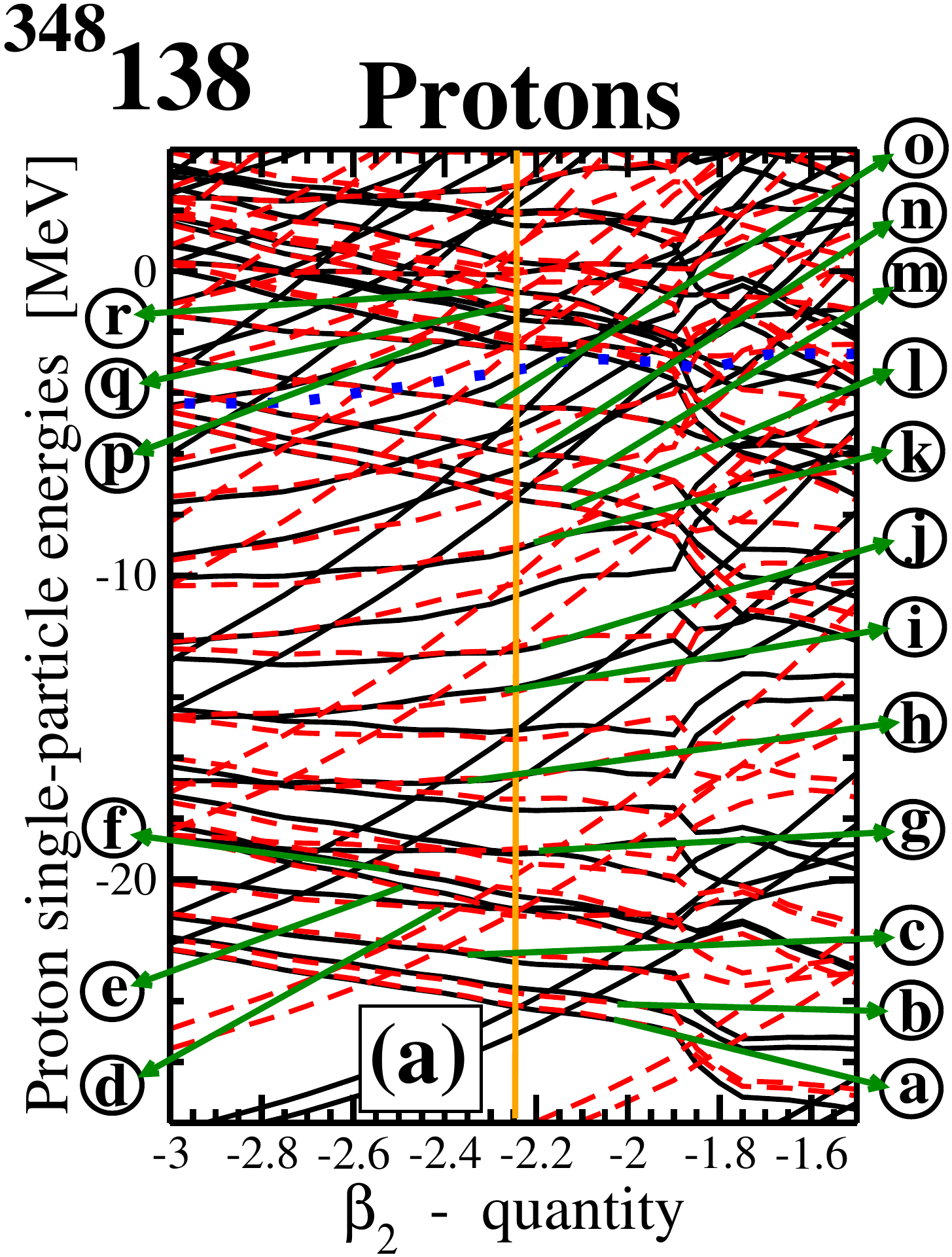}
\includegraphics[angle=0,width=8.5cm]{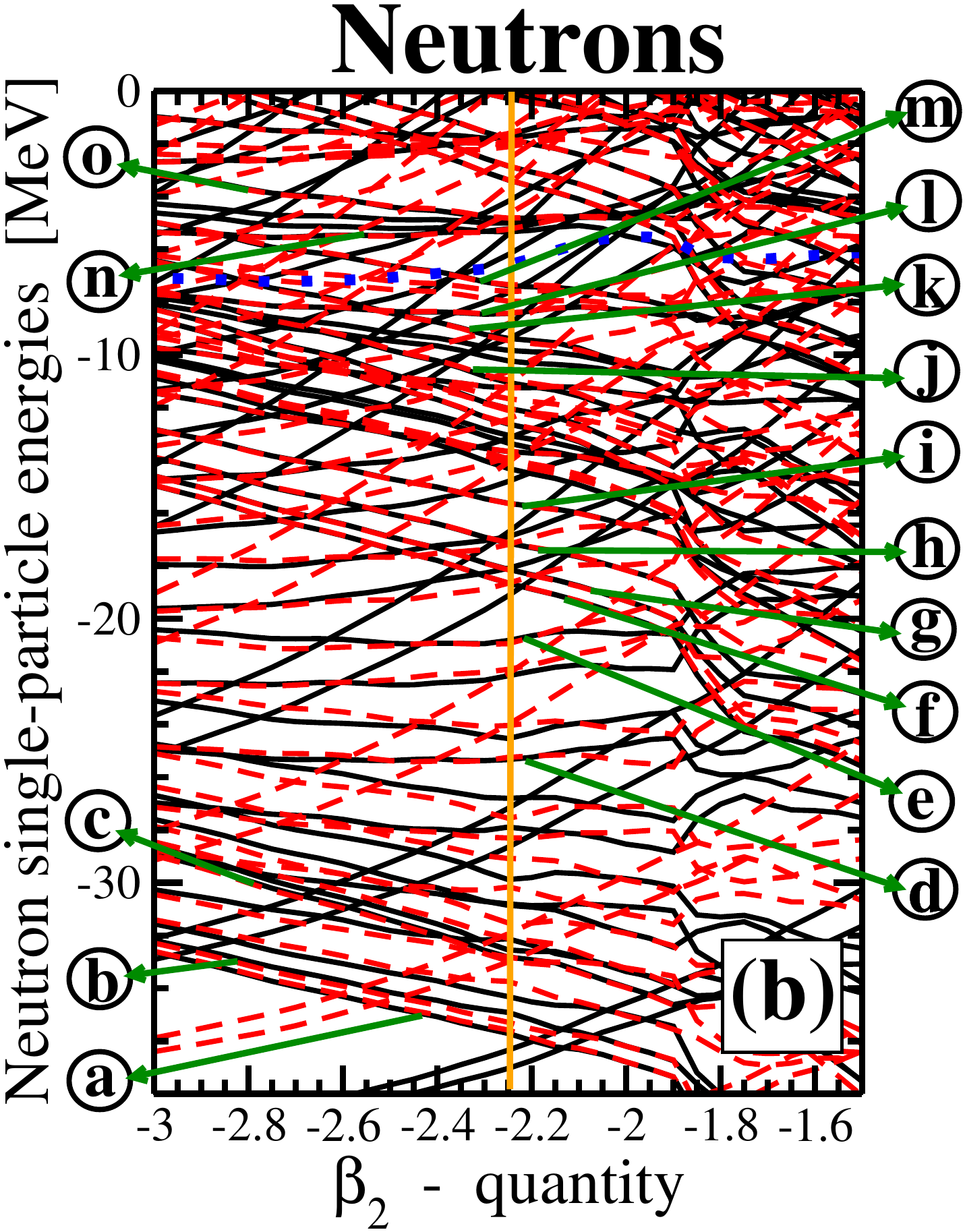}
\caption{The same as Fig.\ \ref{138-348-nil} but for the single-particle
states located in the intermediate energy range of nucleonic potential 
of the $^{348}$138 nucleus and in the $\beta_2$ range from $-3.0$ up to 
$-1.5$. Encircled letters and green arrows are used to indicate the pairs of 
single-particle states of opposite parity which are almost  degenerate in energy. 
The structure of  these states  is shown in Table  \ref{table-mid-energy}.}
\label{138-348-nils-med-energy}
\end{figure*}

\begin{table}[h]
\begin{center}
\caption{The dominant components of the wave functions of the
pairs of proton and neutron single-particle states of opposite
parity which are almost  degenerate in energy.  
The states forming 
the pair are shown in the columns labelled as "parity$=+$"  and  "parity$=-$". 
These pairs are indicated in Fig.\ 
\ref{138-348-nils-med-energy}.  Note that
the dominant components of the wave functions 
are defined at the $\beta_2$ 
value corresponding to LEMAS. 
\label{table-mid-energy}
}
\begin{tabular}{|c|c|c|c|c|} \hline
       &\multicolumn{2}{|c|}{Neutron(${\nu}$)}&\multicolumn{2}{|c|}{Proton(${\pi}$)}\\\hline
 1    &         2     &        3     &       4       &       5       \\ \hline 
        &  parity$= +$& parity$=-$  & parity$=+$ & parity$=-$     \\\hline 
  (a)  & 1/2[8,0,0]    & 1/2[ 9,0,1]  & 1/2[8,0,0]   & 1/2[ 9,0,1] \\\hline
  (b)  & 3/2[8,0,2]    & 3/2[ 9,0,1]  & 3/2[6,1,1]   & 3/2[ 9,0,1] \\\hline
  (c)  & 1/2[6,1,1]    & 1/2[ 7,1,0]  & 5/2[8,0,2]   & 5/2[ 7,1,2] \\\hline 
  (d)  &13/2[10,0,6] &13/2[ 9,1,6] & 7/2[8,1,3]   & 7/2[ 9,0,3] \\\hline 
  (e)  &13/2[8,1,7]   &13/2[11,0,7] & 1/2[6,1,1]  & 1/2[ 7,1,0] \\\hline  
  (f)   & 1/2[10,1,1]  & 1/2[ 9,1,0]  & 3/2[6,1,1]   & 3/2[ 7,1,2] \\\hline
  (g)  & 3/2[10,1,1]  & 3/2[11,0,1]  &9/2[10,0,4]  & 9/2[ 7,1,4] \\\hline 
  (h)  & 5/2[10,1,3]  & 5/2[ 9,1,2]  &11/2[8,1,5]   &11/2[ 9,0,5] \\\hline 
  (i)   & 7/2[10,1,3]  & 7/2[11,0,3]  &13/2[10,0,6] &13/2[ 9,1,6] \\\hline 
  (j)   & 5/2[10,1,3]  & 5/2[ 9,2,3]   &11/2[10,0,6] &11/2[ 9,1,6] \\\hline
  (k)  & 7/2[8,2,4]    & 7/2[11,1,4]  &13/2[8,1,7]  &13/2[11,0,7] \\\hline
  (l)   &13/2[10,2,6] &13/2[11,1,6]  & 1/2[10,1,1] & 1/2[ 9,1,0] \\\hline
  (m) & 9/2[10,1,5]  & 9/2[ 9,2,5]   & 3/2[10,1,1] & 3/2[11,0,1] \\\hline
  (n)  &15/2[12,1,7] &15/2[11,2,7]  & 5/2[10,1,3] & 5/2[ 9,1,2] \\\hline
  (o)  &11/2[10,2,6] &11/2[11,1,6]  & 7/2[10,1,3] & 7/2[11,0,3] \\\hline
  (p)  &                    &                    & 9/2[12,0,4] & 9/2[11,1,4] \\\hline
  (q)  &                    &                    & 1/2[10,1,1] & 1/2[ 7,2,1] \\\hline
  (r)   &                    &                    & 3/2[8,2,2]   & 3/2[ 9,1,2] \\\hline
\end{tabular}
\end{center}
\end{table}

  The bunching of the pairs of the orbitals of the same parity with 
dominant structure of $\Omega[N,n_z,\Lambda]$ and  
$(\Omega+1) [N,n_z,\Lambda]$ with $N\geq 9$ and $n_z=0$
(see Table \ref{138-348-structure-bottom-pot})\footnote{The
  only exception is the last pair of the states
  shown in Table \ref{138-348-structure-bottom-pot} for which
  the $2^d$ state has $n_z=1$.}
 leads to the appearance of toroidal shell gaps at particle numbers 6, 10, 14, 18,
22, 26, 30, 34, 38, 42, 46 at the bottom of proton and neutron potentials 
(see Fig.\ \ref{138-348-bot-poten}). These 
gaps exist in a large range of the $\beta_2$ values; this is  contrary to 
the case of shell gaps for ellipsoid-like shapes which are localized in 
deformation.
They  are also consistent with the ones obtained in the study of
toroidal shapes in light nuclei within  toroidal harmonic
oscillator shell model\footnote{This type of the model has been described
before either as shell model based on radially displaced harmonic oscillator
potential \cite{Wong.73,SW.15} or harmonic oscillator toroidal shell model
\cite{WS.18}.
We abbreviate it here as toroidal harmonic oscillator shell model in order
to stress that the basis of it is formed by the eigenvectors of  radially 
displaced (toroidal) harmonic oscillator potential and that in this respect 
it differs from standard shell model which uses traditional harmonic oscillator
for basis  set expansion.} and Skyrme DFT
(see Figs. 1, 5 and 6 in Ref.\ \cite{SW.15} and Fig.\ 12 in Ref.\ \cite{Wong.73}).
The energies of these pairs of orbitals generally decrease with increasing the absolute
value of $\beta_2$; the only exception from this rule are several lowest
pairs of orbitals located at the bottom of neutron and proton potentials
(see Fig.\ \ref{138-348-bot-poten}). Note that the pairs of the orbitals with 
dominant structure of  $\Omega[N,n_z,\Lambda]$ and  $(\Omega+1) [N,n_z,\Lambda]$ 
are almost near-degenerate in energy  at the bottom of  potential and that this 
near-degenaracy increases with increasing absolute value of $\beta_2$.  There is 
also an alternation of the pairs of the states with positive and negative parities with 
increasing energy (see Fig.\ \ref{138-348-bot-poten}).  These features
of the shell structure dominate the physics of toroidal shapes in light
to medium mass nuclei (see Refs.\ \cite{Wong.73,SW.15}).

\begin{figure*}[htb]
\centering
\includegraphics[angle=0,width=8.5cm]{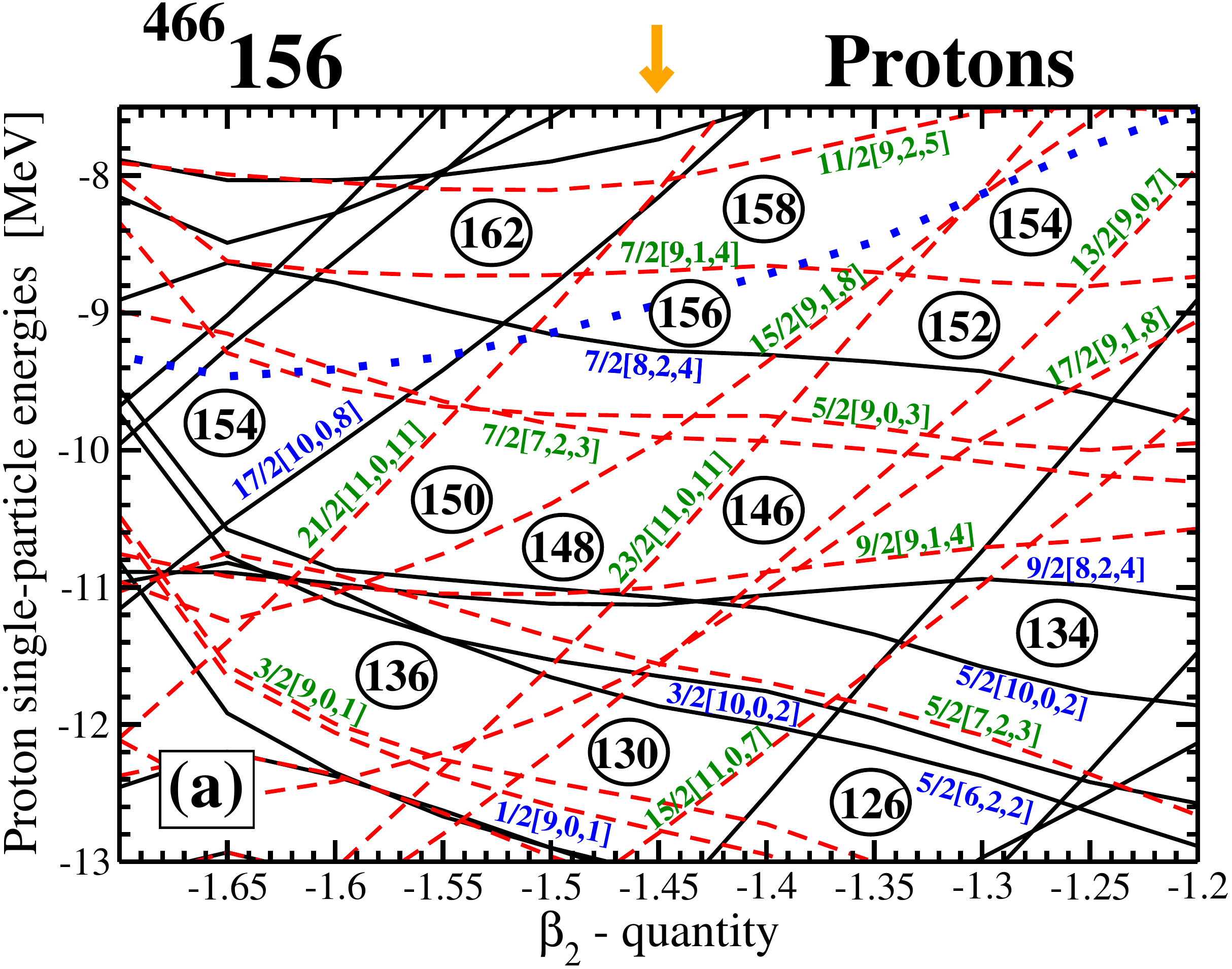}
\includegraphics[angle=0,width=8.5cm]{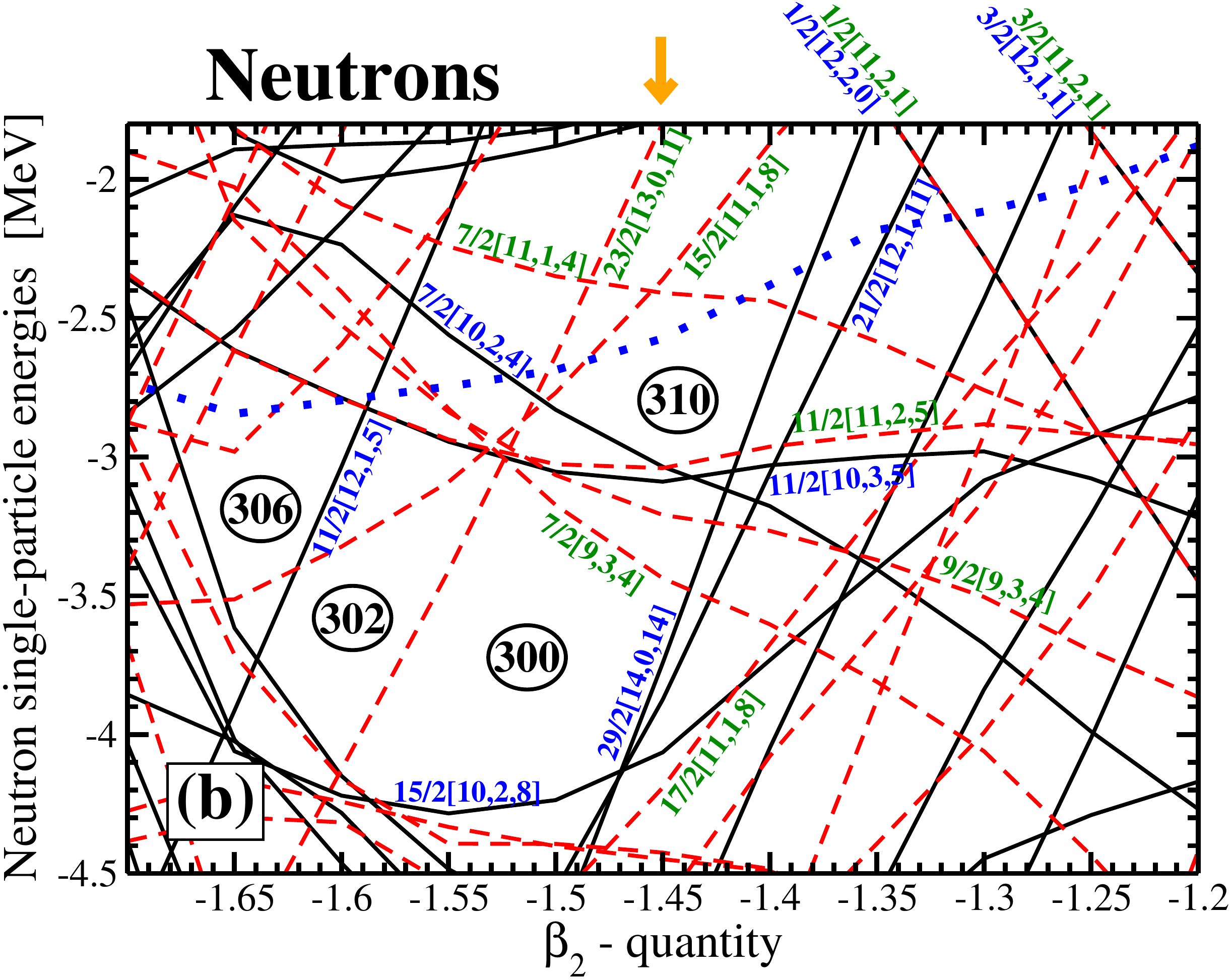}
\caption{The same as Fig.\ \protect\ref{156-466-Nil} but for the single-particle 
states active in the vicinity of the Fermi levels corresponding to the minimum B of 
Fig.\ \protect\ref{156-466-pot-b}.
}
\label{156-466-Nil-min-B}
\end{figure*}

  The general features of the pairs of orbitals with dominant structure of
$\Omega[N,n_z,\Lambda]$ and  $(\Omega+1) [N,n_z,\Lambda]$ with
$n_z=0$ changes drastically in the energy ranges between -20 MeV 
and 0 MeV for protons and between -25 MeV and  0 MeV for neutrons
(see Figs.\ \ref{138-348-nil}, \ref{156-466-Nil} and \ref{138-348-nils-med-energy})
for $\beta_2 \leq -1.8$.  First, their energies decrease
almost linearly with increasing absolute value of $\beta_2$. 
Second, there is a periodic pattern in the change of the orbitals:
with increasing energy two positive parity orbitals are followed
by two negative parity orbitals and then by two positive parity 
orbitals and so on.  Third, these orbitals form the grating-like structure 
with almost  equidistant in energy spacing between them.

  In the same energy range as discussed in previous paragraph, there are other 
single-particle structures dictated by the symmetries of the toroid.  These are almost 
degenerate in energy single-particle states of opposite parity (see Fig.
\ref{138-348-nils-med-energy} and Figs.\ \ref{138-348-nil} and 
\ref{156-466-Nil}) with dominant 
structures of the wave functions given by $\Omega[N,n_z,\Lambda]$ and  
$\Omega [N',n'_z,\Lambda']$ where the following conditions $N'=N\pm1$,
$|\Lambda'-\Lambda|=0$ or 1 and $|n'_z-n_z|=0$ or 1 are typically satified
(see Table \ref{table-mid-energy}). These states change their energy very
slowly when the $\beta_2$ value is varied. Note that such pairs of the
states are also present in the Skyrme DFT calculations of toroidal shapes 
in the $^{304}$120 nucleus (see Figs. 3 and 4 in Ref.\ \cite{SWK.17}).

   The presence of the two types of the single-particle states discussed in previous 
two items is mostly responsible for the shell structure and shell gaps in the 
intermediate energy range of proton and neutron potentials.  This leads to the 
existence of many gaps in the single-particle spectra which are  quite large.  
These are proton $Z=120$, 130, 134, 138, 140, 144 and 148 shell gaps with 
typical size of approximately 1 MeV  and neutron $N=206$, 210 
and 214 shell gaps which are larger than 1 MeV in the $^{348}$138 nucleus
[see Figs.\ \ref{138-348-nil}a and (b)]. Similar situation is also seen in the 
$^{466}$156 nucleus. In this nucleus  the bands of proton 
$(Z=130, 132, 134, 136)$, $(Z=142, 144, 146, 148)$, $(Z=156, 158)$ 
and $(Z=168, 170)$ shell  gaps are formed because of the presence of the 
bunches of four single-particle states with relatively low $\Lambda$ values
located between them.  The energies of
these bunches slightly decrease with increasing absolute value of $\beta_2$ 
[see Fig.\ \ref{156-466-Nil}(a)]. Note that some of these gaps reach almost  2 
MeV in size. Smaller neutron gaps with size of around 1 MeV and below are seen at 
$N=294$, 296, 302, 314, 318 and contrary to proton subsystem they do not
form the bands of shell gaps [see Fig.\ \ref{156-466-Nil}(b)].

   The obtained results for shell structure of toroidal nuclei allow us to understand
its contribution into the stability of toroidal shapes with respect of breathing deformations. 
For example, LEMAS in the $^{348}$138 nucleus corresponds to the situation in which 
proton and neutron  Fermi levels are located in the middle of the region of low density of 
single-particle states in the vicinity of the $Z=134$ and $N=210$ gaps, respectively [see 
Figs.\ \ref{138-348-nil}(a) and (b)]. Any increase or decrease of the
$\beta_2$ value from the one corresponding to LEMAS will lead to the increase
of the density of the single-particle states in the vicinities of respective Fermi levels. 
This effect is especially pronounced for the neutron subsystem. As a consequence, 
the LEMAS corresponds to the largest or near-largest
(in absolute sense) negative proton and neutron shell correction energies, 
while the deviation (in terms of $\beta_2$) from total energy minimum will lead to 
the reduction of these energies. This contributes to the stability of toroidal shapes with  respect of breathing 
deformations. However, as illustrated by the case of the $^{466}$156 nucleus, the 
contribution of shell correction effects  to the stability of the nuclei
is expected to depend on proton and neutron numbers. In this nucleus, the neutron 
Fermi level at LEMAS is located at high density of the neutron single-particle states
[see Fig.\ \ref{156-466-Nil}(b)], which likely leads to positive neutron shell correction energies. 
On the  contrary, shell correction energies will be large and negative in the proton subsystem 
since the proton  Fermi level is located in the vicinity of large $Z=156$ gap 
[see Fig.\ \ref{156-466-Nil}(a)]. Note that this
gap is so large that proton pairing collapses  at the $\beta_2$ values near LEMAS; this
is seen from the fact that the energy of the proton Fermi level coincides with the energy
of the single-particle state located below the $Z=156$ gap.
 
   The same features are also active in respect of the stability of toroidal nuclei in sausage 
deformation  degree of freedom. This is because of two factors. First,  the shell gaps in 
breathing degree of freedom are also the shell gaps in the sausage degree of freedom  (see Sect. IVD 
in Ref.\ \cite{Wong.73}). Second, as shown in toroidal harmonic oscillator shell model for particle 
numbers of interest, the increase of sausage deformations $\sigma_{\lambda}$ of 
multipolarities $\lambda=1$, 2 and 3 from zero to some finite values leads to washing 
out of these shell gaps and an increase of the density of the single-particle states 
in the vicinity of the Fermi level (see Figs. 21, 22 and 23 in Ref.\ \cite{Wong.73}). Let us 
consider the nuclei in which the 
proton and neutron  Fermi levels of the LEMAS solution are located in the region of low 
density of the  single-particle states. In these nuclei the shell correction energy is negative 
at $\sigma_{\lambda}=0$  but it will either be reduced in absolute value or become positive 
when sausage deformations become non-zero. Thus, the instability in the breathing degree
of freedom which exists on the level of liquid drop is counterbalanced in these nuclei 
by the quantum shell effects. The balance of these two contributions defines whether the
toroidal nucleus is stable with respect of sausage deformations  or not. Fully quantum mechanical 
calculations based on DFT are needed to establish the stability of a given nucleus with respect 
of sausage deformations.  However, the analysis of the shell  structure and the level density of 
the single-particle states in the vicinity of the proton and neutron Fermi levels provides a useful 
information on whether a given nucleus could potentially be stable with respect of sausage 
deformations. For example, as discussed 
above such an analysis for the $^{348}$138 nucleus shows low densities of the 
single-particle states in the vicinity of proton and neutron Fermi levels and indeed 
the RMF+BCS calculations of Refs.\ \cite{AAG.18,AATG.19} reveal the stability of toroidal 
shapes in  this nucleus with respect of even-multipole sausage deformations.

\begin{figure}[htb]
\centering
\includegraphics[angle=0,width=8.5cm]{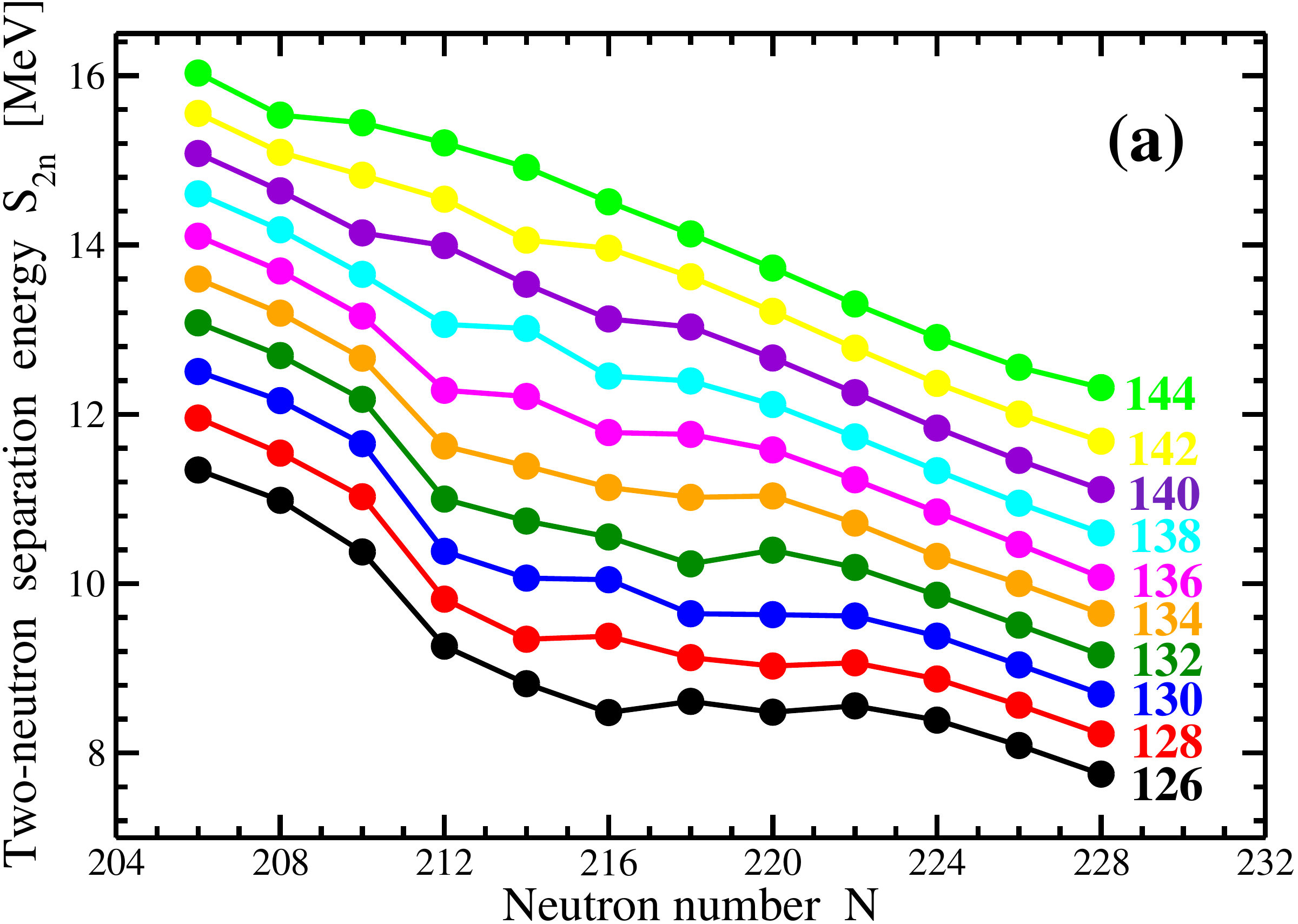}
\includegraphics[angle=0,width=8.5cm]{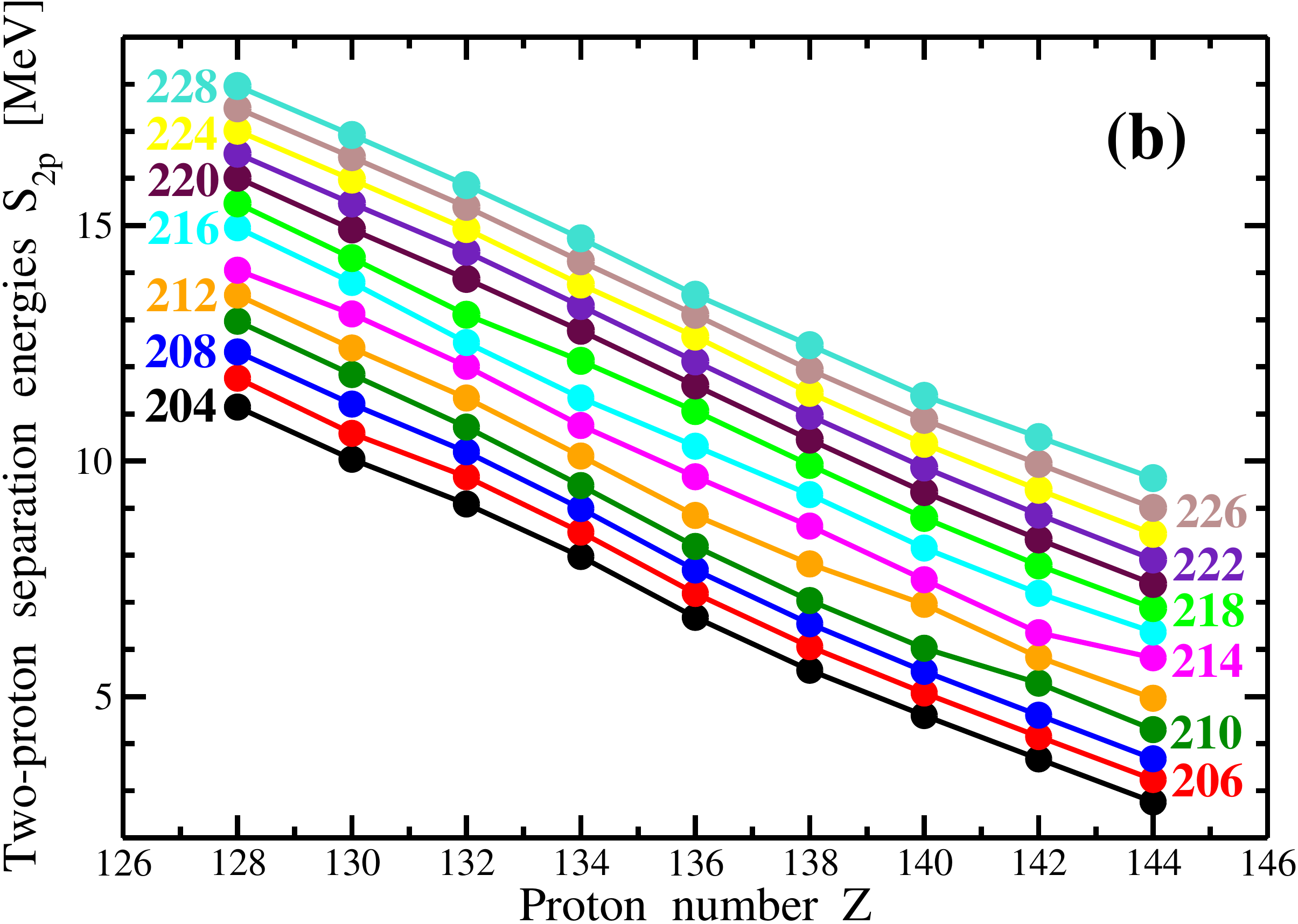}
\caption{Two-neutron $(S_{2n}(Z,N))$ and two-proton $(S_{2p}(Z,N))$ 
separation energies  given for different isotopic/isotonic chains as a function of the 
neutron/proton number. The lines are labeled by respective proton [panel (a)] and 
neutron [panel (b)]  numbers.
\label{sep-Energy}
}
\end{figure}

\begin{figure}[htb]
\centering
\includegraphics[angle=0,width=8.5cm]{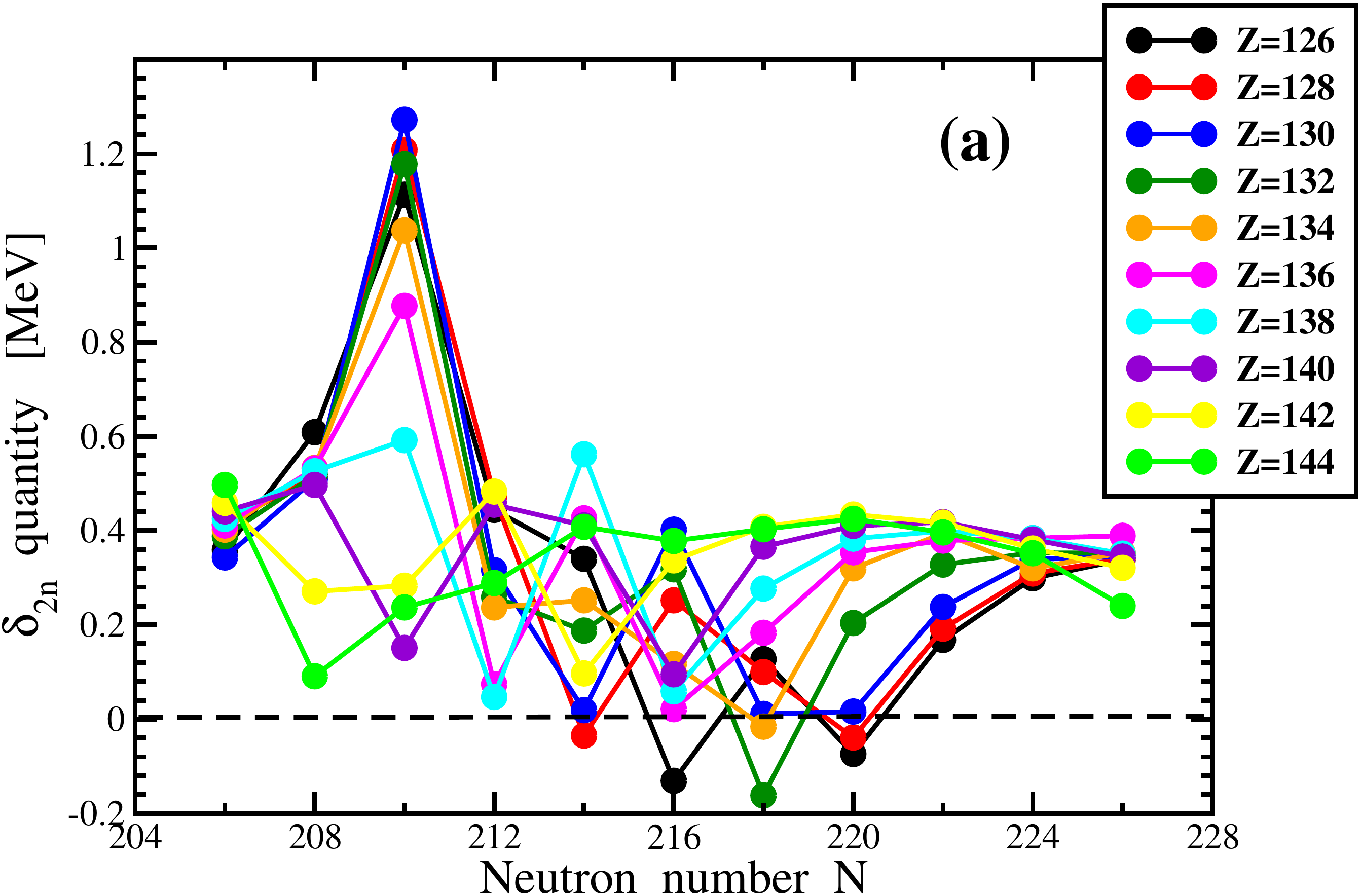}
\includegraphics[angle=0,width=8.5cm]{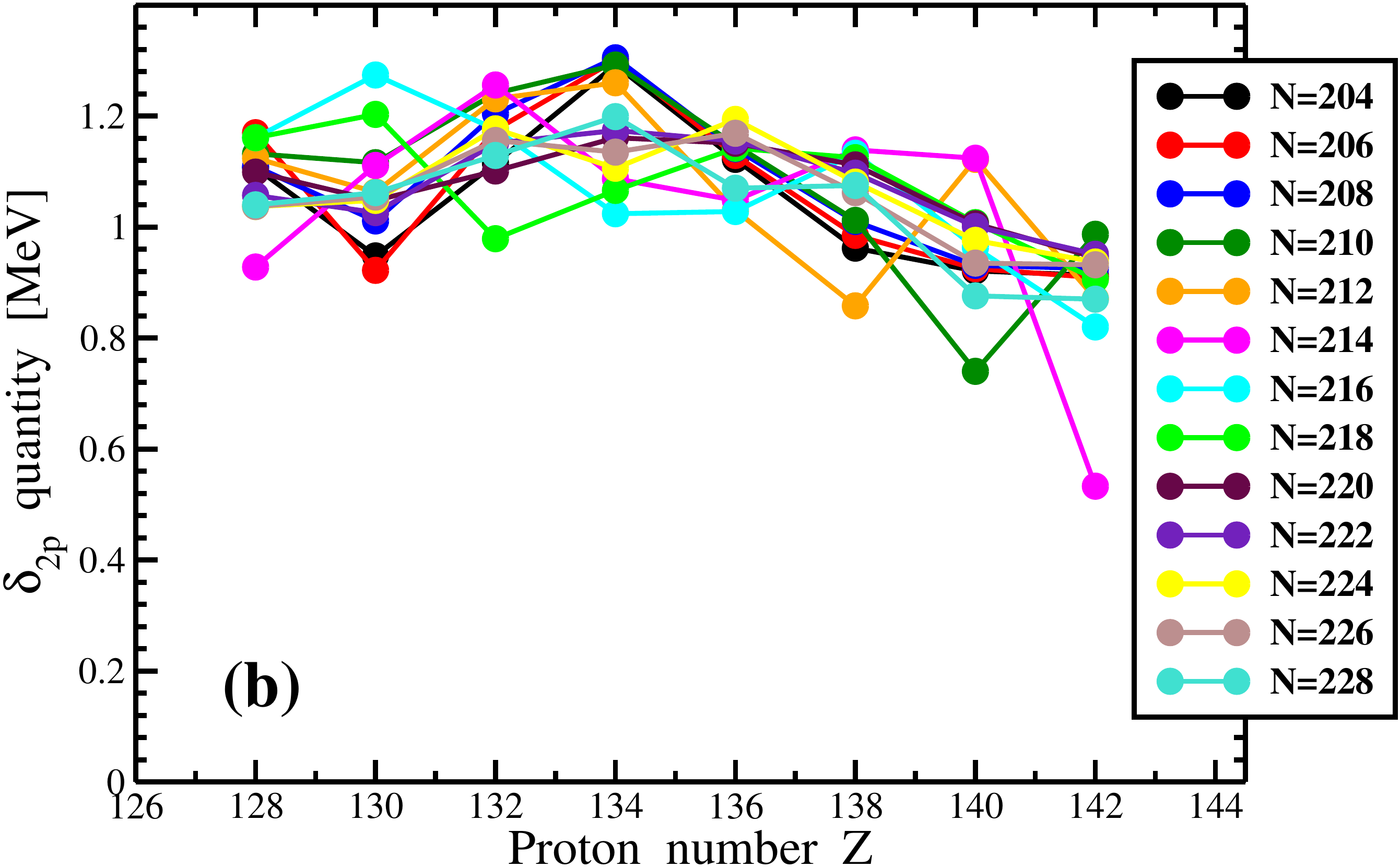}
\caption{The $\delta_{2n}(Z,N)$ and $\delta_{2p}(Z,N)$  
quantities for the toroidal nuclei in the  
$(Z=126-144, N=204-228)$ region.
\label{delta-quantity}
}
\end{figure}

 Figures \ref{138-348-nil} and \ref{156-466-Nil} reveal some global bunching 
of the pairs of almost degenerate in energy single-particle states of opposite parities.
For example, such bunches of the states are seen in the proton 
subsystem of the $^{348}$138 nucleus at the energies  $\approx -19$ MeV, $\approx -14$ 
MeV, $\approx 0$ MeV and $\approx 4$ MeV for $\beta_2=-5.0$
(see Fig.\ \ref{138-348-nil}(c) and (e)). The
density of the single-particle states is high in these bunches and thus it is reasonable to 
expect that the shell correction energy $E_{shell}$ will be positive when the Fermi level is located 
near or within these bunches. For such a situation it is reasonable to expect that the quantum 
shell effects will not help to stabilize toroidal shapes with respect of sausage deformations.
With decreasing absolute value of $\beta_2$ the energies  of these bunches of  the single-particle 
states go down (see Fig.\ \ref{138-348-nil}(c) and (e)). However, these bunches and the low 
density single-particle structure between them persist down to $\beta_2$ values corresponding 
to the transition  from toroidal to concave disk shapes.  The density of the single-particle states 
is low between these  bunches and it is reasonable to expect that for the majority of the 
combinations of particle number and $\beta_2$ the $E_{shell}$ values will be negative when 
the Fermi level is located in this region.   These features are the manifestation of so-called
supershell structure which has been discussed  in the case of ellipsoidal-like shapes  in 
Ref.\ \cite{RNS.78}.   

   There is a drastic difference in the behavior of neutron and proton Fermi levels as a function 
of  the $\beta_2$ value (see Fig.\ \ref{138-348-nil} and Fig.\ \ref{156-466-Nil}).  The neutron 
Fermi level is more or less constant as a function of $\beta_2$.  As a consequence, the calculated 
two-neutron  drip line for toroidal shapes is close to the extrapolation of this line for ellipsoidal-like 
shapes (see Fig.\ \ref{landscape}).  On the contrary, the proton Fermi level dives deeper into 
nucleonic potential with increasing absolute value of $\beta_2$; it is lower by approximately
5 MeV  for  toroidal shapes with large aspect ratio as compared with its position for
biconcave disk shapes.
As a consequence, the transition to toroidal shapes in hyperheavy nuclei creates a substantial expansion 
[the area between black solid and orange dashed lines in Fig.\ \ref{landscape}] 
of the nuclear landscape.

   There are drastic changes in the single-particle structure of the $^{348}$138 
nucleus at $\beta_2 \approx -1.15$ and $\beta_2\approx -1.85$ (see Figs.\ 
\ref{138-348-nil}(c), (d), (e) and (f)). The first change is related 
to the transition from biconcave disk shape to toroidal one (which is equivalent to
an opening of the hole in the center of biconcave disk shape). The second one is 
associated with the redistribution of the proton density in the torus caused by the 
change of the occupation of the single-particle orbitals.  This density is asymmetric 
with respect of the axis of torus tube and has a maximum closer  to an outer 
edge of the torus for the $\beta_2$ values ranging from $\approx -1.15$ down to 
$\approx -1.85$. However, it becomes almost symmetric with respect of the axis 
of the torus tube for $\beta_2 \leq -1.85$. Note that similar changes in the single-particle 
structure are seen at $\beta_2 \approx -1.1$,  $\beta_2\approx -1.8$ and  
$\beta_2\approx -2.7$ in the $^{466}$156 nucleus (see Figs.\ 
\ref{156-466-Nil}(c), (d), (e) and (f)) and their origins are similar to the ones
discussed above in the $^{348}$138 nucleus.

  In order to find potentially most stable toroidal nuclei, 
two-proton $S_{2p}(Z,N)$  and  two-neutron $S_{2n}(Z,N)$ 
separation energies 
\begin{eqnarray}
S_{2n}(Z,N) = B(Z,N) - B(Z,N-2), \nonumber \\
S_{2p}(Z,N) = B(Z,N) - B(Z-2,N),
\end{eqnarray}
and the $\delta_{2n}(Z,N)$ and  $\delta_{2p}(Z,N)$
quantities defined as 
\begin{eqnarray}
\delta_{2n}(Z,N) = S_{2n}(Z,N) - S_{2n}(Z,N-2), \nonumber \\
\delta_{2p}(Z,N) = S_{2p}(Z,N) - S_{2p}(Z-2,N),
\end{eqnarray}
are plotted  in Fig.\ \ref{sep-Energy}  for the region with 
$(Z=132-144, N=204-228)$.  Here $B(Z,N)$ is the binding energy. The 
separation energies show a sudden drop at the shell gaps, if they are 
large. If the variations of the level density are less pronounced, the 
$\delta_{2n}(Z,N)$ and  $\delta_{2p}(Z,N)$ quantities related 
to the derivatives of the separation energies are more sensitive 
indicators of the localizations of the shell gaps (see discussion
in Appendix of Ref.\ \cite{A250}).  They also provide the information
on average density of the single-particle states.

 The presence of the neutron gap at $N=210$ for toroidal shapes is 
visible in Figs.\ \ref{sep-Energy}(a)   and Fig.\ \ref{delta-quantity}(a)
in the $Z=126-136$ nuclei.
The $\delta_{2n}(Z,N)$ values for neutron numbers away from 
$N=210$ are low which are indicative of high density of neutron
single-particle states below and above the $N=210$ shell gap.  These 
features correlate with the ones seen in the Nilsson diagram [see 
Fig.\ \ref{138-348-nil}(b)].

  On the contrary, the $S_{2p}(Z,N)$ and $\delta_{2p}(Z,N)$ values
(see Figs.\ \ref{sep-Energy}(b)  and Fig.\ \ref{delta-quantity}(b)) are 
relatively smooth functions of proton number which indicates that the average
density of proton single-particle states remains more or less constant.
However,  on average the $\delta_{2p}(Z,N)$ values are substantially 
higher than the $\delta_{2n}(Z,N)$ ones; only in the region of the peak 
of  $\delta_{2n}(Z,N)$ at $N=210$ they are comparable (see Fig.\ 
\ref{delta-quantity}).
This clearly indicates that the density of proton single-particle states is low 
in a wide range of proton numbers and this observation is  supported by 
the comparison of  Figs.\ \ref{138-348-nil}(a) and (b).  Note that the peak 
of $\delta_{2p}(Z,N)\approx 1.3$ MeV is seen for neutron numbers 
$N=204-212$ (see Fig.\ \ref{delta-quantity}(b)) suggesting an extra stability
of these nuclei.

   The combination of proton and neutron shell effects should lead to 
an enhanced stability of specific nuclei.  As a result, discussed above 
features are most likely reasons why fission barrier is higher in the 
$N=210$  $^{348}$138 nucleus as compared with the $N=220$ 
$^{354}$134  one.

\subsection{Functional dependence of the results}
\label{diff-funct}
   
 When considering the predictions for toroidal hyperheavy nuclei and their shell
structure it is important to evaluate their dependence on the employed functional.
So far all predictions for such nuclei presented in Refs.\ \cite{AAG.18,AATG.19} and
in the present paper were obtained with the CEDF DD-PC1.  To study functional
dependence of the predictions we perform additional calculations for the $^{348}$138
and $^{466}156$ nuclei with the NL3* \cite{NL3}, PC-PK1 \cite{PC-PK1}, DD-ME2 
\cite{DD-ME2}  and DD-ME$\delta$ \cite{DD-MEdelta} functionals and compare their 
results with the ones obtained with DD-PC1 earlier. These five state-of-the-art functionals represent
three major classes of CDFT models \cite{AARR.14} and have been globally tested in 
Refs.\ \cite{AARR.14,AANR.15,AA.16,LLLYM.15,TAA.20}. Note that in this set of the 
functionals the CEDF DD-PC1 and PC-PK1 provide better description of binding 
energies on a global scale as compared with other functionals.
 
\begin{figure}[htb]
\centering
\includegraphics[angle=0,width=8.5cm]{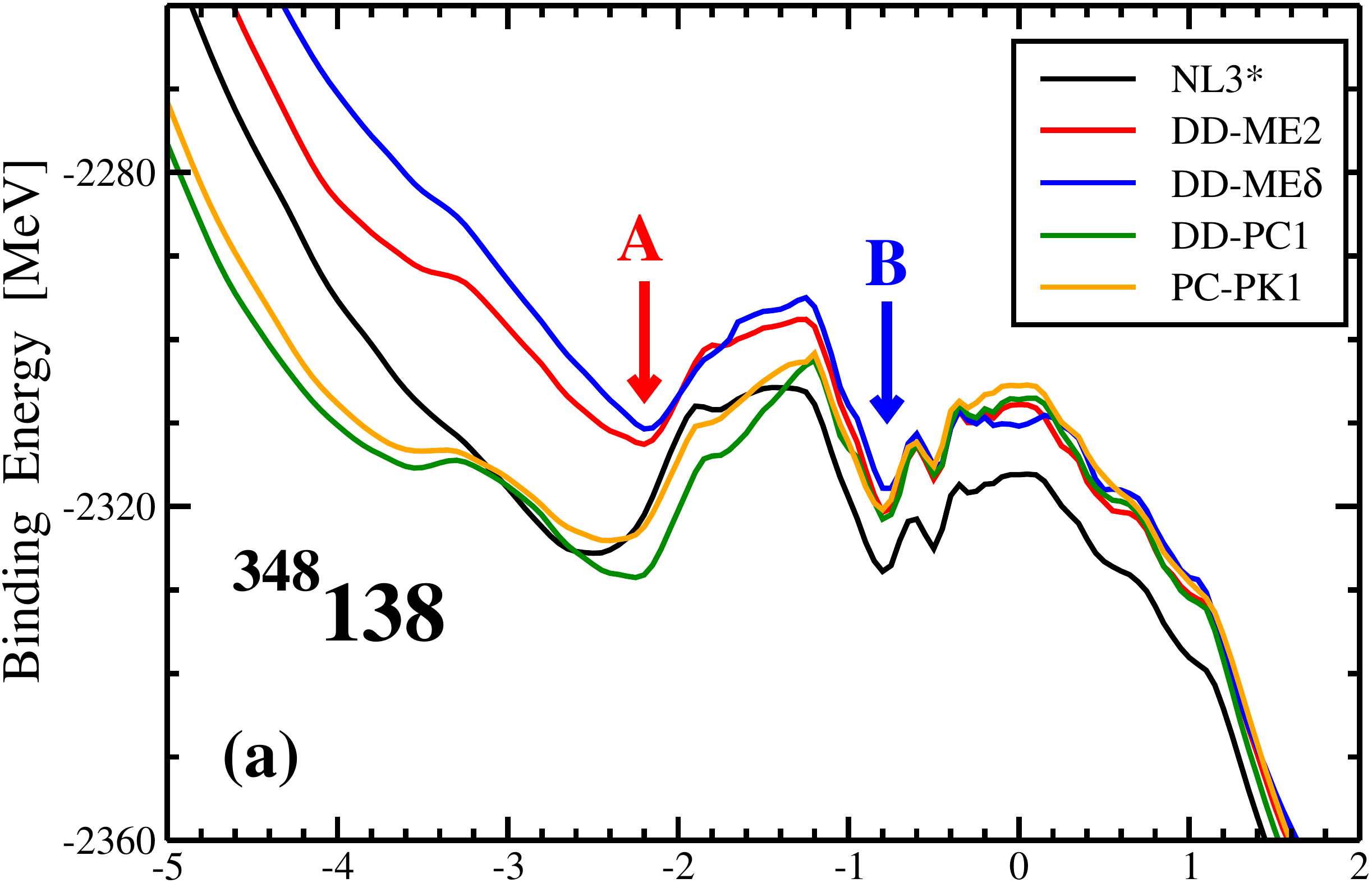}
\includegraphics[angle=0,width=8.5cm]{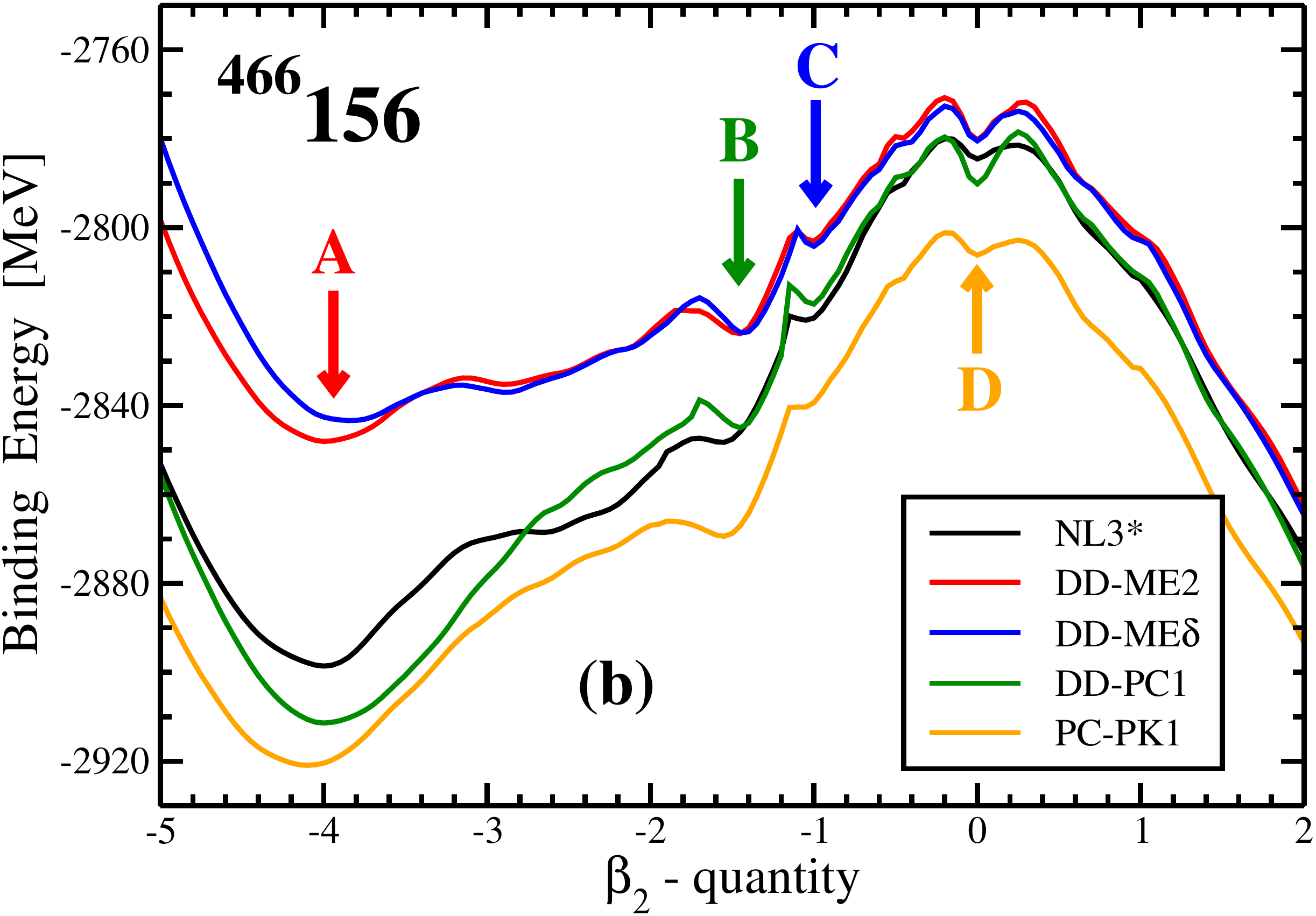}
\caption{The deformation energy curves  obtained in axial RHB calculations with 
indicated CEDFs. The local and global minima are indicated by the arrows with letters. 
The same labelling of minima as shown in Fig.\  \protect\ref{156-466-pot-b} 
is used for the $^{466}$156 nucleus. 
}
\label{Dif-energy-curves}
\end{figure}

  The deformation energy curves obtained with these functionals are presented 
in Fig.\ \ref{Dif-energy-curves}.  In both nuclei and in terms of relative energies of
the minima corresponding to toroidal and ellipsoidal-like shapes there is a large 
similarity of the results obtained with point-coupling models DD-PC1 and PC-PK1
as well as with nonlinear meson-nucleon coupling model NL3* on the one hand and 
those obtained with density-dependent meson-exchange models DD-ME2 and
DD-ME$\delta$ on the other hand. In the latter type of the models, the toroidal 
shapes are less energetically favored with respect of ellipsoidal-like shapes as 
compared with former models.  For example, in the $^{348}$138 nucleus the fat toroidal shapes
corresponding to the minimum A are more (less) energetically favored as compared
with biconcave disk shapes corresponding to minimum B in the calculations
with DD-PC1 and PC-PK1 (DD-ME2 and DD-ME$\delta$)  functionals. Note that 
these two minima are located at approximately the same energies in the calculations with the NL3* 
functional [see Fig.\ \ref{Dif-energy-curves}(a)].  However, this difference in the predictions
of relative energies of the minima A and B is not principal because the minimum B
is not stable with respect of triaxial distortions in the calculations with DD-PC1
functional (see Ref.\ \cite{AAG.18}) and the same situation is expected for other functionals 
because of the similarity of underlying shell structure.  On the other hand, the minimum A is 
relatively stable with respect of even-multipole sausage deformations in the calculations with 
DD-PC1 (see Refs.\ \cite{AAG.18,AATG.19}) and because of similarity of underlying toroidal 
shell structure (see discussion of Fig.\ \ref{138-348-Nil-other-func} below) it is reasonable to
expect that this is also the case for remaining functionals.

  Similar situation to the $^{348}$138 nucleus holds also in the $^{466}$156 one. 
This is because toroidal shapes are more energetically favored as compared
with ellipsoidal-like ones in the calculations with CEDFs DD-PC1, PC-PK1 and 
NL3* than in those employing DD-ME2 and DD-ME$\delta$  functionals 
(see Fig.\  \ref{Dif-energy-curves}(b)). For example, the energy difference $\Delta E_{diff}$
between the minimum A 
corresponding to thin toroidal shapes and the minimum D corresponding to spherical 
shapes is approximately  117 MeV in the calculations with the first group of 
the functionals and only approximately 67 MeV in the calculations with the second 
group.   Note that these differences cannot be explained by the differences in
nuclear matter properties of the functionals since they are similar (quite different) in 
the pair of the DD-PC1 and DD-ME2  (DD-PC1 and PC-PK1) functionals (see Ref.\ 
\cite{AA.16}) which provide the $\Delta E_{diff}$ 
values which differ by 53.4 MeV (by only 6.5 MeV).

  These differences between the functionals, related to the relative energies of the minima 
corresponding to toroidal and ellipsoidal-like shapes,  are expected  to affect the position
of the boundary between ellipsoidal-like and toroidal shapes in the nuclear landscape (see
Fig.\ \ref{landscape} in the present manuscript and the discussion in Sec. XII of Ref.\ \cite{AATG.19}).
However, this boundary depends not only on relative energies of these two types of the
shapes but also on the stability of ellipsoidal-like shapes with respect of fission (see Ref.\ 
\cite{AATG.19}). There is a quite substantial dependence of the fission barrier heights for 
ellipsoidal-like shapes on CEDF with the PC-PK1 and NL3* (DD-ME2 and DD-PC1) functionals 
providing the lowest (highest) barrier heights for superheavy nuclei among the  CEDFs
considered in Ref.\ \cite{AAT.20} and a similar situation is also expected in the 
hyperheavy nuclei.
  
   Despite above mentioned differences there are large similarities between the results
of the calculations obtained with five functionals. For the first time, the results presented
in Fig.\ \ref{Dif-energy-curves} confirm that the transition from ellipsoidal-like to toroidal
shapes with increasing proton number $Z$ does not depend on CEDF. The presence of 
similar local minima in deformation energy curves (such as the minima A, B, C and D in
the $^{466}$156 nucleus and the minima A and B in the $^{348}$138 nucleus) with similar 
equilibrium $\beta_2$ values presented in Fig.\ \ref{Dif-energy-curves} clearly suggest the 
similarity of underlying shell structure in all employed functionals. Note that in a few cases
such minima are shoulder-like in deformation energy curves without a sufficient barrier
on one side:
these are the minimum C in the calculations with NL3* and PC-PK1 and the minimum B
in the calculations with NL3* [(see Fig.\ \ref{Dif-energy-curves}(b)].

   The analysis of toroidal shell structure of the $^{348}$138 and $^{466}$156 nuclei 
obtained  with the NL3*, PC-PK1, DD-ME2 and  DD-ME$\delta$ functionals reveals the same 
general features as those discussed in Sec.\ \ref{Tori-shell-structure} for the DD-PC1 functional. 
Thus we will focus on fine details of the shell structure of these nuclei in the vicinity of the 
respective  Fermi levels at the LEMAS of the minimum A in these two nuclei (see 
Fig.\ \ref{Dif-energy-curves}) since they are responsible for potential stability of respective toroidal 
shapes. The Nilsson diagrams for these four CEDFs are shown in Figs.\ \ref{138-348-Nil-other-func} 
and \ref{156-466-Nil-other-func}; they can be compared with those obtained for DD-PC1 
and presented  in Figs.\ \ref{138-348-nil}(a) and (b) and Figs.\ \ref{156-466-Nil}(a) and (b).
This comparison reveals significant similarities between the results of the calculations obtained
with different functionals. 

For example, in the $^{348}$138 nucleus the proton Fermi level $E_F$ 
at LEMAS is located in the region  of reduced density of proton single-particle states 
between shell gaps at $Z=134$ and $Z=140$ (see Fig.\ \ref{138-348-nil}(a)) in the calculations 
with the DD-PC1 functionals. Similar situation exists also in the calculations with NL3*, PC-PK1, 
DD-ME2 and DD-ME$\delta$ CEDFs [see Figs.\ \ref{138-348-Nil-other-func}(a), (c), (e) and (g)].
In this nucleus, the neutron Fermi level is located in the middle of substantial $N=210$ toroidal 
shell gap in the calculations with DD-PC1 [see Fig. \ref{138-348-nil}(a)],  DD-ME2 and 
DD-ME$\delta$ (see Figs.\ \ref{138-348-Nil-other-func}(e) and (g)] but it is shifted to the 
region of somewhat higher density of the neutron single-particle states below the $N=214$ 
toroidal shell gap in the calculations with NL3* and PC-PK1 [see Fig.\ \ref{138-348-Nil-other-func}(b) 
and (d)]. These results suggest that two-proton separation energies $S_{2p}(Z,N)$ and 
the $\delta_{2p}(Z,N)$ quantities (see the discussion in the end of Sec.\ \ref{Tori-shell-structure}) 
should be very similar for all five employed functionals. The same is true for related neutron
$S_{2n}(Z,N)$ and $\delta_{2n}(Z,N)$ values obtained in the calculations with DD-PC1,
DD-ME2 and DD-ME$\delta$ which are expected to reveal the presence of the $N=210$
toroidal shell gap [see Fig.\ \ref{delta-quantity}(b)]. However, it is quite likely that the peak in 
the $\delta_{2n}(Z,N)$ values visible at $N=210$ in the calculations with DD-PC1 [see Fig.\ 
\ref{delta-quantity}(a)] will be moved to $N\approx 214$ and substantially washed out in 
the calculations with NL3* and PC-PK1.

\begin{figure*}[htb]
\centering
\includegraphics[angle=0,width=7.5cm]{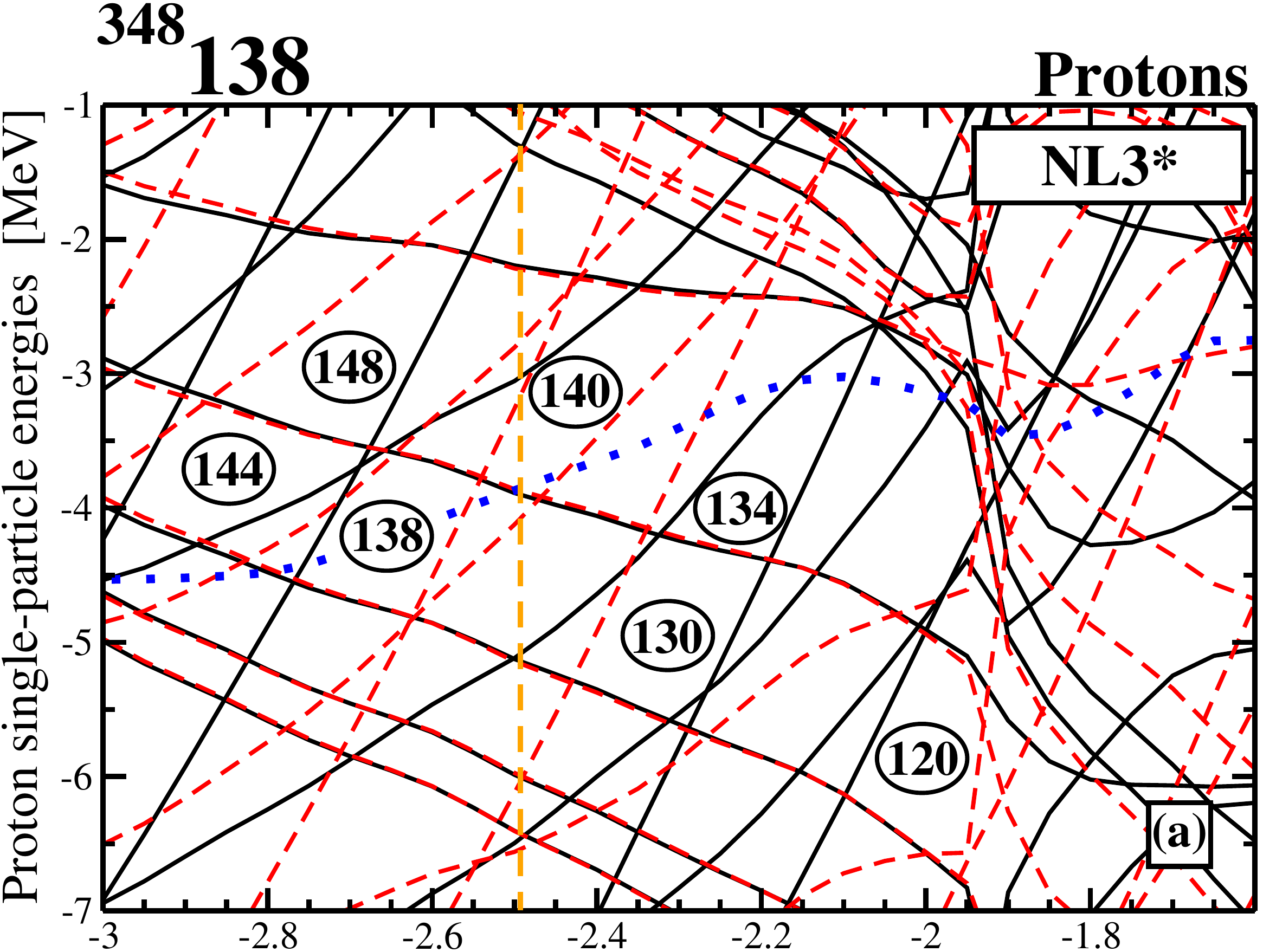}
\includegraphics[angle=0,width=7.5cm]{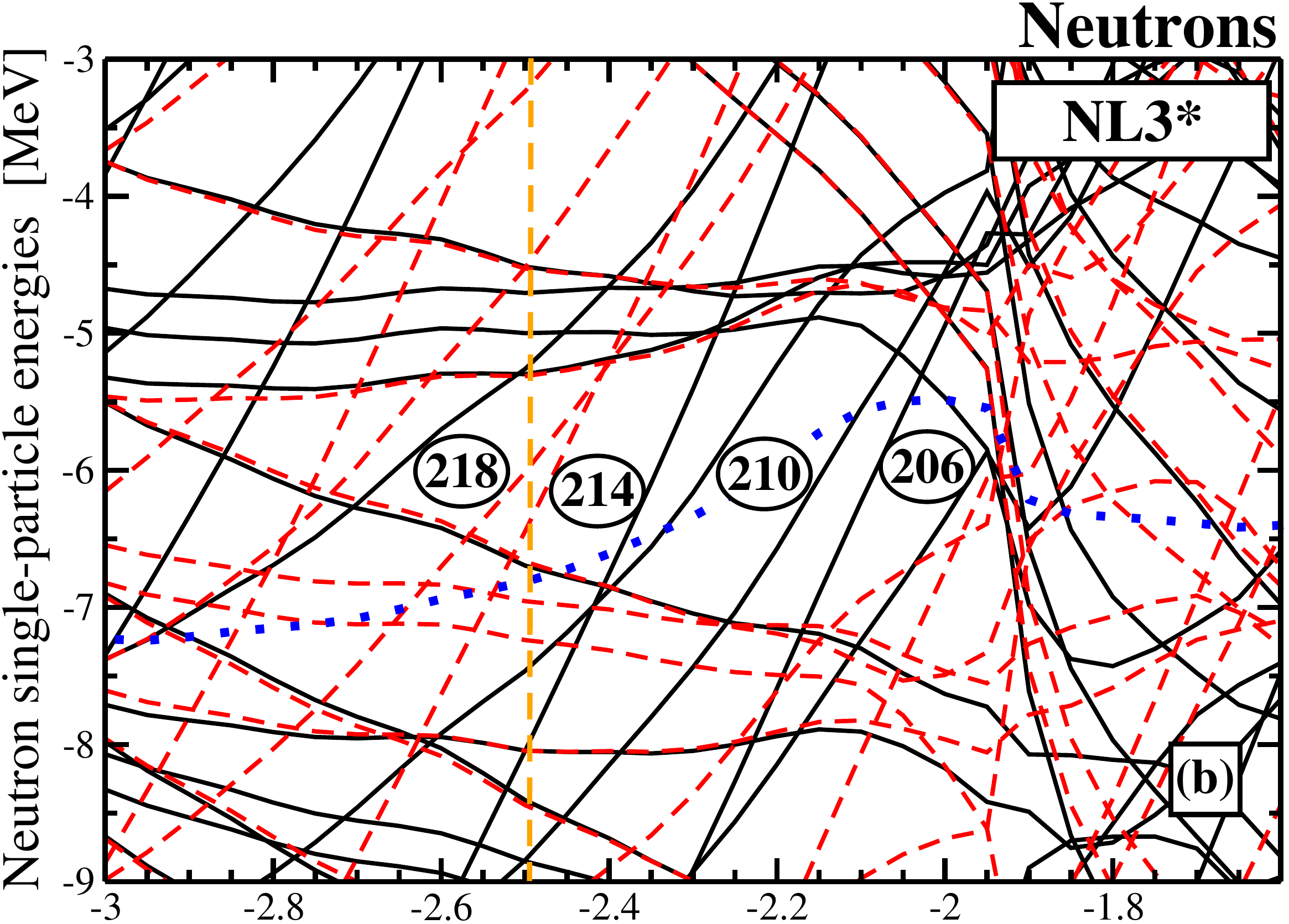}
\includegraphics[angle=0,width=7.5cm]{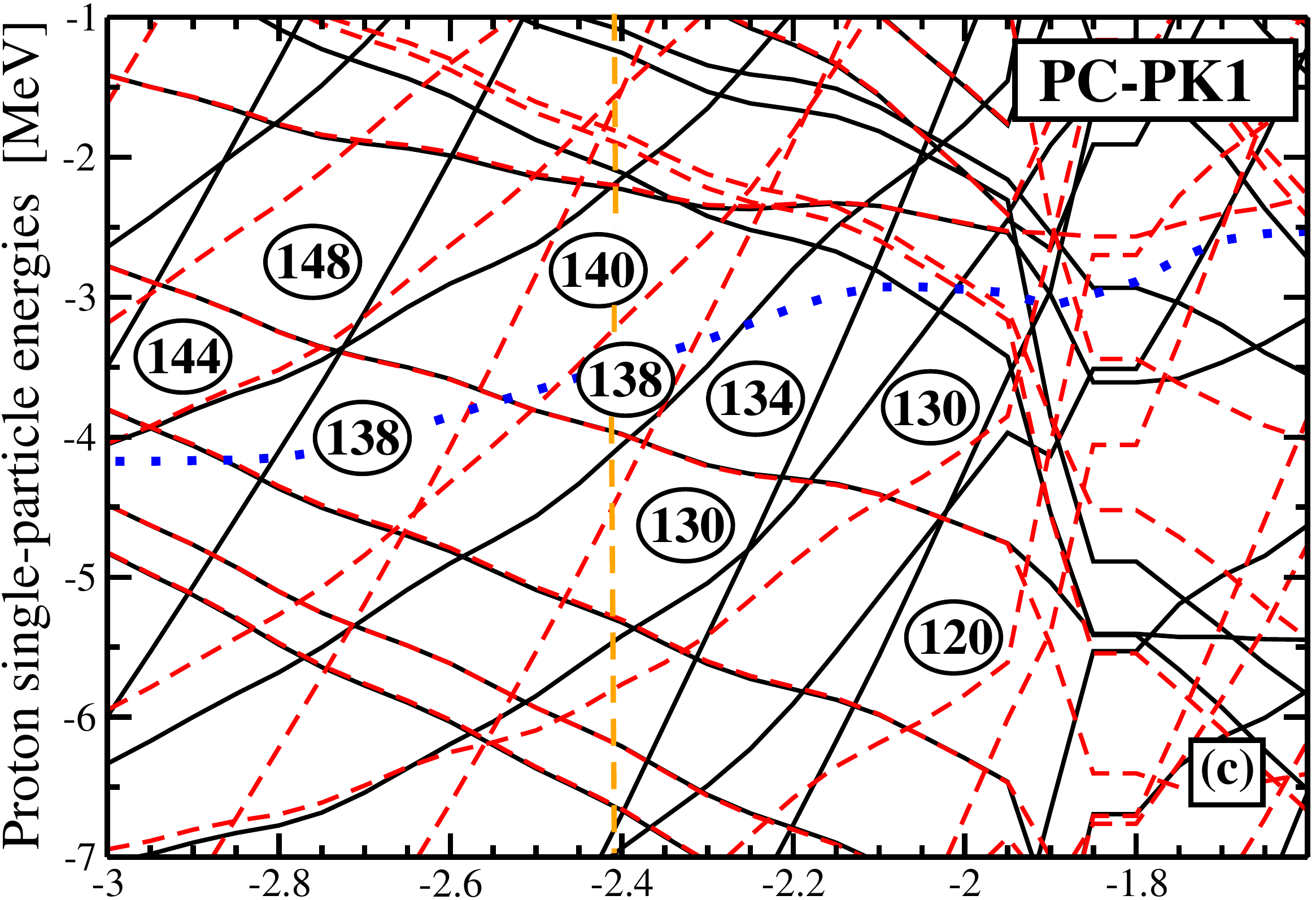}
\includegraphics[angle=0,width=7.5cm]{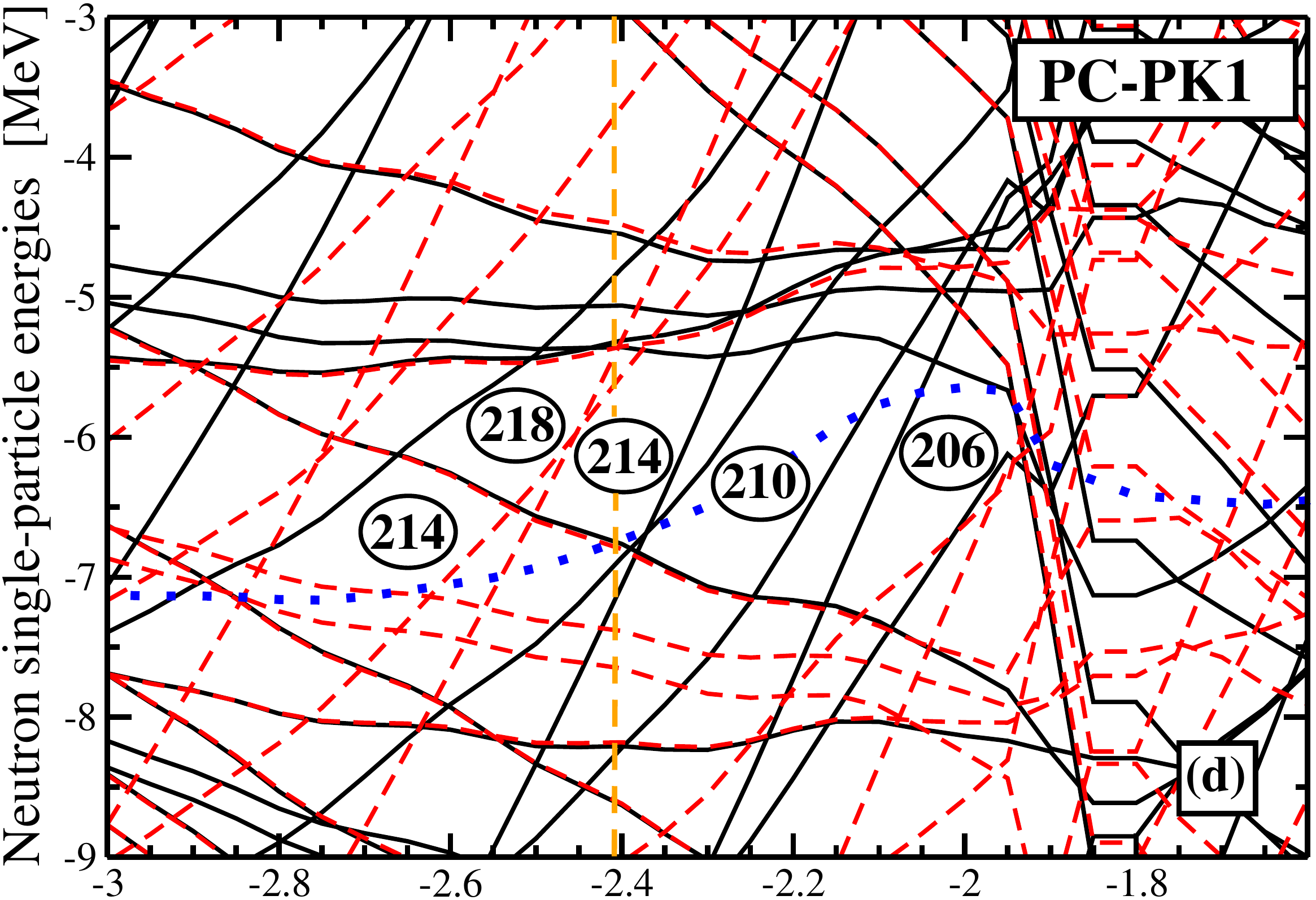}
\includegraphics[angle=0,width=7.5cm]{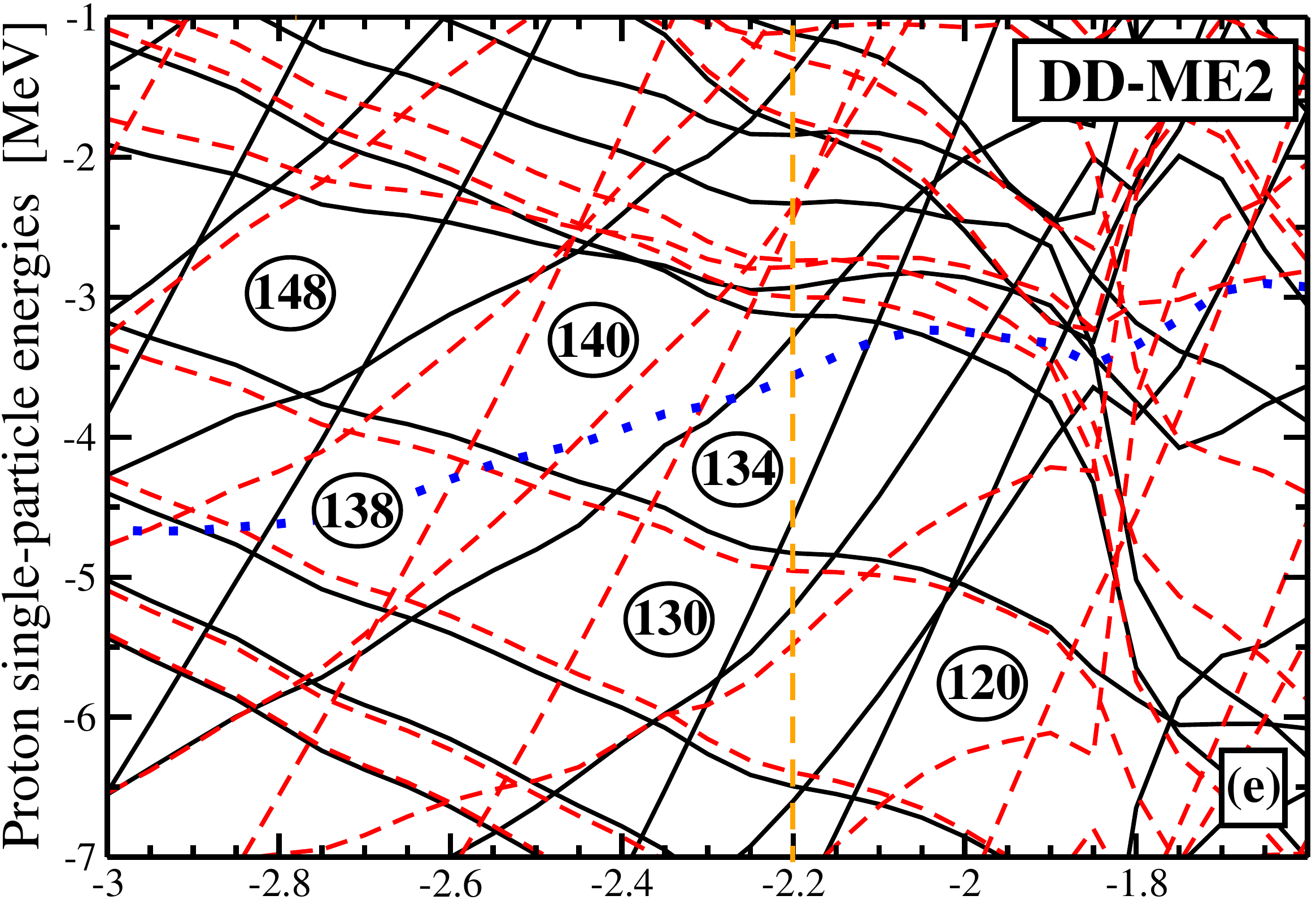}
\includegraphics[angle=0,width=7.5cm]{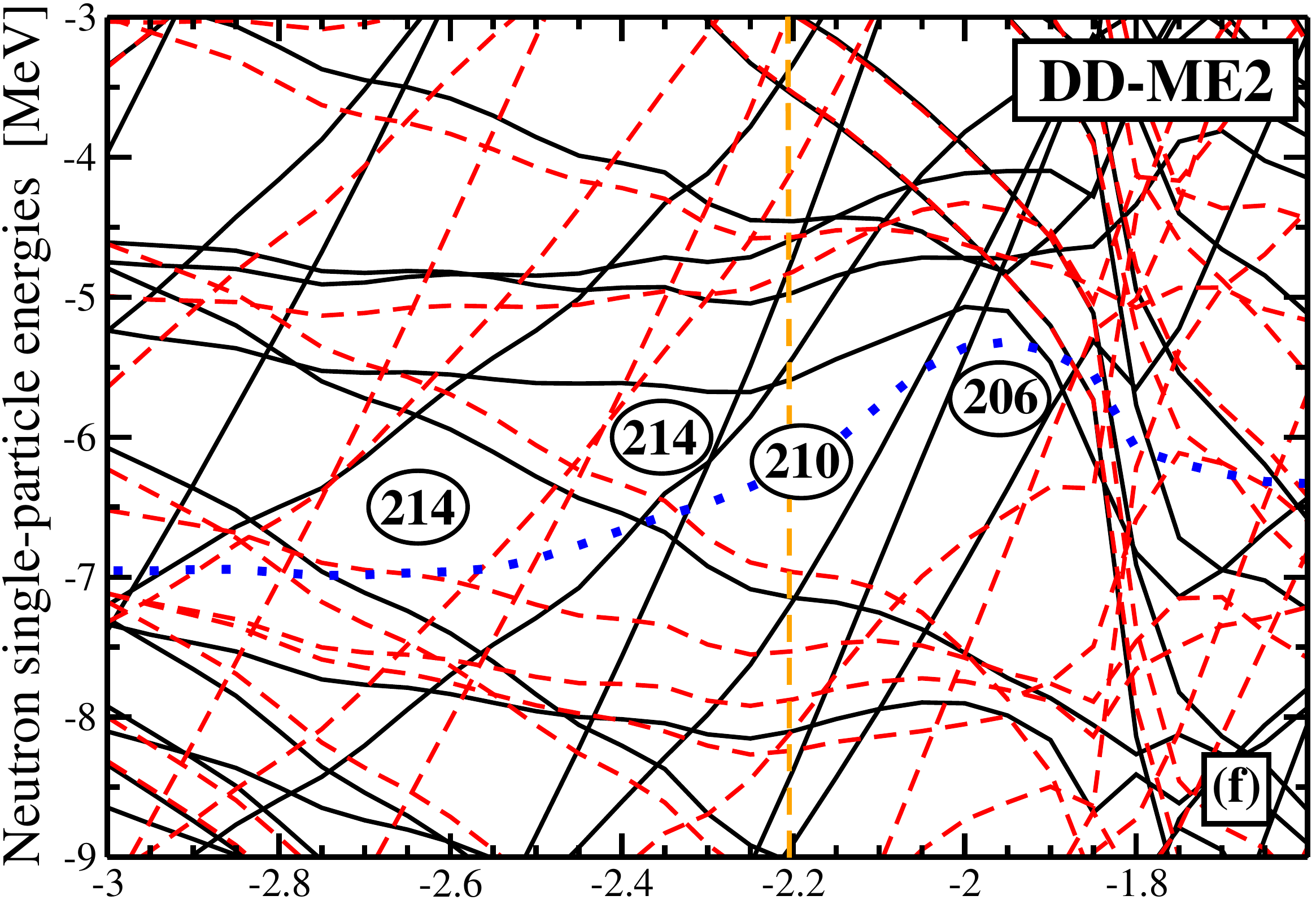}
\includegraphics[angle=0,width=7.5cm]{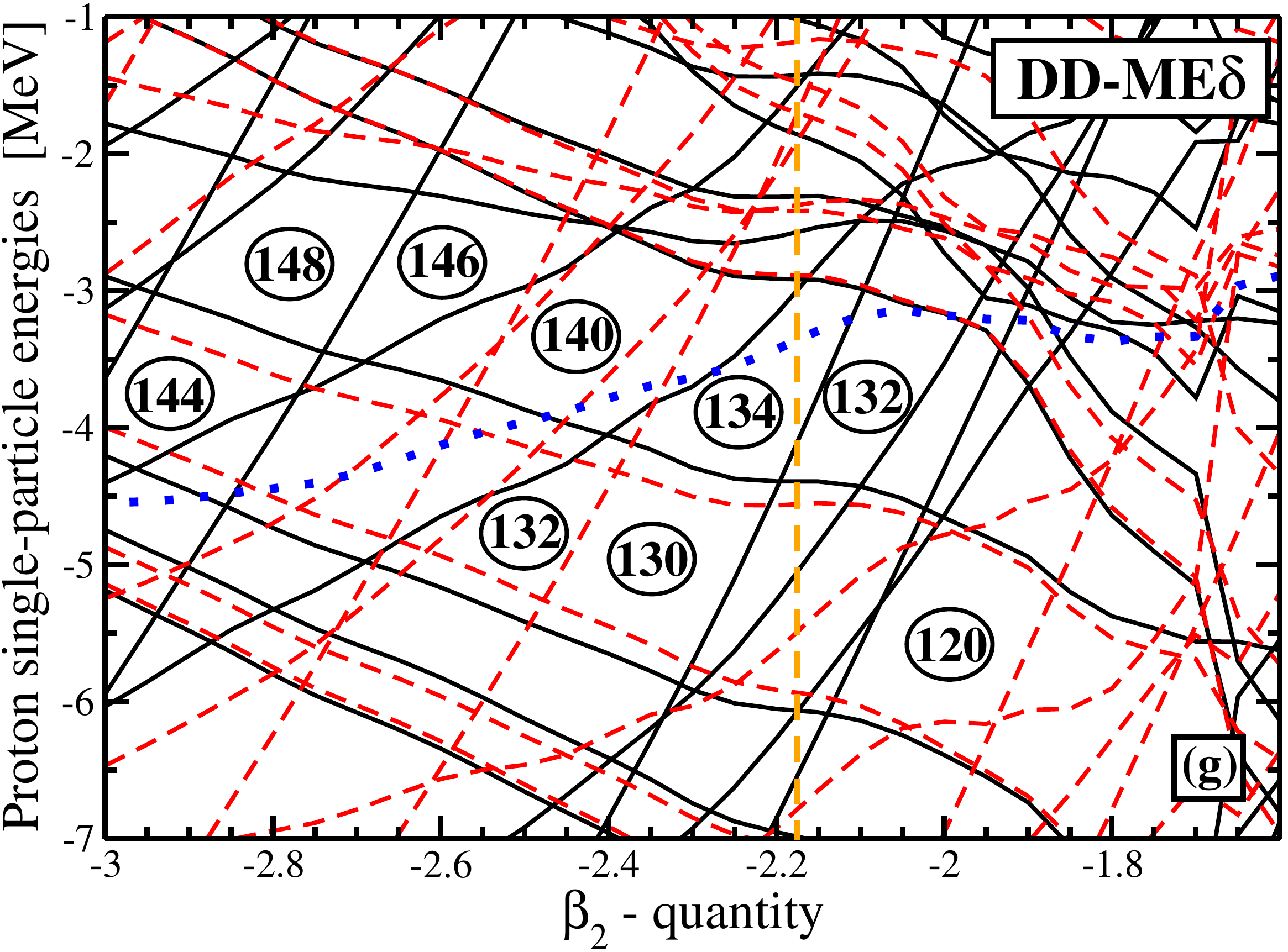}
\includegraphics[angle=0,width=7.5cm]{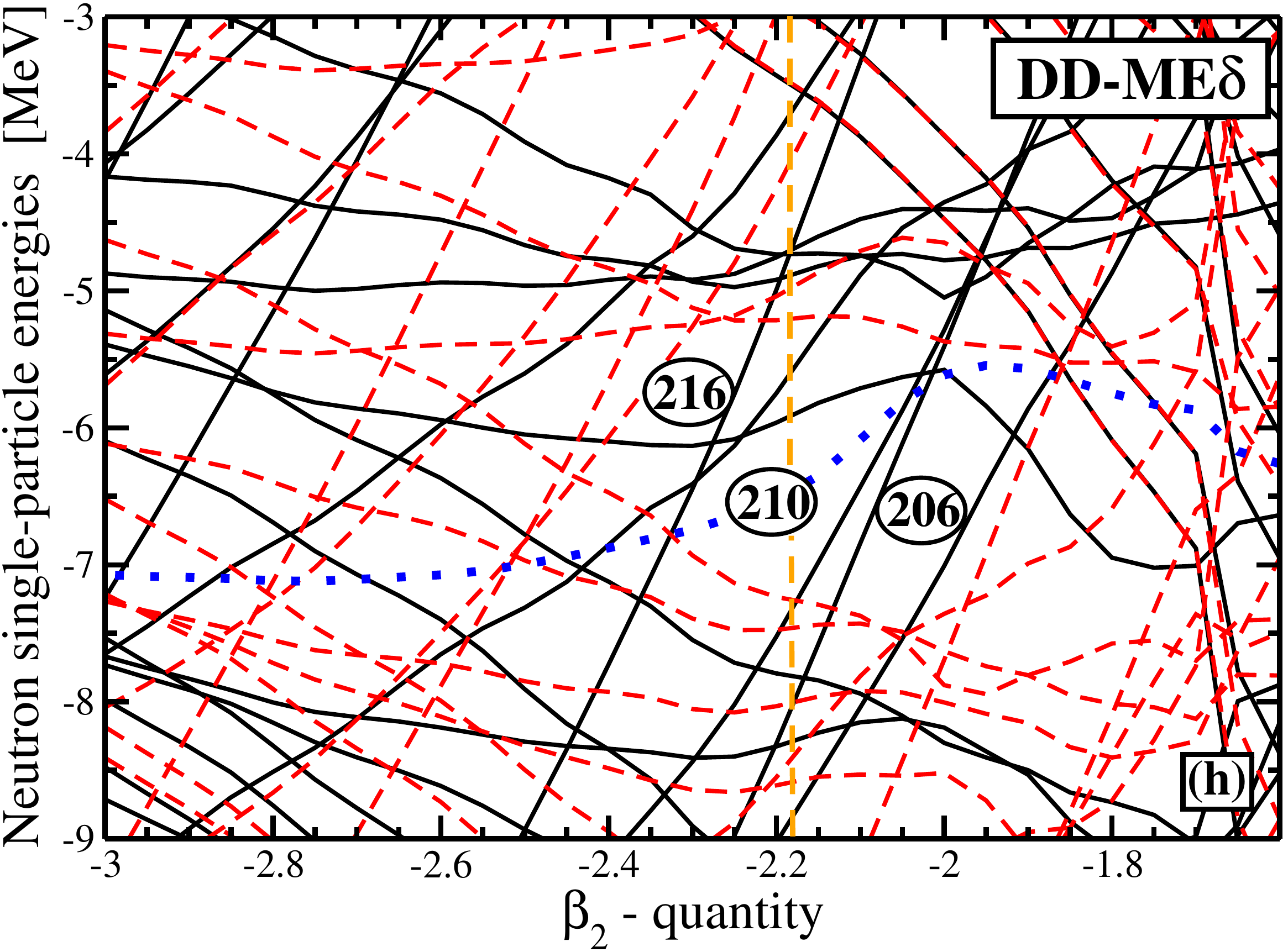}
\caption{The same as Figs.\ \protect\ref{138-348-nil}(a) and (b) but for
the results obtained with indicated CEDFs. The vertical dashed orange
lines are drawn at the $\beta_2$ values corresponding to LEMAS.
}
\label{138-348-Nil-other-func}
\end{figure*}

   Similar situation exists also in the $^{466}$156 nucleus.
The bands of proton (Z = 130; 132; 134; 136), (Z =142; 144; 146; 148), 
(Z = 156; 158; 160) and (Z = 168; 170) shell gaps, formed because of 
the presence of the bunches of single-particle states with relatively 
low $\Lambda$  values located between them, exist in all five functionals
[see Figs.\ \ref{156-466-Nil}(a) and \ref{156-466-Nil-other-func}(a), (c), (e) 
and (g)].  Note that some of these gaps reach almost 2 MeV in size.
The proton Fermi level at LEMAS is located either in the middle of 
large $Z=156$ shell gap in the NL3*, PC-PK1, DD-ME2 and DD-ME$\delta$ 
functionals or at the bottom of this shell gap in the DD-PC1 CEDF and thus 
shell correction energies will be large and negative in proton subsystem in
all functionals.

   Smaller neutron shell gaps with the size of around 1 MeV and below are seen at 
              $N = 294, 296, 302, 314, 318$            in DD-PC1 [Fig.\ \ref{156-466-Nil}(a)], at
$N = 278, 290, 294, 300, 302, 314, 318$          in NL3* [Fig.\ \ref{156-466-Nil-other-func}(b)], at
        $N = 296, 300, 302, 314, 326$                  in PC-PK1 [Fig.\ \ref{156-466-Nil-other-func}(d)], at
        $N = 292, 296, 298, 302, 308$                  in DD-ME2 [Fig.\ \ref{156-466-Nil-other-func}(f)],
        and at
        $N = 282, 298, 304, 306, 312$                  in DD-ME$\delta$ [Fig.\ \ref{156-466-Nil-other-func}(h)]
and contrary to proton subsystem they do not form the bands of shell gaps. Considering relatively 
small size of neutron shell gaps, larger (as compared with proton subsystem) dependence of the 
predictions for neutron shell gaps on the functional is expected.  These differences are not critical
since in all functionals the neutron Fermi level at LEMAS is located at high density of the neutron 
single-particle states, which likely leads to positive neutron shell correction energies.

  The results presented in Fig.\ \ref{Dif-energy-curves} clearly indicate the stability of the 
nuclei under discussion with respect of breathing deformation in all employed functionals.
The similarity of the shell structure in all five functionals strongly suggests that the considerations
provided in Sec.\ \ref{Tori-shell-structure} on potential stability with respect of sausage 
deformations of the nuclei under study in the case of CEDF DD-PC1 are also applicable for 
the NL3*, PC-PK1,  DD-ME2 and DD-ME$\delta$ functionals.
 
\begin{figure*}[htb]
\centering
\includegraphics[angle=0,width=7.5cm]{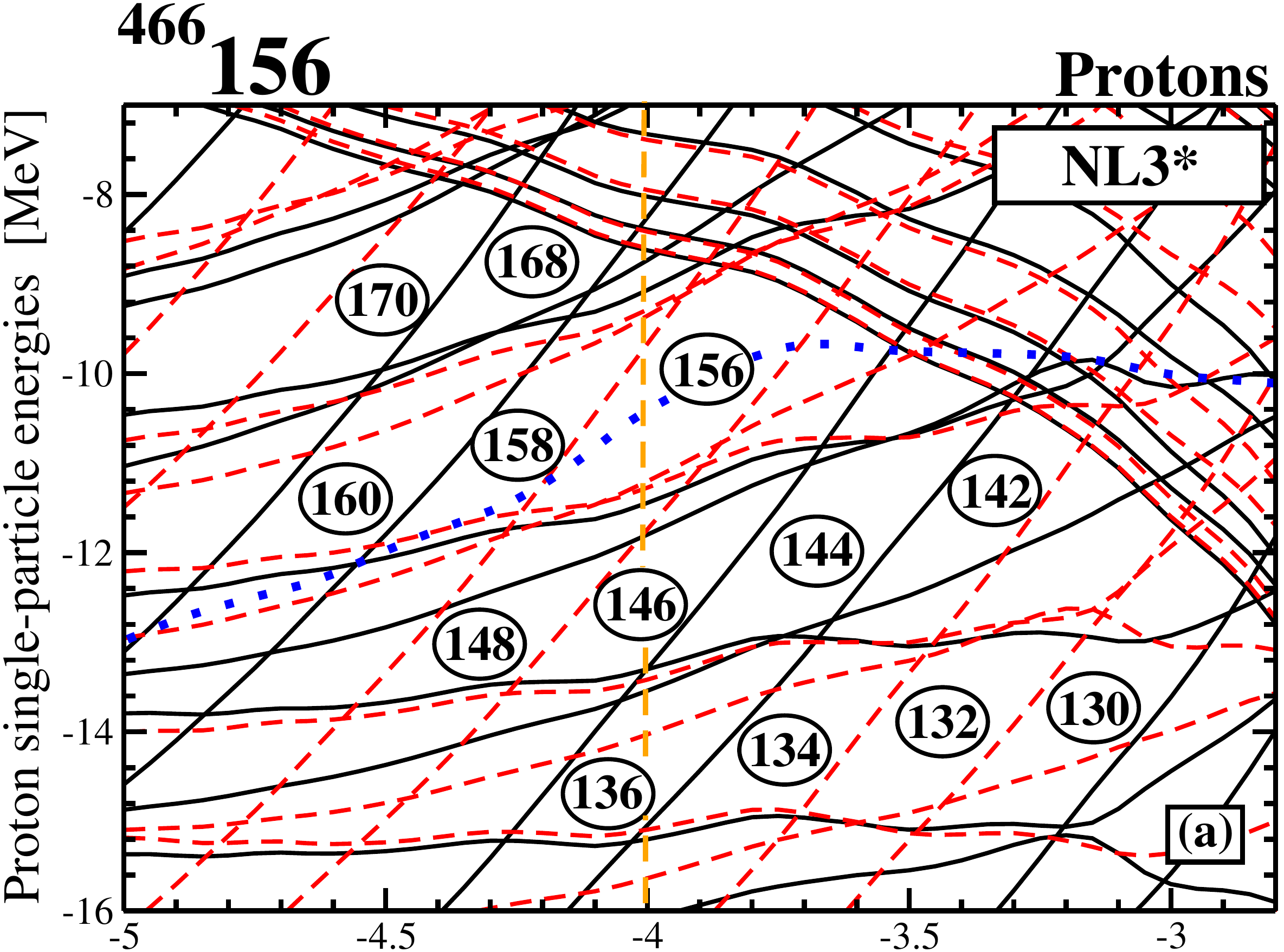}
\includegraphics[angle=0,width=7.5cm]{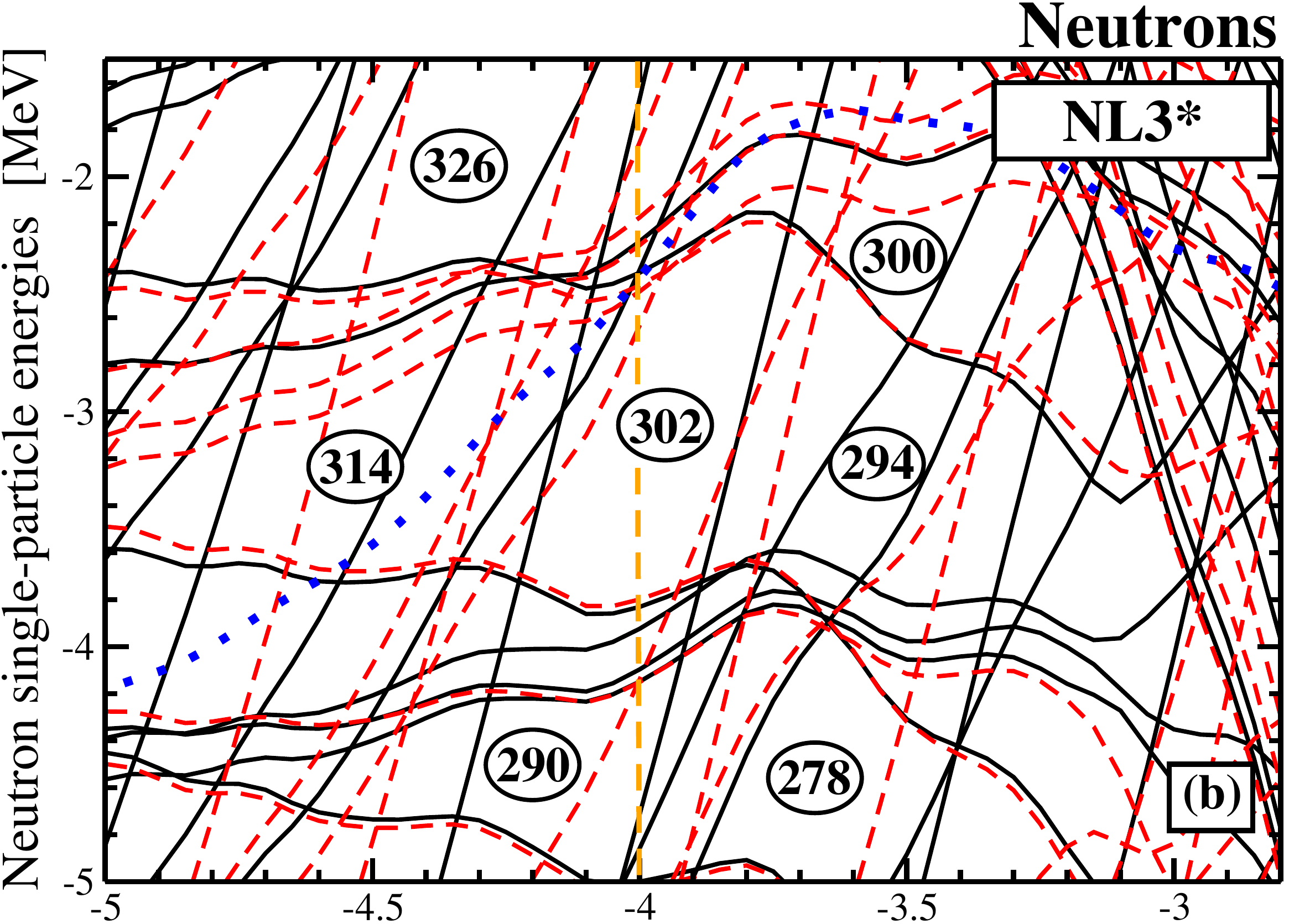}
\includegraphics[angle=0,width=7.5cm]{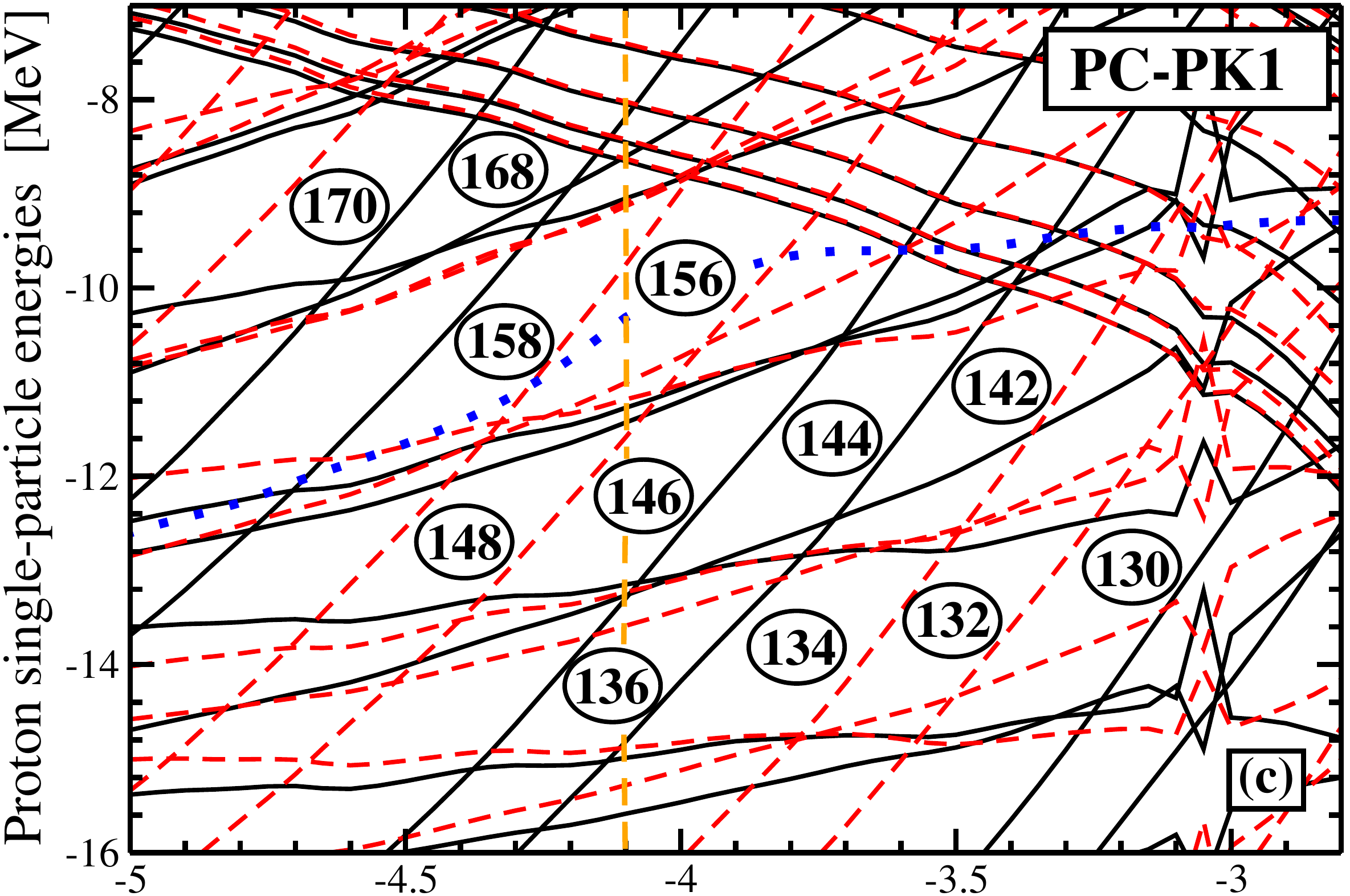}
\includegraphics[angle=0,width=7.5cm]{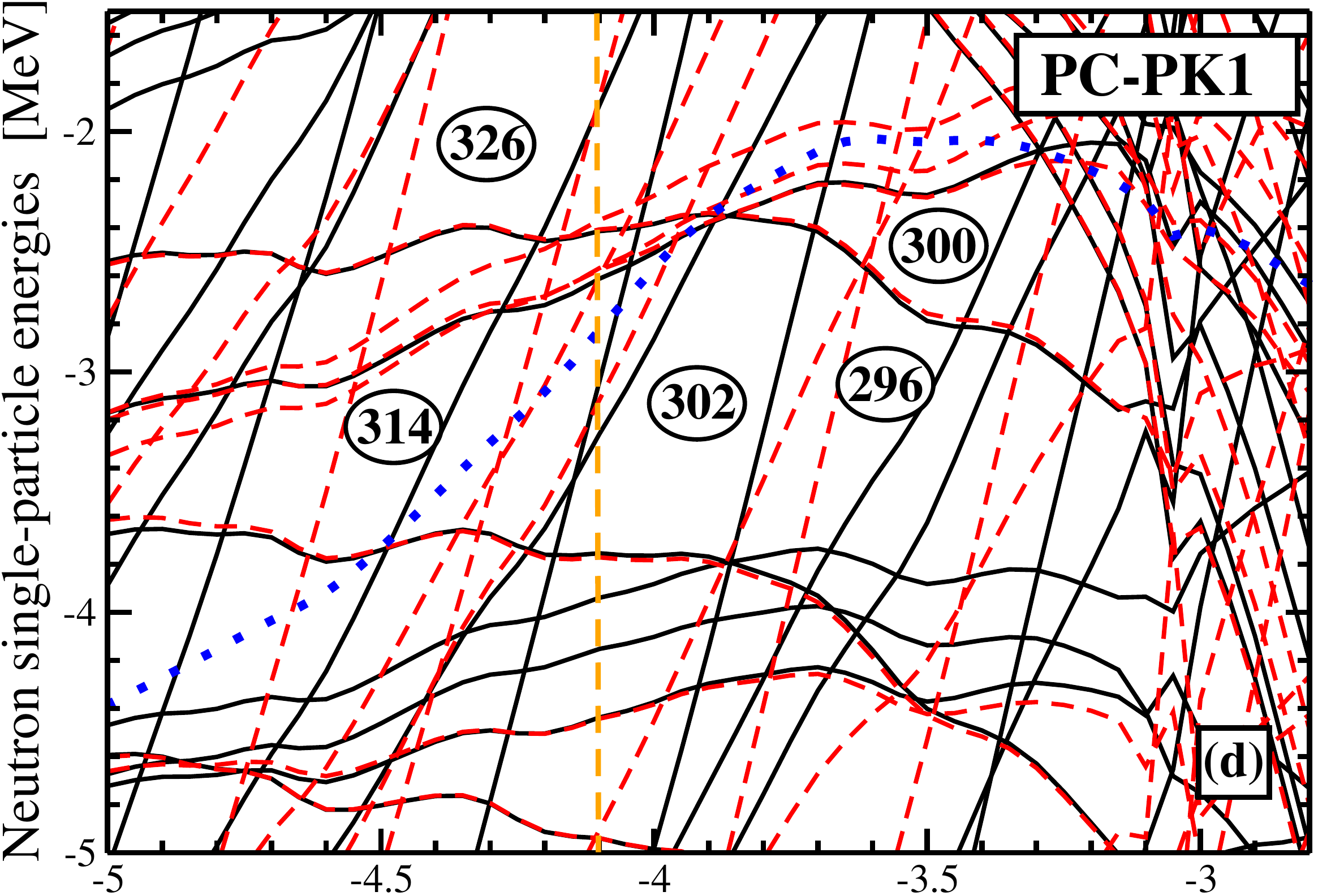}
\includegraphics[angle=0,width=7.5cm]{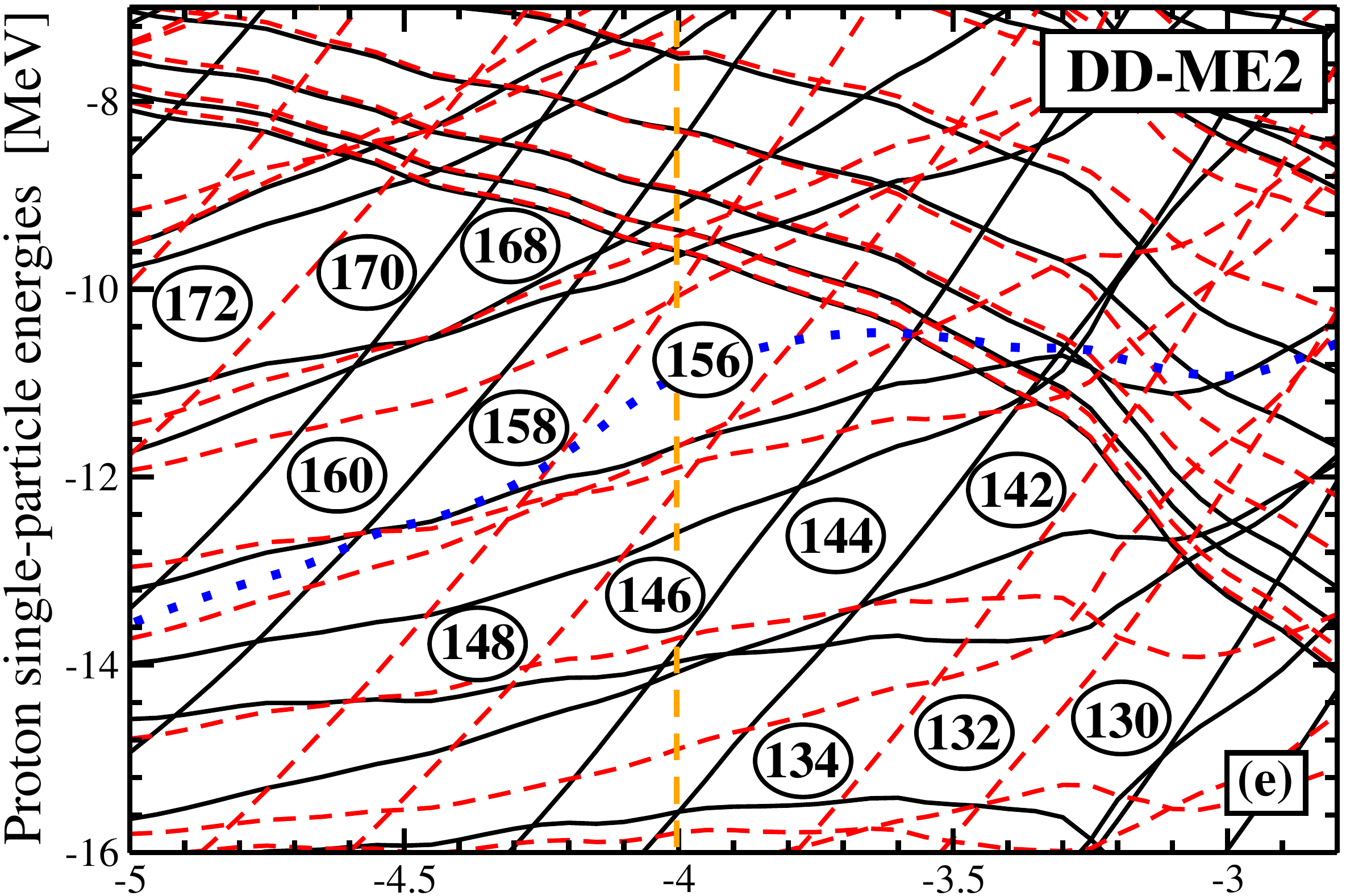}
\includegraphics[angle=0,width=7.5cm]{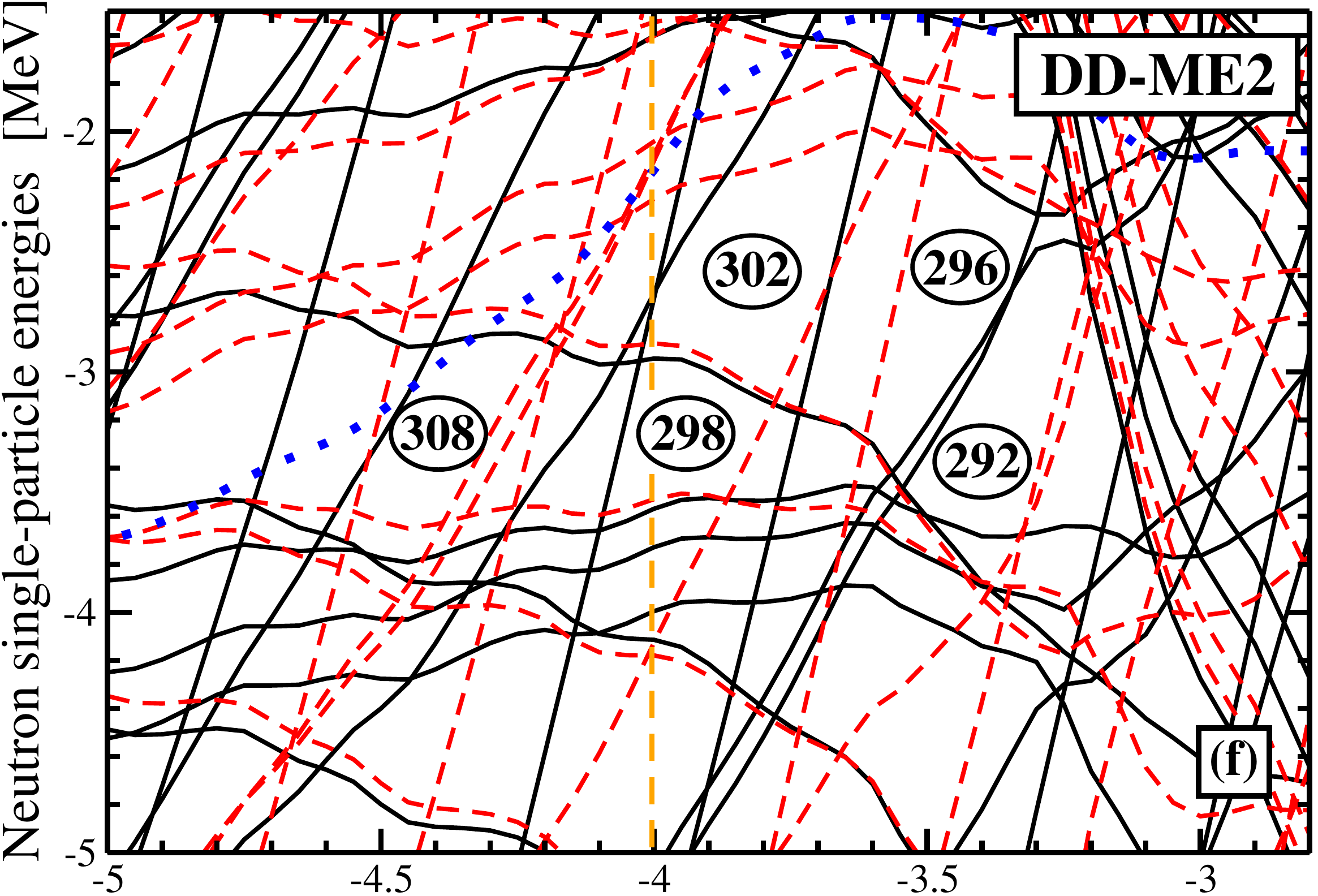}
\includegraphics[angle=0,width=7.5cm]{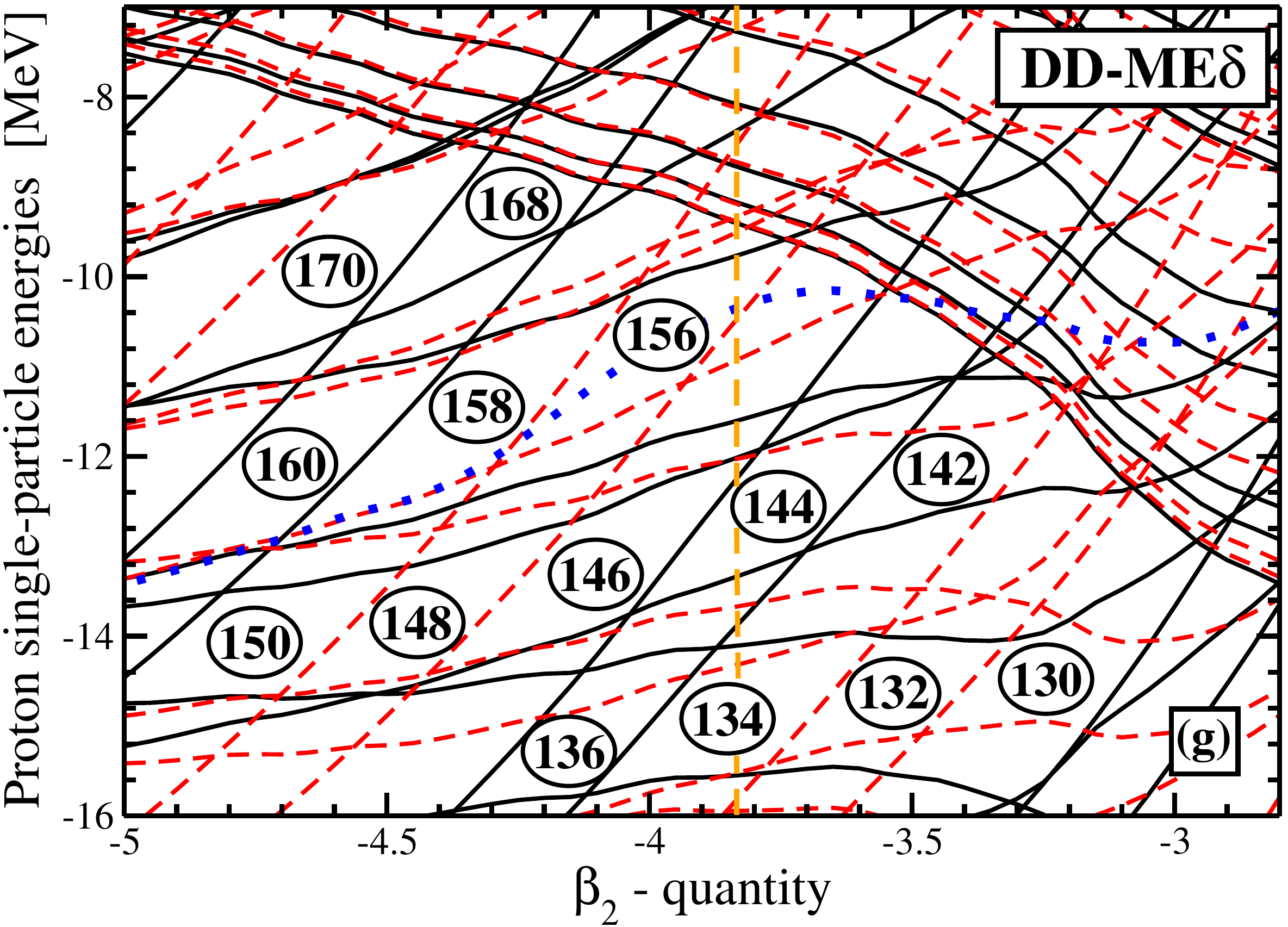}
\includegraphics[angle=0,width=7.5cm]{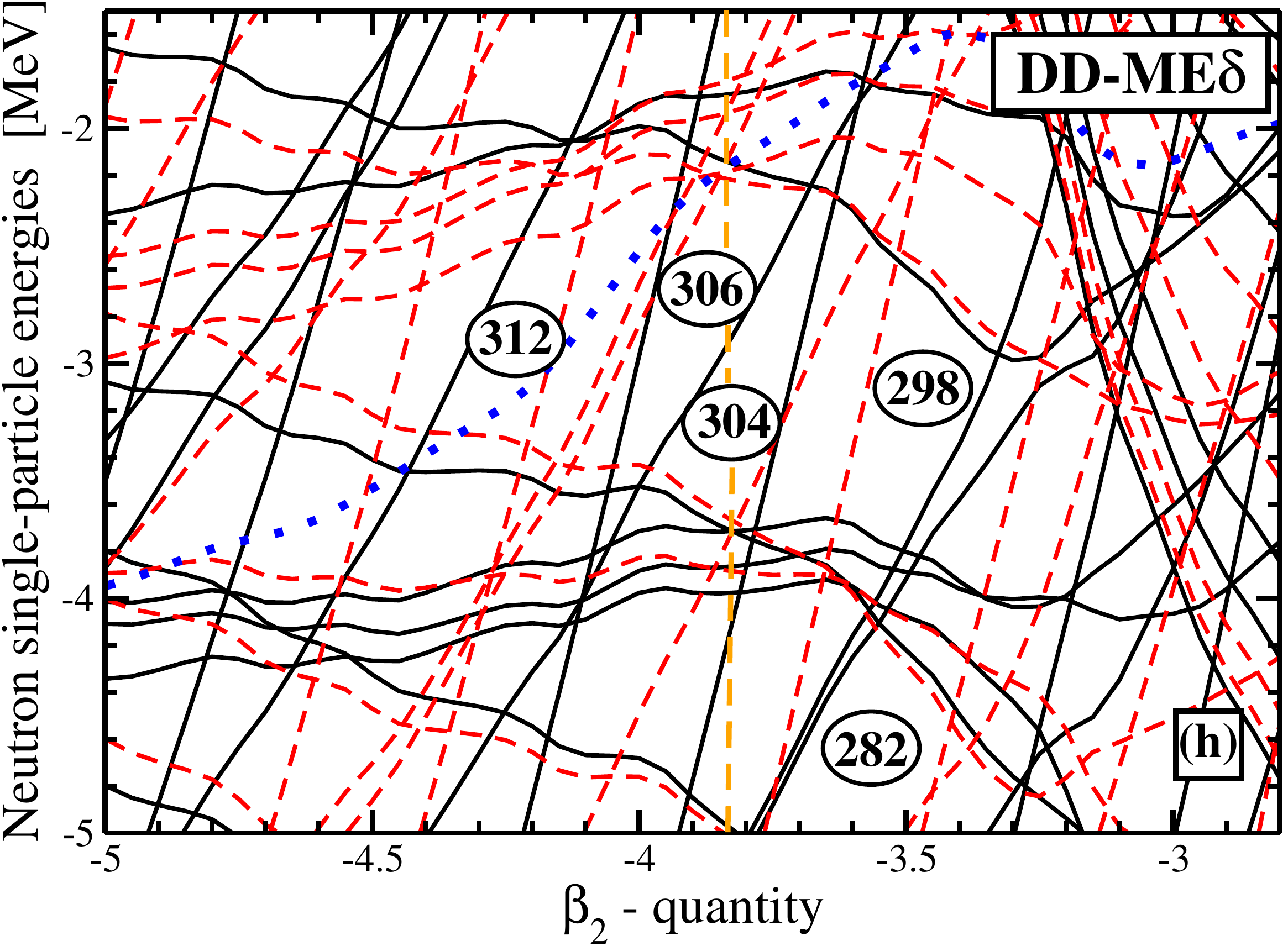}
\caption{The same as Figs.\ \protect\ref{156-466-Nil}(a) and (b) but for
the results obtained with indicated CEDFs. The vertical dashed orange
lines are drawn at the $\beta_2$ values corresponding to LEMAS.
}
\label{156-466-Nil-other-func}
\end{figure*}

\section{Conclusions}
\label{summary}

  In conclusion, the detailed investigation of the properties of spherical 
and toroidal hyperheavy even-even nuclei and their underlying shell structure have 
been performed in the framework of covariant density functional theory. 
The following conclusions have been obtained:

\begin{itemize}
      
\item 
Proton $Z=154$, 186 and neutron $N=228$, 308 and 406 spherical shell gaps exist 
in all employed CEDFs. Their combinations define the islands of stability of spherical
hyperheavy nuclei. The sizes of these gaps (both actual $\Delta E_{gap}$ and scaled  
$\Delta E_{gap} A^{1/3}$) are larger than those of $Z=120$ and $N=184$ 
in superheavy nuclei. This suggests that some spherical hyperheavy nuclei may be more
stable than superheavy ones. Systematic theoretical uncertainties in the predictions of 
the sizes of spherical shell gaps in hyperheavy nuclei are smaller than those in 
superheavy nuclei and experimentally known nuclei.

\item
Detailed calculations in extremely large basis have allowed to establish for the 
first time the general trends of the evolution of toroidal shapes in the 
$Z\approx 130-180$ region of nuclear chart. Although they have been performed 
only for selected $Z=136$, 146, 156, 166 and 
176 nuclei with the step in neutron number of $\Delta N=10$, their distribution 
in the nuclear chart between two-proton and two-neutron drip lines and 
deformation energy curves of these nuclei are such that they allow to safely extrapolate  major
conclusions to all nuclei in above mentioned region. The most compact
fat toroidal nuclei are located in the $Z \approx 136, N\approx 206$ region
(see Fig.\ \ref{torus-def}). Thin toroidal nuclei with large $R/d$ aspect ratio become
dominant with increasing proton number and on moving towards proton and neutron
drip lines.

\item
  All the nuclei in the $Z\approx 130-180$ region located between neutron and 
proton drip lines are expected to be stable with respect of breathing deformations. 
Because of numerical difficulties it is much more problematic to answer the 
question on their stability with respect of sausage deformations.
However, the analysis of theoretical and experimental studies of toroidal liquid 
droplets as well as the results on the stability of the $^{354}134$ and
$^{348}138$ nuclei with respect of even-multipole sausage deformations
obtained in Refs.\ \cite{AAG.18,AATG.19} suggest that fat toroidal nuclei located
in the $Z \approx 136, N\approx 210$ region are potentially more stable with 
respect of sausage  deformations than thin toroidal nuclei located outside of 
this region.  Nevertheless, future fully quantum mechanical calculations based 
on DFT  are needed to establish the stability of specific toroidal nuclei since the 
quantum shell effects can counterbalance the instabilities with respect of
sausage deformations \cite{Wong.73}.

\item
   Toroidal shell structure (especially the one for the shapes with large $R/d$
aspect ratio) has much more pronounced regular features as compared with
the shell structure of deformed ellipsoidal-like nuclei. Global bunching of the
pairs of almost degenerate single-particle states of opposite parities leads
to an appearance of supershell structure. These features are mostly driven 
by the existence of the two classes of the pairs of the orbitals at toroidal shapes.
The pairs of the orbitals with dominant structure of $\Omega[N,n_z,\Lambda]$ and  
$(\Omega+1) [N,n_z,\Lambda]$ with $n_z=0$ belong to the first class. The second 
class is formed by almost degenerate in energy single-particle states 
of opposite parities with dominant structures of the wave functions given by 
$\Omega[N,n_z,\Lambda]$ and  $\Omega [N',n'_z,\Lambda']$ for which the 
conditions $N'=N\pm1$, $|\Lambda'-\Lambda|=0$ or 1 and $|n'_z-n_z|=0$ or 1 are 
typically satisfied.

\item 
  As illustrated by discussed cases, at LEMAS large shell gaps and/or low density of the 
single-particle states  appear at least in one of the subsystems (proton and/or neutron)
in the vicinity of its Fermi level.
These shell gaps are also the gaps in 
breathing and sausage degrees of freedom \cite{Wong.73}.  If the Fermi level 
in a given subsystem is located in the vicinity of the large shell gap or low density 
of the single-particle states, quantum shell effects will act against the instabilities
in breathing and sausage deformations. These stabilizing effects
will be definitely enhanced if both proton and
neutron subsystems are characterized by such features.

\item
However,  the analysis of the Nilsson diagrams for all nuclei calculated in 
Fig.\ \ref{torus-def} shows that in many of these nuclei the level densities are high near the proton 
and neutron Fermi levels at LEMAS.  In reality, such a situation becomes much more frequent with 
increasing proton and neutron numbers and respective rise of the single-particle level 
densities. The reason is quite simple: the $\beta_2$ value of LEMAS is defined mostly by the
competition of nuclear surface tension and Coulomb interaction and the shell correction effects
play only a secondary role here.   As a result, such nuclei are expected to be unstable 
with respect of sausage deformations. Thus,  it is reasonable to expect the existence of the 
"continent" of stability of toroidal nuclei in low-$Z$ systems which is replaced by the "isolated islands" 
of their stability in higher-$Z$ nuclei located in the "sea of the instability".

\end{itemize}

  The problem of the stability of toroidal nuclei with respect of sausage deformations 
emerges as a major obstacle in their study. There are 
several possible ways to investigate such instabilities.  One is based on the analysis of
time evolution of the toroidal nucleus after some external disturbance of 
equilibrium shape in time-dependent Hartree-(Fock)-Bogoliubov framework
formulated in coordinate representation. However, the sizes of thin toroidal 
nuclei are significantly  larger that those of ellipsoidal ones and the tube
of the torus of such nuclei is characterized by a small radius and rapid change
of the densities.  These factors would require very large three-dimensional box with
small step in each direction. At present, it is not clear whether such calculations
are numerically feasible. 

  An alternative possibility is to rewrite existing RHB 
computer codes in the basis of toroidal harmonic oscillator potential and to study "fission"
barriers in respective sausage deformations. Since this is a native basis for toroidal 
shapes, it is reasonable to expect that sufficient numerical accuracy could be 
achieved at significantly lower size of the toroidal harmonic oscillator basis as 
compared with existing computers codes formulated in the traditional harmonic 
oscillator basis which is more suitable for ellipsoidal-like shapes.
For example,  in the latter codes the $N_F=20$ fermionic shells are sufficient for the
description of  spherical and ellipsoidal shapes in the $^{466}$156 nucleus but
$N_F=30$ is required for the description of toroidal shapes \cite{AATG.19}. The use 
of toroidal harmonic oscillator basis would reverse the situation and hopefully 
the basis with  $N_F=20$ will be sufficient for the description of toroidal shapes near 
LEMAS and  their instabilities with respect of sausage deformations. Our experience tells us that 
numerical calculations in such a basis are feasible with existing high performance 
computers.

  The instabilities of toroidal nuclei with respect of sausage deformations can 
potentially be studied by means of three-dimensitional  lattice (3D lattice) method 
suggested  in Ref.\ \cite{RZM.17}. For example, this method has been used  
for the investigation of  the stability of linear chain structure of three  $\alpha$ 
clusters in $^{12}$C against bending and fission in the framework of cranking 
CDFT in Ref.\ \cite{RZZIMM.19}.

\section{ACKNOWLEDGMENTS}

 This material is based upon work supported by the U.S. Department of Energy,  Office of Science, 
Office of Nuclear Physics under Award No. DE-SC0013037.

\bibliography{references-28-PRC-HHE-shell-structure}
\end{document}